\pdfoutput=1
\documentclass[aps,pre,twocolumn,amssymb,amsmath]{revtex4}
\usepackage{graphicx,latexsym}
\usepackage{times}
\newcommand{\rme}{{\mathrm{e}}}
\newcommand{\rmd}{{\mathrm{d}}}
\newcommand{\sym}{\ {\mbox{Sym}\ }}
\newcommand{\half}{\frac12}
\newcommand{\nn}{\nonumber}

\newcommand{\fig}[2]{\includegraphics[width=#1\columnwidth]{./figures/#2}}
\newcommand{\Fig}[1]{\includegraphics[width=\columnwidth]{./figures/#1}}
\newlength{\bilderlength}
\newcommand{\bilderscale}{0.35}
\newcommand{\storebilderscale}{\bilderscale}
\newcommand{\bilderskip}{\hspace*{0.8ex}}
\newcommand{\textdiagram}[1]{%
\renewcommand{\bilderscale}{0.25}%
\diagram{#1}\renewcommand{\bilderscale}{\storebilderscale}}
\newcommand{\diagram}[1]{%
\settowidth{\bilderlength}{\bilderskip%
\includegraphics[scale=\bilderscale]{./figures/#1}\bilderskip}%
\parbox{\bilderlength}{\bilderskip%
\includegraphics[scale=\bilderscale]{./figures/#1}\bilderskip}}

\begin{document}

\title{Size distributions of shocks and static avalanches from the Functional Renormalization Group\parbox{0mm}{\raisebox{8mm}[0mm][0mm]{\hspace*{-2cm}\normalsize \mbox{LPTENS 08/63}}}}

\author{Pierre Le Doussal and Kay J\"org Wiese}
\address{CNRS-Laboratoire de Physique Th\'eorique de l'Ecole
Normale Sup\'erieure, 24 rue Lhomond, 75231 Paris
Cedex-France
%\thanks{LPTENS is a Unit\'e Propre du C.N.R.S.
%associ\'ee \`a l'Ecole Normale Sup\'erieure et \`a l'Universit\'e Paris Sud}
}

\begin{abstract}
Interfaces pinned by quenched disorder are often used to model jerky self-organized critical motion. We study static avalanches, or shocks, defined here as jumps between distinct global minima upon changing an external field. We show how the full statistics of these jumps is encoded in the functional-renormalization-group fixed-point functions. This allows us to obtain the size distribution $P(S)$ of static avalanches in an expansion in the internal dimension $d$ of the interface. Near and above $d=4$ this yields the mean-field distribution $P(S) \sim S^{-3/2} \rme^{-S/4 S_m}$ where $S_m$ is a large-scale cutoff, in some cases calculable. Resumming all 1-loop contributions, we find
$P(S) \sim S^{-\tau} \exp\!\left(C (S/S_m)^{1/2} -
\frac{B}{4} (S/S_m)^\delta \right)$ where $B,C,\delta,\tau$ are obtained to first order in $\epsilon=4-d$. Our result is consistent to $O(\epsilon)$ with the relation $\tau=\tau_\zeta:=2-\frac{2}{d+\zeta}$, where $\zeta$ is the static roughness exponent, often conjectured to hold at depinning. Our calculation applies to all static universality classes, including random-bond, random-field and random-periodic disorder. Extended to long-range elastic systems, it yields a different size distribution for the case of contact-line elasticity, with an exponent compatible with $\tau=2-\frac{1}{d+\zeta}$ to $O(\epsilon=2-d)$. We discuss consequences for avalanches at depinning and for sandpile models, relations to Burgers turbulence and the possibility that the relation $\tau=\tau_\zeta$ be violated to higher loop order. 
Finally, we show that the avalanche-size distribution on a hyper-plane of co-dimension one is in mean-field (valid close to and above $d=4$) given by $P(S) \sim  K_{\frac{1}{3}}( S )/S $, where $K$ is the Bessel-$K$ function, thus $\tau_{\mathrm{hyper~plane}} = \frac 4 3$.
\end{abstract}

\maketitle

%\narrowtext

%\deuxcol

%\begin{widetext}

\section{Introduction}

A hallmark of complex non-linear systems, as well as systems with quenched disorder or inhomogeneities, is that the response to an applied field is very often not smooth but involves jumps, bursts or avalanches. This is true for domain walls in a magnet responding to a change in external magnetic field, leading to the Barkhausen noise \cite{UrbachMadisonMarkert1995,Repain2004}, the flux lattice in type-II superconductors upon varying the field \cite{Nori1995} the contact line of a liquid partially wetting a disordered substrate when emptying the container \cite{MoulinetGuthmannRolley2002,MoulinetRossoKrauthRolley2004}, or piles of granular material when adding grains \cite{EmigClaudinBouchaud1999}.
Other examples are motion of cracks in brittle materials, dry friction \cite{CuleHwa1998} and earthquakes \cite{FisherDahmenRamanathanBen-Zion1997,DSFisher1998}. This jerky behaviour often arises as a non-equilibrium phenomenon, but it may also occur at equilibrium, in systems with many metastable states, as they switch from one global minimum to another when an external perturbation is applied. The statistics of these jumps ubiquitously exhibits scale invariance and self-organized criticality as in sandpiles\cite{BakTangWiesenfeld1987}, with power-law tails for the probability of rare large events: if one defines an event size $S$, the  probability distribution behaves as $P(S) \sim S^{- \tau}$, up to some large-size cutoff $S_m$, e.g.\ imposed by the finite system size.

An outstanding question is the degree of universality of the jump statistics. This can in principle be answered in particular prototype models, but even then it turns out to be rather difficult to obtain analytical results. One such class of models are sandpile automata which are dynamical systems where small events can trigger large avalanches \cite{BakTangWiesenfeld1987,IvashkevichPriezzhev1998,Dhar1999,Dhar1999b}. There, some beautiful results have been obtained analytically, but the full avalanche statistics, including the distribution of avalanche sizes $P(S)$, has not yet been obtained in the cases of physical interest,  such as spatial dimensions $d=2,3$. Mean-field theories \cite{StapletonChristensen2005,DharMajumdar1990} predict $\tau=3/2$ and various scaling arguments have been constructed \cite{BakTangWiesenfeld1987,IvashkevichPriezzhev1998,Dhar1999,Dhar1999b, LeeaGohaKahngKima2004,Krug1993}, not all in mutual, agreement especially for $d=2$. Among these, the conjecture $\tau=2 - 2/d$ seems to be the best guess \cite{DharDhar1997,AgrawalDhar2001} in $d=3$, but most often one has resorted to extensive numerics.

Another class of prototype models are random field Ising magnets (RFIM) in an external field $H$. There one may study either the changes in the ground state as $H$ is varied or the non-equilibrium zero-temperature dynamical evolution from an initial state. Tuning the system near the critical values of field and disorder where the macroscopic magnetization jump vanishes, one can study finite-size avalanches, and these are found to exhibit self-organized criticality. Extensive numerical work has been carried out to determine avalanche-cluster statistics \cite{BanerjeeSantraBose1995}. In parallel, the avalanche size distribution was studied  \cite{DahmenSethna1996} using a a field theoretic RG approach in an expansion in $d=6-\epsilon$. However, since the RG used there is based on the dimensional reduction property, which is well known to fail for even simpler disordered problems, as well as for the static RFIM, the status of the result of Ref.\ \cite{DahmenSethna1996}, $\tau=3/2+O(\epsilon^2)$ remains to be clarified. Furthermore it has been recently argued, mostly on the basis of numerics, that the avalanche statistics for the RFIM ground state and for its non-equilibrium dynamics are not distinct but belong to the same universality class \cite{LiuDahmen2006}. Hence, another outstanding question is to clarify the possible differences between equilibrium and non-equilibrium avalanches in the RFIM, and in a broader class of models.

Elastic systems pinned by a quenched random substrate provide yet another frequently used model for avalanche phenomena. In the statics they are known to exhibit glassy phases where the Gibbs measure, while localized near a minimal energy configuration, does exhibit jumps at low temperature due to the presence of many low-lying metastable states. As an external force $f$ is applied, the system exhibits a depinning transition: at $T=0$ it starts moving at non-zero velocity only for $f>f_c$. The motion near the depinning threshold $f_c$ has been studied extensively numerically and is known to proceed by avalanches \cite{MiddletonFisher1993,NarayanMiddleton1994,LuebeckUsadel1997,LeDoussalMiddletonWiese2008}. These are actually not obvious to define non-ambiguously in the moving phase, hence some of the literature on the subject is qualitative. Understanding these avalanche statistics is however a challenging question, since elastic plates driven by springs exhibit stick-slip motion, which provide a first step to model complex systems such as earthquakes \citation{FisherDahmenRamanathanBen-Zion1997} In Ref.\ \cite{DSFisher1998} a very simple model for avalanches, mean-field in inspiration, was proposed and solved, and yields again the value $\tau=3/2$. Surprisingly, there is to this day no first-principle derivation of even this simple mean-field behaviour, within this class of models. Besides extensive numerics, the main theoretical result is a conjecture for the size exponent near depinning, equivalent to $\tau=\tau_\zeta=2-2/(d+\zeta)$, based on a scaling argument and some unproven assumptions \cite{NarayanDSFisher1993a,ZapperiCizeauDurinStanley1998}. Thus there is the  need for an analytic tool to approach this problem. The Functional RG (FRG), a powerful field-theoretic method to deal with disordered elastic systems in the statics \cite{Fisher1985b,DSFisher1986,BalentsDSFisher1993,ChauveLeDoussalWiese2000a,ScheidlDincer2000,LeDoussalWiese2001,ChauveLeDoussal2001,LeDoussalWieseChauve2003,LeDoussalWiese2003b,LeDoussalWiese2004a,LeDoussalWiese2005a,BalentsLeDoussal2002,BalentsLeDoussal2003,BalentsLeDoussal2004,FedorenkoLeDoussalWiese2006b,Wiese2004}
and in the driven dynamics
\cite{NattermanStepanowTangLeschhorn1992,NarayanDSFisher1992b,NarayanDSFisher1993a,ChauveGiamarchiLeDoussal1998,ChauveGiamarchiLeDoussal2000,LeDoussalWieseChauve2002,LeDoussalWiese2002a,LeDoussalWiese2003a,LeDoussalWieseRaphaelGolestanian2004,FedorenkoLeDoussalWiese2006}, as well as with random field models
\cite{Feldman2000,Feldman2001,Feldman2002,TarjusTissier2004,LeDoussalWiese2005b,LeDoussalWiese2006b,TarjusTissier2005,TarjusTissier2006}, has been quite succesful at computing for instance the roughness properties of the displacement field $u(x)$, in a dimensional $d=4-\epsilon$ expansion. Following the procedure proposed in \cite{LeDoussal2006b,LeDoussal2008long,LeDoussalWiese2006a}, the
numerical determination of the predicted FRG fixed-point functions in dimensions $d=1,2,3$ have confirmed the theory to high precision, up to two-loop accuracy both in the statics \cite{MiddletonLeDoussalWiese2006} and at quasi-static depinning
\cite{RossoLeDoussalWiese2006a}. This was an important test, given that the field theory is quite non-standard: the coupling constant is a function of the field, $\Delta(u)=-R''(u)$, the renormalized second cumulant of the pinning force, and the zero-temperature effective action is non-analytic, e.g.\ $\Delta(u)$ exhibits a cusp at $u=0$. However, until now no calculation of avalanche distributions has been even attempted using FRG. In fact, possible difficulties in handling the fast-jump motion within the field theory have been emphasized \cite{NarayanDSFisher1992b}, and it was unclear whether it was possible at all.

The aim of this paper is to show that one can extract the avalanche statistics from the FRG in a controlled way. It is a priori quite involved as it requires the calculation of the non-analytic part (the cusp) of {\it all} cumulants of the renormalized disorder, which encode the jump distribution. However, it turns out to be feasible in the end, as we find that remarkably simple self-consistent equations are obeyed by suitable generating functions. As an application we compute the distribution of avalanche sizes from first principles in an expansion in $d=4-\epsilon$. We derive the mean-field result and obtain a 1-loop, i.e.\ $O(\epsilon)$ prediction for $P(S)$. Here we study the case of {\it static avalanches}, also called {\it shocks}, which are defined as jumps between distinct global minima of the energy upon changing an external field. The most convenient setting is to add an external harmonic well with a variable central  position, i.e.\ consider an elastic interface in a random potential tied to a harmonic spring. As a function of the spring position, the center of mass of the interface changes in discrete jumps. We study all static universality classes including random-bond, random-field and random-periodic disorder, as well as long-range elastic systems. We obtain not only the exponent $\tau$ but the full scaling function of the avalanche distribution. The latter is (almost fully) universal with respect to small scales, and the dependence in the large scales, i.e.\ in $S/S_m$, is calculable. A short summary of some of our main results, together with a parallel numerical study, has recently appeared \cite{LeDoussalMiddletonWiese2008} and here we provide the necessary details of the approach and of the calculation.
Although we do not directly study depinning here, there are connections between static and dynamical FRG, the main idea of the method being similar. As found in a companion study \cite{FedorenkoLeDoussalWiesePREP} of avalanches at depinning, the results of the present paper are most likely to hold also for depinning to one loop. Since for some observables, such as the roughness exponent $\zeta$, differences between statics and depinning within the FRG appear only at two loop \cite{ChauveLeDoussalWiese2000a}, the question of the difference between non-equilibrium and static avalanches remains open and requires a 2-loop study \cite{FedorenkoLeDoussalWiesePREP}. Note that to our knowledge, while avalanches near depinning have received a lot of attention, the static avalanches and shocks defined here have not. Their
connection to the Burgers equation \cite{LeDoussal2006b,LeDoussal2008long}, which provides their $d=0$ limit, helped formulate the problem in a way amenable to FRG calculations. 

Let us also point out that there are interesting relations between the depinning in the random periodic class, i.e.\ charge density waves (CDW), and sandpile automata\cite{NarayanMiddleton1994,Alava2002}. Sandpile automata are also related to spanning trees and loop erased random walks \cite{MajumdarDhar1992,DharDhar1997,AgrawalDhar2001,Priezzhev1999, PoghosyanGrigorevPriezzhevRuelle2008,MoghimiAraghiRajabpourRouhani2005,PirouxRuelle2005,Jeng2005}, and this has lead to recent FRG predictions for the latter \cite{FedorenkoLeDoussalWiese2008a}. Hence the present work is also relevant to sandpile models. An important issue is the status of the conjecture $\tau=\tau_\zeta$, which in sandpile models reads $\tau=\tau_{\zeta=0}$ consistent with a value $\zeta=0$ for CDW depinning. Of course, in the present paper the question becomes whether such a relation also exists in the statics. Finally, let us note that our results extend beyond disordered systems. Indeed in some complex dynamical systems, the randomness lies in the initial condition, and avalanches or shocks appear in the course of a non-linear deterministic evolution. This is the case for the sandpile automata (e.g.\ in the conserved energy ensemble) as well as for the problem of decaying Burgers turbulence. The latter is indeed intimately connected to the pinning problem, and to the FRG, as is recalled below. Therefore the methods introduced in this article should  find applications in a much broader class of non-linear complex systems, possibly including turbulence and spin glasses.

The outline of the article is as follows. In Section \ref{s:model} we define the model of an elastic manifold in a random potential and a harmonic well. In Section \ref{s:Observables} we give the expected scaling form for shock ditributions. In Section \ref{s:Connection with FRG functions} we detail the connection to FRG functions. In Section \ref{sec:tree} we present the tree-level calculation which leads to the mean-field result. In Sections \ref{a22} and \ref{a40} we present the 1-loop resummation and in Section \ref {s:1-loop-res} we discuss the result for $P(S)$. 
The {\em local} avalanche-size distribution is discussed in section \ref{s:spatial}. 
Extensions to non-local elasticity are given in Section \ref {s:non-local}.
More technical material is relegated to various appendices. 

Note that we will not review  manifolds in random potentials, for which we
refer to \cite{BucheliWagnerGeshkenbeinLarkinBlatter1998,NattermannScheidl2000}, nor static FRG methods for which we refer to \cite{WieseLeDoussal2006} for a pedagogical introduction and to \cite{LeDoussalWieseChauve2003,LeDoussal2008long} for details of the FRG calculations.

\section{Model}\label{s:model}
We consider an elastic interface parameterized by a one component ($N=1$) displacement field noted
$u(x) \equiv u_x$. It is subjected to a random potential $V(u,x)$ and to a harmonic well centered at $u_x=w_x$. It is defined by the Hamiltonian:
\begin{eqnarray}  \label{model0}
{\cal H}[u;w] = \frac{1}{2} \int_q g_q^{-1} |u_q-w_q|^2 + \int_x V(u_x,x)
\end{eqnarray}
where we denote $\int_x=\int d^dx$ and the Fourier transform $u_q=\int_x \rme^{i q x} u_x$ and $\int_q = \int \frac{d^dq}{(2 \pi)^d}$. A cutoff at small scale $x \sim a$ or $q \sim \Lambda$, is implicit everywhere. In most of the paper we focus on the choice of a standard local elasticity:
\begin{eqnarray}  \label{model0g}
g_q^{-1} = K q^2 + m^2
\end{eqnarray}
and often set $K=1$ for convenience. The bare disorder is assumed to be short range in internal space, statistically translationnaly invariant, with a bare second cumulant:
\begin{eqnarray}
 \overline{ V(u,x) V(u',0) }^c = \delta(x-x') R_0(u-u')
\end{eqnarray}
whose precise form is unimportant, apart from some global features which determine the universality classes, mainly (i) random bond (RB) with $R_0(u)$ a short range (SR) function, (ii) random field (RF) with $R_0(u) \sim - \sigma |u|$ a long range (LR) function, (iii) random periodic (RP), with $R_0(u)$ a periodic function of period (arbitrarily) set to unity.

To study the equilibrium statics we define the renormalized potential $\hat V$ as the free energy of the system, and the renormalized force as its functional derivative w.r.t $w_x$:
\begin{eqnarray}  \label{deffunct} 
&& \rme^{- \hat V[w]/T} = \int Du_x \rme^{- {\cal H}[u;w]/T} \label{renpot} \\
&& \hat V'_x[w] = \int_{x'} g^{-1}_{xx'} (w_{x'} - \langle u_{x'} \rangle) \\
&&  g_{xx'} = \int_{k} g_{k}\rme^{ik(x-x')} 
 \end{eqnarray}
where $\langle \ldots \rangle$ denotes thermal averages over ${\cal H}[u;w]$ in a given disorder realization and $[\ldots]$ is reserved for arguments of functionals, while $(\ldots)$ is used for arguments of functions. For further details and notations we refer to \cite{LeDoussal2006b,LeDoussal2008long,MiddletonLeDoussalWiese2006,LeDoussalWiese2006a,RossoLeDoussalWiese2006a}
where this model and the observables where introduced and studied.

In this article we are mostly interested in energy minimization as $w$ is varied in a given realization of the random potential $V(x,u)$, i.e.\ the $T=0$ problem. At $T=0$ the minimum-energy configuration is denoted $u_x[w]$; and $u_x(w)$ for $w$ uniform, thus dropping the expectation value. $\hat V[w]$ becomes the minimum energy, 
\begin{eqnarray} 
 \hat V[w] &=& \min_{u_x} {\cal H}[u;w] \ .
 \end{eqnarray}
and for a uniform $w_x=w$ one defines the ground state energy of the system per unit volume:
 \begin{eqnarray} 
\hat V(w)&: =&L^{-d} \hat V[\{w_x=w \}] 
 \end{eqnarray}
 $L^d$ being the volume of the system. Its derivative w.r.t $w$:
 \begin{eqnarray} 
\hat V'(w) &=& m^2 (w- 
u(w))\\
u(w) &=& L^{-d} \int_x u_x(w)
\end{eqnarray}
coincides with the force per unit volume exerted by the spring. We are ultimately interested in the small-mass limit $m \to 0$ where scale invariance becomes manifest. As shown previously \cite{DSFisher1986, BucheliWagnerGeshkenbeinLarkinBlatter1998, NattermannScheidl2000,Middleton1995,NohRieger2001,MiddletonLeDoussalWiese2006}, the
optimal interface is statistically self-affine with $\overline{(u(x)-u(0))^2} \sim |x|^{2\zeta}$ and a roughness exponent $\zeta$ which depends on the class of disorder, and with a $\epsilon=4-d$ expansion
\cite{ChauveLeDoussalWiese2000a}:  $\zeta=\epsilon/3$ for RF, $\zeta=0$ for RP, and $\zeta= 0.2083 \epsilon+ 0.006 86 \epsilon^2$ for RB ($\zeta=2/3$ in $d=1$). This holds for scales $L_c<L<L_m$, where $L_c$ is the Larkin length (here of the order of the microscopic cutoff) and $L_m \sim 1/m$, the large scale cutoff induced by the harmonic well. It is useful to picture the interface as a collection of $(L/L_m)^d$ regions pinned almost independently.

Note that in this article we study the static problem where the interface finds the global energy minimum for each $w$. The function $u(w)$ is then a single valued monotonically increasing function of $w$.
One can also study the quasi-static dynamic problem where $w(t)$ grows very slowly as a function of time, and the interface visits a deterministic sequence of metastable states \cite{LeDoussalWiese2006a}. In that case $u(w)$ is history dependent, although due to the no-passing theorem \cite{Middleton1992} different initial conditions converge to the same asymptotic trajectory  as $w$ is increased by an amount larger than $L_m^\zeta$ \cite{RossoLeDoussalWiese2006a}. In the $m \to 0$ limit this is a method to study avalanches near the depinning transition. It is different from the more standard method where the force is increased infinitesimally while remaining below threshold $f<f_c$, and in a sense it is a cleaner method since it produces a steady state for avalanches without getting closer to threshold, i.e.\ without changing the cutoff length which remains $L_m \sim 1/m$.
The study of this case requires dynamical FRG and will be performed in \cite{FedorenkoLeDoussalWiesePREP}.

\section{Observables}
\label{s:Observables}
We now define a few useful observables and notations, focusing on the simplest case of a parabola centered around a uniform $w_x=w$. We also discuss their expected scaling form in the limit $m \to 0$, confirmed later via the FRG analysis.

\subsection{Energy fluctuations}

The renormalized potential defined in (\ref{renpot}) is a random function of $w$ and of the bare random potential $V(u,x)$. Hence one can define its cumulants as averages over the bare random potential:
\begin{eqnarray}
 \overline{\hat V(w_1) \hat V(w_2)}^c &=& L^{-d} R(w_1-w_2) \\
 \overline{\hat V(w_1)\ldots  \hat V(w_n)}^c &=& L^{-(n-1) d} (-1)^n \hat S^{(n)}(w_1,\ldots w_n)\ .\nn
\end{eqnarray}
We denote $\bar S^{(n)}(w_1,\ldots, w_n)$ the same expectation values for non-connected averages and use that $\hat R=R$. Note that
connected averages scale with system size as indicated, while non-connected do not. In the limit $m \to 0$ one expects that they take the scaling form
$\hat S^{(n)}(w_1,\ldots, w_n) \sim m^{d -  n \theta } \hat s^{(n)}(m^\zeta w_1,\ldots, m^\zeta w_n)$
with $\theta=d-2+2 \zeta$.

\subsection{Force fluctuations}
One defines the cumulants of the renormalized pinning force:
\begin{align}
& m^4 \overline{(u(w_1)-w_1) (u(w_2)-w_2)}^c  \nn\\
&~~=  L^{-d} \Delta(w_1-w_2) \label{deltadef} \\
& { m^{2 n} \overline{(u(w_1)-w_1)\ldots (u(w_n)-w_n)}^c } \nn\\
& ~~=(-1)^n \overline{\hat V'(w_1),\ldots,  \hat V'(w_n)}^c \nn\\
& ~~= L^{-(n-1)
d} (-1)^n \hat C^{(n)}(w_1,\ldots ,w_n)  \label{chatdef}
\end{align}
and similarly with $\bar C^{(n)}$ for non-connected averages. (The same remark as above applies for the system size dependence). In the first line of Eq.\ (\ref{deltadef}) the connected
average can be replaced by a non-connected average, since $\overline{u(w_1)-w_1}=0$ (as shown by parity and statistical translational invariance of the disorder).
One has the relation $\hat C^{(n)}(w_1,\ldots, w_n) = (-1)^n \partial_{w_1}\ldots \partial_{w_n} \hat S^{(n)}(w_1,\ldots, w_n)$
and $\Delta(w)=-R''(w)$.
In the limit $m \to 0$ one expects that they take the scaling form
$\hat C^{(n)}(w_1,\ldots, w_n) \sim m^{2 n - (n-1) d -  n \zeta } \hat c^{(n)}(m^\zeta w_1,\ldots ,m^\zeta w_n)$, and
\begin{eqnarray}
\Delta(w) = A_d m^{\epsilon - 2 \zeta} \tilde \Delta(m^\zeta w) \label{fixeddelta}
\ ,
\end{eqnarray}
where a convenient choice of the constant $A_d$ is given below, see (\ref{defresc}). 

\subsection{Shock observables}

\label{sec:shocks}

Consider a uniform $w$. It is reasonable to assume (and confirmed by numerical studies\cite{MiddletonLeDoussalWiese2006,LeDoussalMiddletonWiese2008}) that $u_x(w)$ consists of smooth parts, which become constant in the scaling limit $u_x(w) \sim m^{-\zeta}$ as $m \to 0$, and jumps, also called shocks, or static avalanches, that it can be decomposed as
\begin{eqnarray}
 u_x(w) &=& \sum_i S_i^x \theta(w-w_i)  \label{shockdec} \\
  u(w) &=& \frac{1}{L^d} \sum_i S_i \theta(w-w_i) \label{shockdec2} \\
  S_i &=& \int_x S_i^x\ .
\end{eqnarray}
Here $S_i$ is the ``size'' of the shock labelled $i$, and
$\theta(x)$ the unit-step function. To each environment corresponds a unique set of
$u_x(w)$, and a unique set
$\{(w_i,S_i)\}$. From this one defines the normalized probability of an (infinite)
sequence $p(\ldots ; w_1,S_1;\ldots w_n,S_n;\ldots )$. One can also define the 1-point probability-density
\begin{equation}\label{b1}
\rho(S,w) = \overline{ \sum_i \delta(S-S_i) \delta(w-w_i) }\ ,
\end{equation}
so that the total number of shocks in an interval of size $w$ is $N_w=\int \rmd S \int_0^w \rho(S,w)$. We will
assume that $N_w $ is proportional to $w$. This is equivalent to
\begin{equation}
 \rho(S,w) = \rho_0 P(S) = \rho(S) \quad , \quad \rho_0 = \int^\infty_{S_{\mathrm{min}}} \rmd S \rho(S)
\end{equation}
where $P(S) \rmd S$ is the normalized probability that a given shock has a
size in the interval $[S,S+\rmd S]$ and $\rho_0 \rmd w$ is the average total number
of shocks in an interval $\rmd w$, i.e.\ the shock density, assumed here to be a finite number. Note that the mass $m$ provides a scale ensuring convergence at large $S$, while the notation $S_{\mathrm{min}}$ refers to the avalanche size at the small scale cutoff of the model, which in some cases will be explicitly needed to ensure convergence at small $S$ of our scaling forms, see below. The statistical translational invariance of the disorder, together with parity implies:
\begin{equation}\label{sts}
 \overline{u'(w)} \equiv \frac{1}{L^d} \overline{\sum_i S_i
\delta(w-w_i)} \equiv \frac{1}{L^d} \int_{S_{\mathrm{min}}}^\infty \rmd S S \rho(S)
= 1\ .
\end{equation}
This also gives
\begin{equation}\label{sts2}
L^{-d} \rho_0 \langle S \rangle  = 1  \ ,
\end{equation}
where here and below $\langle S^n \rangle:= \int \rmd S S^n P(S)$ are
normalized moments.

We now consider the $m \to 0$ limit. One expects, and later verifies that the
shock distribution takes the following scaling form (see \cite{NarayanMiddleton1994,LeDoussalMiddletonWiese2008}):
\begin{equation}\label{b2c}
 \rho(S)=L^d m^\rho S^{- \tau} \tilde \rho(S m^{d+\zeta})\ .
\end{equation}
where $\tilde \rho(s=0)$ is a constant and $\tilde \rho(s)$ has a fast decay to zero at large $s$, suppressing large avalanches. This form involves a priori two exponents $\rho$ and $\tau$. However if one assumes $\tau <2$, which is found to always hold here, the constraint (\ref{sts}) implies a relation between the two exponents obtained by writing, as $m \to 0$:
\begin{align}\label{b3}
&\frac{1}{L^d} \int_{S_{\mathrm{min}}}^\infty \rmd S S \rho(S) = m^{\rho - (2-\tau) (d+\zeta)}
 \int_{0}^\infty ds s s^{- \tau} \tilde \rho(s)=1 \nn \\
&  \rho = (2-\tau) (d+\zeta) \quad , \quad \int_{0}^\infty ds s^{1- \tau} \tilde \rho(s) = 1\ ,
\end{align}
It also implies a constraint on the scaling function. From now on we denote $S_m$ the scale at which the
avalanche sizes are cut by the mass $m$. It behaves at small mass as
\begin{equation} \label{b2b}
S_m = c m^{-(d+\zeta)}\ .
\end{equation}
For now it is defined up to a multiplicative constant $c$, for which we make a convenient choice below.

We must now distinguish two cases where the distribution is qualitatively different:

\medskip

\leftline{{\it (i) $\tau < 1$ }:}

In that case the total shock density
\begin{eqnarray}\label{b2h}
 \rho_0 = L^d m^{(d+\zeta)} \int_{0}^\infty ds s^{- \tau} \tilde \rho(s)
\end{eqnarray}
is given by a convergent integral at small $s=c S/S_m$ (it always converges at large $s$).
Hence the normalized avalanche size distribution $P(S)$ takes the simple scaling form
\footnote{Note however that even if the weight in probability of avalanches of the size of the UV cutoff  is negligible as $m \to 0$, these may still control some (negative) moments of the probability distribution.}:
\begin{eqnarray}\label{b2d}
 P(S)&=&S_m^{-1} p(S/S_m) \\
 p(x) &=& c^{1-\tau} x^{-\tau} \tilde \rho(c x) \frac{1}{\int_0^\infty dy y^{-\tau} \tilde \rho(y)}
\end{eqnarray}
controlled by a unique scale $S_m$. The scaling function $p(x)$ for the probability satisfies:
\begin{equation}\label{b2e}
 \int_0^\infty p(x) dx = 1  \quad , \quad  \int_0^\infty x p(x) dx =
\frac{1}{c \int_0^\infty dy y^{-\tau} \tilde \rho(y)} 
\end{equation}
It is itself a probability (therefore the notation $p(x)$).

Two known examples with $\tau<1$ are (for review see \cite{LeDoussal2008long}): (i) the Sinai model, which corresponds to the $d=0$ limit of the random field RF case, i.e. $\zeta=4/3$, and also to the decaying Burgers equation with initial uncorrelated velocities. There the scaling function $p(s)$ is known, 
with $\tau=1/2$ and $p(s) \sim s^{5/2} \exp(-\alpha s^3)$ at large $s$ (ii) $d=0$ short range random potential, also known as Kida turbulence in Burgers literature, which corresponds to $\zeta=1$. There $p(s) = \frac{1}{2} s e^{-s^2/4}$. 

\medskip

\leftline{{\it (ii) $2> \tau > 1$}: }

Here the total shock density in the small-mass limit $S_{\mathrm{min}} \ll S_m$ is controlled by small shocks:
\begin{equation} \label{b4}
 \rho_0 = \int_{S_{\mathrm{min}}}^\infty \rho(S) \rmd S \approx L^d m^\rho \frac{S_{\mathrm{min}}^{1-\tau}}{1-\tau}\ .
\end{equation}
The normalized size-distribution can then be written for $S \gg S_{\mathrm{min}}$ as:
\begin{eqnarray} \label{b5}
 P(S) &=& C \langle S \rangle S_m^{-2} (S/S_m)^{-\tau} f(S/S_m) \\
 C^{-1} &=& \int_0^\infty dx x^{1-\tau} f(x) \label{b5bis}
\end{eqnarray}
with $C f(x)=c^{2-\tau} \tilde \rho(c x)$, using (\ref{sts2}). Since $f$ is not normalizable, it is not a probability, hence the notation. The normalization integral (\ref{b5bis}) of $\int \rmd S S P(S)$ converges at small $S$ and does not depend on the small-scale cutoff \footnote{It depends on it subdominantly, i.e.\ in
$(S_{\mathrm{min}}/S_m)^{2-\tau}$.} However the integral $\int \rmd S P(S)$ is divergent if we extend the form (\ref{b5}) to small $S$. It means that
\begin{equation} \label{b5ter}
 \langle S \rangle \sim k S_m^{2-\tau} S_{\mathrm{min}}^{\tau-1}\ ,
\end{equation}
with $k=(1-\tau)/(C f(0))$. 
This is an approximate value, obtained setting $\int_{S_{\mathrm{min}}}^\infty \rmd S P(S)=1$ and extending the form (\ref{b5}) down to $S_{\mathrm{min}}$. Note that although $\langle S \rangle=\int \rmd S S P(S)$ is a convergent integral at small $S$, its actual value depends on the precise cutoff at small scale. This is because, for $\tau>1$, almost all avalanches are in size of the order of the UV cutoff $S_{\mathrm{min}}$. Despite that fact however, all moments $\left<S^p\right>$
with $p>\tau-1$ are controlled by rare avalanches of size $\sim S_m$, the large-scale cutoff. For $\tau>1$ it is only this part of the distribution, i.e. for $S \gg S_{\mathrm{min}}$, which is universal, up to a multiplicative constant. 

Finally, in the case $\tau=1$ both phenomena are present. We will not study this case here.

\section{Connection with FRG functions}
\label{s:Connection with FRG functions}

The hypothesis of a finite density of shocks implies that the functions
$\hat C^{(n)}(v_1,\ldots, v_n)$ and $\bar C^{(n)}(v_1,\ldots, v_n)$ are
continuous and have no ambiguities at coinciding points; i.e.\ choosing
a given order for its arguments one can take the limit of coinciding arguments, and the result
does not depend on the chosen order.

The derivatives of $\hat C^{(n)}(v_1,\ldots, v_n)$ however are distributions and they do contain information about shock statistics as we now show.

\subsection{Cusp}
Let us start with the second moment (\ref{deltadef}). Taking derivatives w.r.t.\ $w_1$ and $w_2$ one finds:
\begin{align}\label{b6}
&- m^{-4} L^{-d} \Delta''(w_1-w_2) = \overline{u'(w_1) u'(w_2)} + 1 \nn \\
&\qquad =
L^{-2 d} \overline{\sum_i S_i^2 \delta(w_1-w_i)} \delta(w_1-w_2) \nn \\
&\qquad\quad  +
L^{-2 d} \overline{\sum_{i \neq j} S_i S_j \delta(w_1-w_i)
\delta(w_2-w_j)} + 1\ ,
\end{align}
where we have used the decomposition in shocks (\ref{shockdec2}) and separated in the double sum the contributions from the same shock and from different shocks.

Hence the second derivative is the sum of a $\delta$ singularity and a
smooth part:
\begin{eqnarray}
 - \Delta''(w) &=& m^{4} L^{-d} \int \rmd S S^2 \rho(S) \delta(w) \nn \\
&&+ m^{4} L^{-d} \int \rmd S_1 \rmd S_2 S_1 S_2 \rho_c(S_1,S_2,w)\ , \qquad
\end{eqnarray}
where
\begin{align}\label{b7}
& \rho_c(S_1,S_2,w_1-w_2)  \nn \\
&\qquad =\overline{\sum_{i \neq j} \delta(S_1-S_i) \delta(S_2-S_j)
\delta(w_1-w_i) \delta(w_2-w_j)}^c\ .
\end{align}
Note that $\rho(S)$ and the connected joined  2-shock size-density $\rho_c(S_1,S_2)$ are proportional to the size of the system. Integrating around zero yields \footnote{We use that
$\int_{-\infty}^{\infty} \rmd u \, \Delta'' (u) =0$, to conclude that
$-\lim_{\epsilon \to 0}\int_{-\epsilon}^{\epsilon} \rmd u\,\Delta''
(u)=\lim_{\epsilon \to 0}\left[ \int_{-\infty}^{-\epsilon} \rmd u\,\Delta''
(u) + \int_{\epsilon}^{\infty} \rmd u\,\Delta''
(u)\right]= - 2\Delta' (0^{+}) $}:
\begin{equation} \label{secmom}
 - 2 \Delta'(0^+) = m^{4} L^{-d} \int \rmd S\, S^2 \rho(S) = m^4 \frac{\langle S^2 \rangle}{\langle S \rangle} 
\end{equation}
using (\ref{sts2}). This provides a rather nice interpretation of the FRG function $\Delta(w)$ in terms of shocks: the cusp gives directly the second moment of the shock size, and the remaining smooth part of the function gives the correlation of the sizes of two (distinct) shocks at different points. A non local generalization of this formula is given in (\ref{nonlocrel}).

Let us now verify the scaling for small $m$. From the above one finds:
\begin{eqnarray}\label{b8}
- 2 \Delta'(0^+) = \tilde c m^4 S_m = c \tilde c m^{\epsilon -  \zeta}
\end{eqnarray}
with $\tilde c:= c^{-1} \int_0^\infty \rmd s\, s^{2-\tau} \tilde \rho(s)
=\int_0^\infty \rmd s\, s^{2-\tau} f(s)/
\int_0^\infty \rmd s\, s^{1-\tau} f(s)$ (the latter for $\tau>1$ only). This has 
the expected fixed-point scaling
(\ref{fixeddelta}) with $- \tilde \Delta'(0^+) = c \tilde c/(2 A_d)$. Below,
we will make the convenient choice $\tilde c=2$ to fix the arbitrariness in definition of $S_m$. 

This correspondence can be extended to higher moments.
For illustration, let us consider the third moment:
\begin{align}\label{b9}
&\! \overline{u'(w_1) u'(w_2) u'(w_3)}  \nonumber \\
&= L^{-3 d} \int \rmd S\, S^3 \rho(S) \delta(w_1-w_2) \delta(w_1-w_3) \nonumber \\
& + \Big[ \delta(w_1-w_2) L^{-3 d} \overline{\sum_{i \neq j} S_i^2 S_j \delta(w_1-w_i) \delta(w_3-w_j)}\nn \\
&\qquad \qquad  + {\rm 2
perm} \Big] \nn
\\
& + L^{-3 d} \overline{\sum_{i\neq j \neq k \neq i} S_i S_j S_k \delta(w_1-w_i) \delta(w_2-w_j)
\delta(w_3-w_k)}\nn \\
\end{align}
From this expression one can  integrate $\int_0^w \rmd w_1\int_0^w \rmd w_2 \int_0^w \rmd w_3$ and obtain
\begin{eqnarray}
 \overline{[u(w)-u(0)]^3} = L^{-3 d} \int \rmd S\, S^3 \rho(S) w + O(w^2) \ . \label{3mom}
\end{eqnarray}
This can be generalized to any order, as discussed in appendix \ref{app:generalizations}.

Note that in dimension $d=0$ the field $m^2(w-u(w))$ identifies with the velocity field of a 1-dimensional fluid which evolves in time $t=m^{-2}$ according to the Burgers equation with random initial conditions. The linear cusp of the third moment, Eq.\ (\ref{3mom}) is the analogous for Burgers turbulence to the famous exact result of Kolmogorov for Navier Stokes in the inertial range, both models exhibiting similarities. We will not discuss these connections further here, see Ref.\ \cite{LeDoussal2006b,LeDoussal2008long}, but since the manifold problem is a $d>0$ generalization of the Burgers equation, we will call Kolmogorov moments moments such as (\ref{3mom}) and their generalizations.

\subsection{Kolmogorov cumulants and generating function}
\label{sec:kolmogorov}

We now generalize the results of the previous section and construct a
very useful generating function which resums all shock-size moments.

Since we are interested in the jump sizes, it is convenient to
define the Kolmogorov cumulants (by analogy with the famous K41 paper \cite{Kolmogorov1941}, as discussed above):
\begin{eqnarray} \label{Kndef}
 \overline{[u(w)-w-u(0)]^n}^c &=& m^{-2 n} L^{-(n-1) d} K^{(n)}(w) \qquad \\
 K^{(2)}(w)&=& 2 (\Delta(0)-\Delta(w))\ ,
\end{eqnarray}
where all $K^{(n)}$ have a large-$L$ limit, and $K^{(1)}=0$.  We find that they are proportional to moments of the shock-size distribution:
\begin{eqnarray}
 K^{(n)}(w) &=& m^{2 n} L^{- d} \int \rmd S\, S^{n} \rho(S) |w| ({\rm sign}\ w)^n + O(w^2) \nn \\
&=& m^{2 n} \frac{\langle S^n \rangle }{\langle S \rangle} |w| ({\rm sign}\ w)^n + O(w^2)  \label{relKmom}
\end{eqnarray}
Note that the leading small-$w$  coefficient, the cusp,  is the same for
$\overline{[u(w)-w-u(0)]^n}^c$, $\overline{[u(w)-w-u(0)]^n}$ and
$\overline{[u(w)-u(0)]^n}$. We thus study the generating function:
\begin{align}\label{b10}
& \overline{\rme^{ \lambda L^d [u(w)-w-u(0)] }} - 1 = \sum_{n=2}^\infty \frac{\lambda^n}{n!} L^{n d}
\overline{[u(w)-w-u(0)]^n} \nonumber \\
&= \sum_{n=2}^\infty \frac{\lambda^n}{n!}  \int \rmd S\, S^{n} \rho(S) |w| ({\rm
sign} w)^n + O(w^2),
\end{align}
The following generating function hence has a finite large-$L$ limit:
\begin{eqnarray}\label{gen1}
G(\lambda) &:=& L^{-d} ( \overline{\rme^{ \lambda L^d  [u(w)-w-u(0)]}} - 1 ) \nonumber \\
&=& L^{-d} \int \rho(S) (\cosh(\lambda
S)-1) |w| \nonumber \\
&& + L^{-d} \int \rho(S) (\sinh(\lambda S)- \lambda S ) w + O(w^2) \qquad
\end{eqnarray}
For positive $w>0$ this yields:
\begin{eqnarray}\label{gen2}
&&  G(\lambda) = \hat Z(\lambda) w + O(w^2) \\
&& \hat Z(\lambda) = \frac{1}{\langle S \rangle} (\langle \rme^{\lambda S}
\rangle -1 - \lambda \langle S \rangle) \label{gen2bis}
\end{eqnarray}
Hence by computing the cusp of $G(\lambda)$ one has direct access to the
characteristic function of the shock-size distribution. The formula (\ref{gen1}) is also derived in
appendix \ref{app:generalizations} by another method. In the following we reserve the notation $Z(\lambda):=\lambda + \hat Z(\lambda)$ and the notation $\tilde Z$ for the rescaled version of $Z$ (see below).

\section{Tree and improved tree calculation}
\label{sec:tree} 

We now compute the cumulants $\hat C(w_1,\ldots, w_n)$ and from them,
the Kolmogorov cumulants and the generating function $G(\lambda)$.
We use several methods which give equivalent results: a calculation using replicas, one
without replicas and a graphical representation using static or dynamic graphs.

In this Section we obtain the form of $G(\lambda)$ to lowest order in the $\epsilon$ expansion (i.e.\ zero-th order). It is essentially a tree-level calculation, although some loops can be incorporated, as we discuss. In a diagrammatic language, one  defines suitable trees where the building blocks contain loops,  which are resummed here. Since we restrict to a uniform $w$, i.e.\ to zero external momentum, all trees carry zero momentum $q=0$. This is why the calculations to this order look very similar to calculations on a $d=0$ toy model.

\subsection{Method using replicas}

The cumulants $\hat C$ are
contained in the generating functional $W[w]$ for connected correlations (for more definitions see Ref.\ \cite{LeDoussal2006b}).
We focus on a uniform $w_x=w$, in which case $W[w]=L^d W(w)$ and
\begin{eqnarray}
W(w)&=&\frac{m^2}{2 T} \sum_a w_a^2 + \frac{1}{2! T^2} \sum_{ab} R(w_{ab}) \nonumber \\
&&+ \sum_{p=3}^\infty  \frac{1}{p! T^p}
\sum_{a_1\ldots a_p} \hat S^{(p)}(w_{a_1},\ldots, w_{a_p})\ ,
\end{eqnarray}
where $w_{ab}=w_a-w_b$. It can be obtained via a Legendre transform from the effective action per unit volume, i.e.\ $\Gamma(u) = L^{-d} \Gamma[u]$ for a uniform field $u_x=u$:
\begin{eqnarray}
W(w) &+& \Gamma(u) = \frac{m^2}{T} \sum_a u_a w_a \\
\Gamma(u)&=&\frac{m^2}{2 T} \sum_a u_a^2 - \frac{1}{2! T^2} \sum_{ab} R(u_{ab}) \nonumber \\
&& - \sum_{p=3}^\infty \frac{1}{p! T^p} \sum_{a_1\ldots a_p} S^{(p)}(u_{a_1},\ldots u_{a_p})  \label{effective}
\end{eqnarray}
The $\Gamma$-cumulants are naturally computed in the field theory as the sum of all 1-particle irreducible (1PI) graphs, as in \cite{LeDoussalWieseChauve2003}. Each $S^{(p)}$ can be
computed in an expansion in powers of $R$ (at $T=0$ it is the usual loop expansion). They
have the property that to lowest order $S^{(3)} \sim R^3$ and $S^{(p)} \sim R^p$. Hence near $d=4$ where $R
\sim \epsilon$, $S^{(p)} \sim \epsilon^p$. Note that the $\hat S^{(p)}$ have a different counting, e.g.\ $S^{(3)} \sim \epsilon^2$. An important property, arising from the Legendre transform, is that $W(w)$ can be written as a sum of tree diagrams with vertices made of $\Gamma$.

The {\it improved tree approximation} consists in setting the higher cumulants of the effective action $\Gamma$ to zero,  $S^{(p)}=0$ for $p>2$,  and then perform the Legendre
transform to obtain the effective action $W(w)$ and $\hat S^{(p)}$, i.e.\ to use
\begin{eqnarray}
\Gamma_{\mathrm{tree}}(u)=\frac{m^2}{2 T} \sum_a u_a^2 - \frac{1}{2! T^2} \sum_{ab} R(u_{ab})\ .
\end{eqnarray}
It is ``improved'' as one keeps the exact two-replica part $R$, which itself has a loop expansion. For instance, from the above discussion one has $\Gamma=\Gamma_{\mathrm{tree}}+O(\epsilon^3)$.

To perform the Legendre transform one must invert the relation $m^2 w_a/T = \Gamma_{\mathrm{tree}}'(u_a)$, i.e.\ find the
function $u_a(w)$, solution of
\begin{eqnarray}
u_a(w) = w_a  + \frac{1}{T m^2} \sum_{b} R'(u_a(w)-u_b(w))\ .  \label{invert}
\end{eqnarray}
This solution can be expanded in the number of free replica sums,
\begin{equation}
u_a(w) = w_a  + \sum_{a_1} u^{(1)}_{a,a_1}(w) + \sum_{a_1, a_2} u^{(2)}_{a,a_1,a_2}(w) + \ldots  \label{exprepsum}\ ,
\end{equation}
which is also an expansion in powers of $R$.
Inserting in (\ref{invert}) generates recursion relations:
\begin{align}\label{b13}
& u^{(1)}_{a,a_1} = \frac{1}{T m^2} R'(w_{a a_1}) \\
& u^{(2)}_{a,a_1,a_2} = \frac{1}{T m^2} {\rm sym}_{a_1,a_2} R''(w_{a a_1}) (u^{(1)}_{a,a_2} - u^{(1)}_{a_1,a_2}) \\
& u^{(3)}_{a,a_1,a_2,a_3} = \frac{1}{T m^2} {\rm sym}_{a_1,a_2,a_3} \nonumber \\
&\qquad \qquad \quad \Big[ \frac{1}{2}  R'''(w_{a a_1}) (u^{(1)}_{a,a_2} -
u^{(1)}_{a_1,a_2})(u^{(1)}_{a,a_3} - u^{(1)}_{a_1,a_3})\nonumber \\
&\qquad \qquad \quad  + R''(w_{a a_1}) (u^{(2)}_{a,a_2,a_3} - u^{(2)}_{a_1,a_2,a_3})\Big]
\end{align}
This is easy to automatize using Mathematica, the combinatorics being similar to expanding $R'(\sum_p b_p x^p)$, with the additional difficulty of
attributing new labels to repeated indices. Once the terms in the expansion (\ref{exprepsum}) are known, since $m^2 u_a/T = W'(w_a)$, one obtains the
derivatives of the energy cumulants:
\begin{eqnarray}
 u^{(1)}_{a,a_1}(w) &=& \partial_{w_a} R(w_{a a_1}) \\
 u^{(n)}_{a,a_1,\ldots a_n}(w) &=& \frac{1}{(n-1)!} \partial_{w_a} \hat S^{(n+1)}(w_{a},w_{a_1},\ldots w_{a_n})\ .\qquad
\end{eqnarray}
Upon further derivation, we obtain the cumulants of the force, $\hat C^{(n)}(w_1,\ldots w_n) = (-1)^n \partial_{w_1}\ldots \partial_{w_n} \hat S^{(n)}(w_1,\ldots w_n)$ as
\begin{align}\label{b14}
& \hat C^{(n+1)}(w_a,w_{a_1},\ldots ,w_{a_n}) \nonumber \\
&  =(-1)^n (n-1)!
\partial_{w_{a_1}}\ldots \partial_{w_{a_n}} u^{(n)}_{a,a_1,\ldots a_n}(w)
\end{align}
We display the obtained explicit forms for the third and fourth cumulants:
\begin{eqnarray} \label{treecum}
 \hat S^{(3)}(w_1,w_2,w_3) &=& \frac{3}{m^2} {\rm sym}_{123} R'(w_{12}) R'(w_{13}) \\
 \hat S^{(4)}(w_1,w_2,w_3,w_4) &=& \frac{12}{m^4} {\rm sym}_{1234} R''_{12}(R'_{13}-R'_{23}) R'_{14} \nonumber \\
\end{eqnarray}
for the energy, already given in \cite{LeDoussal2006b}. We use the
shorthand notation $R_{12}=R(w_{12})$. They have a simple graphical
representation in terms of tree diagrams with $R$ vertices.
For the corresponding force cumulants this gives (with everywhere $\Delta(u)=-R''(u)$):
\begin{eqnarray}\label{b15}
 - \hat C^{(3)}(w_1,w_2,w_3) &=& \frac{6}{m^2} {\rm sym}_{123} \Delta'(w_{12}) \Delta(w_{13}) \\
 \hat C^{(4)}(w_1,w_2,w_3,w_4) &=& \frac{12}{m^4} {\rm sym}_{1234}\Big[ 2 \Delta_{12} \Delta'_{13} \Delta'_{14} \nonumber \\
&& +
\Delta_{12} \Delta'_{23} \Delta'_{14} + 2 \Delta_{12} \Delta'_{13} \Delta'_{34}\nonumber \\
&& +  \Delta_{13} \Delta_{14}
\Delta''_{12} - \Delta_{13} \Delta_{24} \Delta''_{12} \Big] \nonumber
\end{eqnarray}
with $\Delta_{12}=\Delta(w_{12})$ and so on. Note that this expression for the third force cumulant was tested against numerics near depinning in Ref.\ \cite{RossoLeDoussalWiese2006a} (within the improved tree approximation, the relation (\ref{b15}) is the same for statics and depinning with, however, {\it different} functions $\Delta(w)$ in each case).

The present method allows to compute significantly higher cumulants, using mathematica. We do not display the full form of $\hat C^{(n)}(w_1,\ldots ,w_n)$, which are very tedious, but we give the resulting Kolmogorov cumulants:
\begin{widetext}
\begin{eqnarray}
 \hat K^{(2)}(w) &=& 2 (\Delta(0)-\Delta(w)) \\
 m^2 \hat K^{(3)}(w) &=& - 12 \Delta'(w) (\Delta(0)-\Delta(w)) \\
 m^4 \hat K^{(4)}(w) &=& 120 (\Delta(0)-\Delta(w)) \Delta'(w)^2 - 48 (\Delta(0)-\Delta(w))^2 \Delta''(w) \\
 m^6 \hat K^{(5)}(w) &=& - 80 (\Delta(0)-\Delta(w)) ( 21 \Delta'(w)^3 - 24 (\Delta(0)-\Delta(w)) \Delta'(w)
\Delta''(w) + 2  (\Delta(0)-\Delta(w))^2 \Delta'''(w) ) \\
 m^8 \hat K^{(6)}(w) &=& - 480 (\Delta(0)-\Delta(w)) ( - 63 \Delta'(w)^4 + 138 (\Delta(0)-\Delta(w))
\Delta'(w)^2 \Delta''(w) \nonumber \\
&&  - 22  (\Delta(0)-\Delta(w))^2 \Delta'(w) \Delta'''(w) + (\Delta(0)-\Delta(w))^2 (-18 \Delta''(w)^2 +
(\Delta(0)-\Delta(w)) \Delta''''(w))\qquad   \label{k6}
\end{eqnarray}
\end{widetext}
We expect, from the assumption of a shock density (see the discussion above), that the $\hat C^{(n)}$ are continuous functions of their arguments. This we checked explicitly. Hence there are no ambiguities at coinciding points, i.e.\ to perform the limits one can choose any order for the arguments, with the result being independent of the chosen order.

One first checks that the values at zero vanish:
\begin{eqnarray}
\overline{(u(w)-w)^n}^c = \overline{u(0)^n}^c \propto \hat C^{(n)}(0,0,\ldots, 0) = 0
\label{valuezero}
\end{eqnarray}
Hence the distribution of the center-of-mass position in the quadratic well is Gaussian to lowest order in $\epsilon$, i.e to the improved tree approximation. To this order the calculation is the same in statics and dynamics. As we see below corrections to the Gaussian arise to one loop order (see \cite{LeDoussalWiese2003a,RossoKrauthLeDoussalVannimenusWiese2003}  for some results on the deviations of the distribution of the interface width to the Gaussian at depinning). At quasi-static depinning (\ref{valuezero}) also gives the cumulants of the distribution of the critical force \cite{LeDoussalWiese2006a}. Hence it is also Gaussian to this order. This is consistent with Ref.\ \cite{FedorenkoLeDoussalWiese2006} were deviations from Gaussian were found and computed to one loop. In fact (\ref{valuezero}) validates these calculations within the present well-controlled setting of a quadratic well.

We now expand the above result for $\hat K^{(n)}(w)$ to small argument and find:
\begin{eqnarray}
\hat K^{(n)}(w) = a_n (- \Delta'(0^+))^{n-1} w ({\rm sign} w)^n  m^{4-2 n} \label{final}
\end{eqnarray}
with
\begin{align}\label{series}
& a_2 = 2\ , \quad a_3 = 12\ , \quad a_4 = 120\ , \quad a_5 = 1680\ ,\nonumber \\
&  a_6 = 30240 \ , ...
\end{align}
Note that $a_{n+1}/a_n=4 n -2$.
Defining $b_{n}:=a_{n}/n!$, the first coefficients are
\begin{equation}\label{firstbn}
 b_2 = 1\ , \quad b_3 = 2\ , \quad b_4 = 5\ , \quad b_5 = 14\ , \quad
 b_6 = 42\ .
\end{equation}

\subsection{Method without replicas}

An equivalent method to this order is as
follows. One notices that each cumulant $\hat C^{(n)}$ is computed to
lowest non-vanishing order in $R$ (or $\Delta$). Hence it is formally equivalent to start from a replicated
action ${\cal S}$ containing $R$ {\it only} (i.e.\ a bare disorder with the
substitution $R_0 \to R$) and compute the moments of $u(w)-w$ each to
lowest order in perturbation theory. Hence the calculation of the
previous paragraph is equivalent to the following
one in dimension $d=0$: Denote $u(w)$ the minimum of the toy model:
\begin{eqnarray}\label{a54}
\frac{1}{m^2} H_{\mathrm{toy}} = \frac{1}{2} (u-w)^2 + \frac{\lambda}{m^2} V(u)\ ,
\end{eqnarray}
where $V(u)$ is a Gaussian random potential of correlator $R(u)$. The minimum satisfies:
\begin{equation}\label{a55}
u(w)=w + \frac{\lambda}{m^2} F(u(w))\ ,
\end{equation}
with $F(u)=-V'(u)$ and $\lambda$ has been introduced to count the powers in $V$.
Compute each moment defined as:
\begin{align}\label{a56}
&(-1)^n m^{-2n} L^{-(n-1) d} {\hat C}^{(n)} (w_{1},\dots ,w_{n}) \nonumber \\
& \qquad:= \overline{(u(w_1) - w_1) \ldots  (u(w_n) - w_n)}^c \nonumber \\
&  \qquad = \bigg(\frac{\lambda}{m^2}\bigg)^{\!\!n}\; \overline{F(u(w_1))\ldots  F(u(w_n))}^c
\end{align}
perturbatively in $\lambda$ (in fact in $\lambda/m^2$) to lowest non-trivial order, which is $O(\lambda^{2n -2})$. The factor $L^{-(n-1) d}$ can be omitted since  $d=0$. One uses iteratively (\ref{a55}) and the Wick theorem with $\overline{F(w) F(w')}=\Delta(w-w')$ with all other cumulants of $F(w)$ set to zero (i.e.\ a Gaussian $F$). Using Mathematica we have reproduced most of the results of the previous Section.

Note that this is different from the standard perturbative expansion of the toy model which yields dimensional reduction, i.e.\ a trivial perturbation expansion involving only $\Delta(0)=-R''(0)$. The difference is that one computes here cumulants $\hat C^{(n)}$ at {\em different} points and only at the end the limit of coinciding arguments is taken,  using a {\em non-analytic} $\Delta(u)$ with a cusp. Nevertheless  ambiguous terms at intermediate stages may be generated. To the lowest order studied here, those represent possible contributions to $\bar C^{(n)}$ which depend on a smaller number of points and they cancel in the calculation of the connected correlations $\hat C^{(n)}$.  

%The calculation is done in stat-avalanche-F.nb.

\subsection{Graphical representation}
\label{sec:graphrep}

It is useful for the following to give a graphical representation of the results of the two previous sections. Define:
\begin{align}
& {\cal C}^{(n)} (w_{1},\dots ,w_{n}) = (-1)^n m^{-2 n} {\hat C}^{(n)} (w_{1},\dots ,w_{n}) \nonumber  \\
& = L^{(n-1) d} \, \overline{(u(w_1) - w_1) \ldots  (u(w_n) - w_n)}^c
\end{align}
For instance we write the 2-point correlation as
\begin{equation}\label{a57}
{\cal C}^{(2)} (w_1,w_2) =  \textdiagram{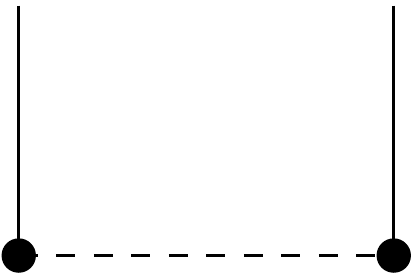} = \frac{1}{m^4} \Delta (w_{1}-w_{2})\ .
\end{equation} 
The graphical notation here and in all diagrams of this type are as follows: there are external legs with points on the top labeled by integers 1 to $n$, corresponding to positions $w_{1}$ to $w_{n}$ and external fields $u(w_i)-w_i$, here $n=2$. The $\Delta$ (or equivalently $R$) vertices are double vertices (non-local in $w$) with two points joined by a dotted line, and can be interpreted equivalently as in the statics or in the dynamics (we have checked equivalence to the order we are working). In the statics, they are $R$ vertices, giving $-R''$ after contraction with the $u$-fields. In the dynamic formulation, they are $\Delta$ vertices and the two lines exiting a vertex are directed to the top, and  end up being equal to a static propagator: in the real dynamics they are response function, which usually are denoted with  an arrow: we do not show the arrow here but they are always implicitly towards the top of the picture. Here they are always evaluated at zero frequency. Thus the lines are static propagators, evaluated here all at $q=0$, hence giving a factor of $1/m^2$. For a generalization to non-zero external momenta see
Appendix \ref{app:spatial}.

With these diagrammatic rules there is a single diagram to represent the third cumulant:
\begin{equation}\label{a2}
{\cal C}^{(3)} = \textdiagram{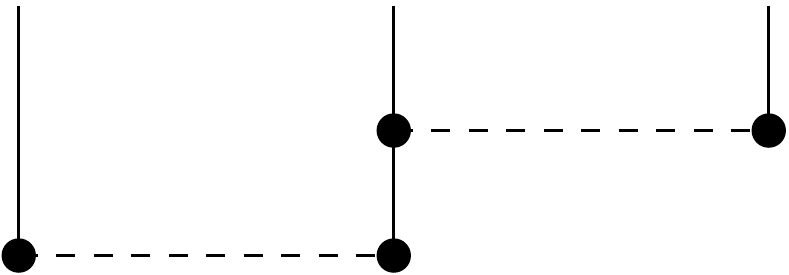} = \frac{6}{m^8} ~\mbox{Sym}~ \Delta_{12} \Delta'_{13}\ ,
\end{equation}
where the combinatoric factor comes from the $6$ inequivalent ways to assign three labels to external legs. The result agrees with (\ref{b15}). Similarly there are five diagrams for the fourth cumulant:
\begin{align}\label{a4}
& {\cal C}^{(4)} = \textdiagram{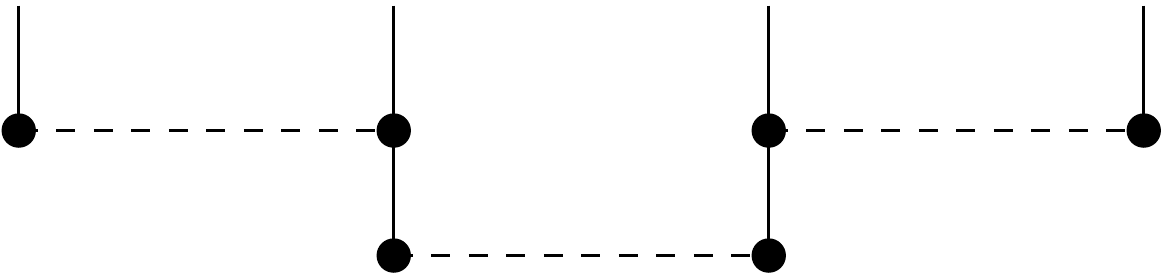} \nonumber \\
&\ \ \ \ \ \ \qquad + \textdiagram{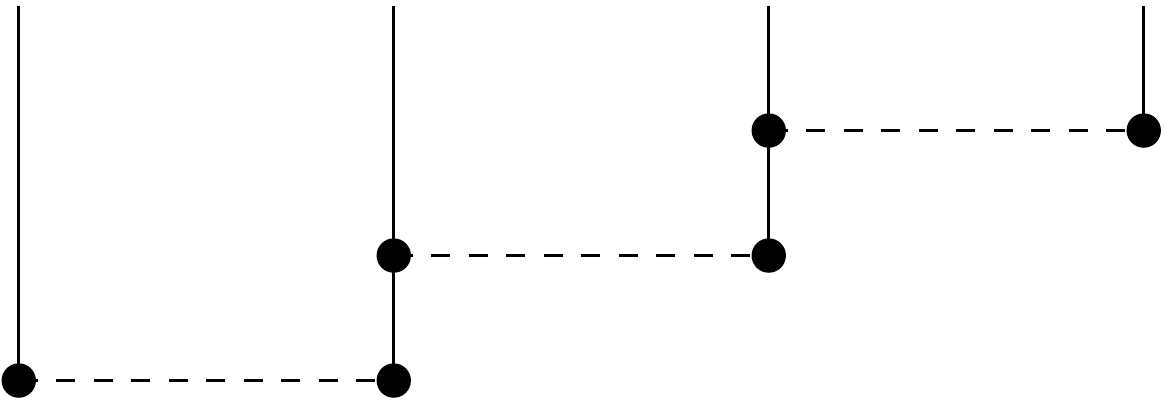}+\textdiagram{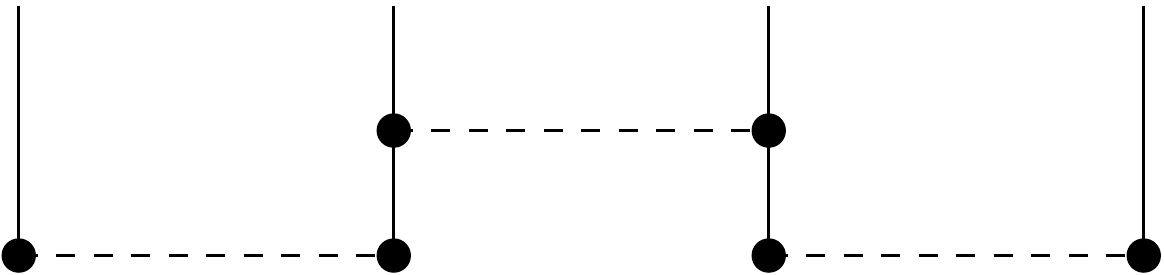}
\nonumber \\
&\ \ \ \ \ \ \qquad + \textdiagram{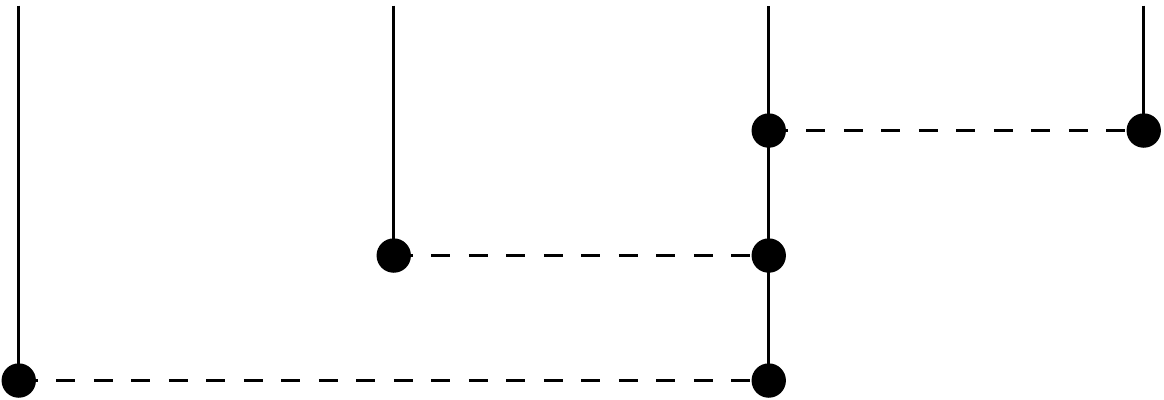}+\textdiagram{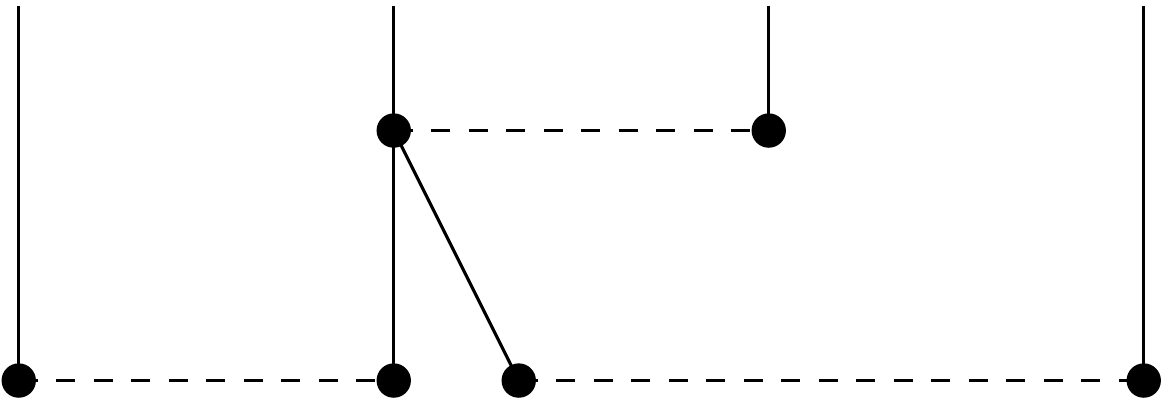}
\end{align}
which reproduce each term in (\ref{b15}), all with a factor $4!=24$ and a factor $1/2$ from the three diagrams symmetric in exchange of a pair among the four labels.

To compute the $n$-th cumulant of $(u(w) - w-u (0))$, we must evaluate
${\cal C}^{(n)}$ for $w_{i}\to w$ minus $w_{i}\to 0$, for each $i$.  Writing this
operation as ${\cal K} f(w_{1},\ldots , w_{n}):= \prod_{i=1}^{n}
\left[(w_{i}\to w )- (w_{i}\to 0)\right] f(w_{1},\ldots , w_{n})$, this
gives
\begin{equation}\label{a1}
\hat K^{(2)}(w) = {\cal K}\textdiagram{cumul2} = 2 \left[\Delta (0)-\Delta (w) \right]
\approx -2 \Delta' (0^{+}) |w|
\end{equation}
where from now on we extract the $m$ dependence from the lines of the graphs, which
are hence set to one. Note that there are four choices to assign $0$ or $w$ to each leg, hence four terms. For the third cumulant one finds:
\begin{eqnarray}\label{a3}
m^2 \hat K^{(3)}(w) &=&
{\cal K}\textdiagram{cumul3} \nonumber \\
&=& -12 \left[\Delta (0)-\Delta (w)
\right]\Delta' (w) \nonumber \\
&\approx& 12 \Delta' (0^{+})^{2} w
\end{eqnarray}
which produces $6 \times 4=24$ terms.

Similarly one finds:
\begin{align} %\label{a4old}
& m^4 \hat K^{(4)}(w) = {\cal K} [{\cal C}^{(4)}] \nn \\
&  =  120\left[\Delta (0)-\Delta (w)\right] \Delta' (w)^{2} - 48
\left[\Delta (0)-\Delta (w) \right]^{2} \Delta'' (w) \nonumber \\
& \approx 120
\Delta' (0^{+}) \,|w| \label{a4old2}
\end{align}
in terms of the five diagrams in (\ref{a4}). These results agree with those of the two previous sections. Here too the contractions are unambigous, as long as all $w_i$ are different. It would appear naively that it is equivalent to apply ${\cal K}$ to an unsymmetrized expression or to its symmetrized form. This is not true in fact because of ambiguities at coinciding points. One must be very careful to apply ${\cal K}$ to the symmetrized expression and then take the limit, i.e.
\begin{align}
& (-1)^n \hat K^{(n)}(w)  = {\cal K} \hat C^{(n)} \\
& = \lim_{\delta_i,\delta'_i \to 0} \nn\\
& \hspace{0.7cm}\prod_{i=1}^{n}
\left[(w_{i}\to w+\delta_i )- (w_{i} \to \delta'_i)\right] \hat C^{(n)}(w_{1},\ldots , w_{n}) \nonumber
\end{align}
where the $\delta_i,\delta'_i$ are taken to zero in such a manner that all arguments remain distinct and the order is fixed, the $\hat C^{(n)}$ being fully symmetric functions of their arguments. We have checked here and in the 1-loop extension given below
that the result does not depend on the order as required by the continuity of the force correlations.

\subsection{Recursion relation and resummation}

It is clear from the above results (\ref{a1})-(\ref{a4old2}) and
(\ref{k6}) that the leading term in the
expansion of $\hat K^{(n)}(w)$ in small $w$ is obtained from those terms which contain a single factor of
$\Delta (w)-\Delta (0)$, since the latter is of order $w$.
There are also terms such as $[\Delta (w)-\Delta (0)]^p$ with $p \geq 2$ in the expressions
for $\hat K^{(n)}(w)$, but they are of higher order (since e.g.\ $\Delta''(w) \to \Delta''(0^+)$ has a finite limits at $w=0^+$). Therefore,  the final result (\ref{final})
for the coefficient of the cusp depends only on the part of $\hat
C^{(n)}$ of the form $[\Delta(w)-\Delta(0)]\Delta'(w)^{n-2}$, and is the diagram
with one and only one terminating (lower) $\Delta$ vertex. This is not
the case e.g.\ for the third and fifth diagrams in (\ref{a4}) which have $p=2$.

Let us call ${\sf C}^{(n)}:=(-1)^n \hat C^{(n)}$ the part which contains only one factor of  $\Delta(w_i)$ for some $i$ (we do not write explicitly the $-\Delta(0)$, but remind that every $\Delta(w)$ comes with a $-\Delta(0)$). If we restrict to
this part we can write a recursion relation using the tree structure. Indeed, to
construct ${\sf C}^{(n)}$ we can either:

(i) take ${\sf C}^{(n-1)}$ and a single (as yet unlabeled) leg $u(w)-w$, and attach a
$\Delta$ vertex at the bottom (hence without derivatives) to each of these two elements. It results in a diagram with again only one lower $\Delta$, e.g.
\begin{equation}\label{a5}
 \textdiagram{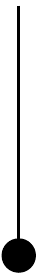}\qquad \textdiagram{cumul3} \ \longrightarrow \ \textdiagram{cumul4b}
\end{equation}
The combinatorial factor is $n$ (here $n=4$) for labeling the newly added leg.

(ii) starting from the fourth cumulant, take two already constructed cumulants
(trees) of size $l \geq 2$ and $(n-l) \geq 2$, and glue them together using again a $\Delta$ vertex at the bottom, e.g.:
\begin{equation}\label{a6}
\textdiagram{cumul2} \qquad \textdiagram{cumul2} \ \longrightarrow \ \textdiagram{cumul4a}
\end{equation}
Now the combinatorial factor is $\left({n \atop l} \right)$, for
choosing the group of indices for the first and second element, with
the restriction that $l<n-l$. If $n=l/2$, an additional factor of
$1/2$ appears. It is more systematic to sum over all pairs, i.e.\
$l=2,\dots ,n-2$, and divide this sum by 2. This can be summarized in the following schematic recursion relation for ${\sf C}^{(n)}(w_1,\ldots, w_n)$, which by construction is a  function of the set of $\Delta_{ij}=\Delta(w_i-w_j)$ and $\Delta'_{ij}=\Delta'(w_i-w_j)$:
\begin{eqnarray}
{\sf C}^{(n)} &=& \sum_{i<j} \frac{\partial {\sf C}^{(n-1)}}{\delta \Delta_{ij}} \Delta'_{ij}
(\Delta_{i,n+1}-\Delta_{j,n+1})\, n \nn  \\
&& + \sum_{l=2}^{n-2} \sum_{i<j} \frac{\partial {\sf C}^{(l)}}{\delta \Delta_{ij}} \sum_{k<m} \frac{\partial
{\sf C}^{(n-l)}}{\delta \Delta_{km}} \Delta'_{ij} \Delta'_{km}
\nonumber \\
&&\qquad \times \left(\Delta_{i,k}+\Delta_{j,m}-\Delta_{i,m}-\Delta_{j,k}\right) \, \frac{1}{2}
\left({n\atop l} \right)\qquad
\label{72}
\end{eqnarray}
We have temporarily set the mass $m=1$ to  not  burden notations.
We remark that the first term has formally the same form as the others, except that one of the derivative-terms does not
exist.  Since $n = \left({n \atop 1} \right)=\left({n \atop n-1} \right) = \frac{1}{2} \left[\left({n \atop 1}
\right)+\left({n \atop n-1} \right) \right]$, and when adding the two new terms $l=1$ and $l=n-1$,  the combinatorics is identical. We can indeed take advantage of this feature. To do so, we go to the Kolmogorov-cumulants and set as in (\ref{final}) and for $w>0$,  $m^{2n-4} \hat{K}^{(n)}(w) \approx  a_n
[- \Delta' (0^{+})]^{n-2}[ \Delta (0)-\Delta(w)]$, i.e.\ $m^{2n-4} \hat{K}^{(l) \prime}(0^+)=a_l [- \Delta' (0^{+})]^{l-1}$. We note that gluing
$\hat{K}^{(l)}$ and $\hat{K}^{(n-l)}$ yields $\hat{K}^{(l) \prime}(0^+)
 \hat{K}^{(n-l) \prime}(0^+) [ \Delta (0)-\Delta(w)]$ times the combinatoric factor
$\left({n\atop l} \right)$. Indeed one glued part should be at $w_i=w$ and the other one at $w=0$. Hence we can convert (\ref{72}) into a recursion relation for the $a_{n}$.
Inserting the ansatz (\ref{final}), the  above recursion relation
becomes:
\begin{equation}\label{a7}
 a_{n} = \sum_{l=1}^{n-1} \left({n\atop l} \right) a_l a_{n-l}\ , \qquad a_1=1\ .
\end{equation}
We note that it correctly  reproduces the series
(\ref{series}). The series $b_n: = a_n/n!$ then
satisfies the recursion relation
\begin{equation}\label{a8}
 b_{n} = \sum_{l=1}^{n-1} b_l b_{n-l}\ .
\end{equation}
Hence the generating function
\begin{equation}\label{a9}
 \tilde Z_0(\lambda) := \sum_{n=1}^\infty b_n \lambda^n
\end{equation}
satisfies the equation:
\begin{equation} \label{a10}
 \tilde Z(\lambda) = \lambda + \tilde Z(\lambda)^2\ .
\end{equation}
It has solution:
\begin{equation}\label{b16}
 \tilde Z(\lambda) = \tilde Z_0(\lambda) := \frac{1}{2} (1 - \sqrt{1 - 4 \lambda})
\end{equation}
Let us now recall that from (\ref{Kndef}) and (\ref{final}) one has at small $w$:
\begin{align} \label{b18}
& \frac{1}{n!} \overline{[u(w)-w-u(0)]^n}^c  \\
& = L^{(1-n) d}\, b_n \left[- \frac{\Delta'(0^+)}{m^4} \right]^{n-1} w ({\rm sign}(w))^{n-1} +O(w^2)  \nonumber
\end{align}
Following the discussion of Section \ref{sec:kolmogorov} we find that:

(i) the
generating function defined in (\ref{gen1}), (\ref{gen2}) for $w>0$ from
$G(\lambda) = \hat Z(\lambda) w + O(w^2)$, satisfies the scaling form
\begin{eqnarray} \label{b18a}
 \hat Z(\lambda)  = \frac{1}{S_m} \tilde Z( \lambda S_m ) - \lambda\ ,
\end{eqnarray}
where from now on we define (recall $\Delta'(0^+)<0$):
\begin{eqnarray} \label{b188a}
 S_m = \frac{|\Delta'(0^+)|}{m^4}\ .
\end{eqnarray}
Since (\ref{secmom}) is exact,  Eqs.\ (\ref{b18a}) and (\ref{b188a}) are {\it valid to all orders}, as they amount to the choice $\tilde Z( \lambda)=\lambda+\lambda^2+O(\lambda^3)$, i.e they fix the coefficient $\lambda^2$ in $\tilde Z$ to one. Scaling means that in the limit $m \to 0$  the function $\tilde Z$ becomes $m$ independent, whose validity is discussed below. In summary, using the exact relation (\ref{secmom}) the definition of $S_m$ chosen in this paper
is:
\begin{eqnarray} \label{b188a}
 S_m: = \frac{\langle S^2 \rangle}{2 \langle S \rangle} 
\end{eqnarray}

(ii) in the (improved) tree level approximation the function $\tilde Z(\lambda)$ satisfies the self-consistent equation (\ref{a10}), with solution $\tilde Z_0(\lambda)$ given in (\ref{b16}), but with $\Delta'(0)$ renormalized. Hence we find the generating function of the avalanche size distribution:
\begin{eqnarray} \label{resmeanfield}\nn
 \hat Z(\lambda) &:=& \frac{1}{\langle S \rangle} \left[\langle \rme^{\lambda S}
\rangle -1 - \lambda \langle S \rangle\right] \\
& =& \frac{1}{S_m}  \bigg[ \frac{1}{2} \left(1 - \sqrt{1 - 4  \lambda S_m }\right) - \lambda S_m \bigg]
\end{eqnarray}
with $S_m$ given by (\ref{b188a}). Note that the presence of the factor $1/\langle S \rangle$ indicates that one can only obtain information about the distribution $P(S)$ up to an unknown multiplicative factor $\left<S\right>$, consistent with the discussion in Section (\ref{sec:shocks}). We now analyze this result.

\subsection{Moments of avalanche sizes and universal ratios}

The easiest quantities to extract from (\ref{b18}) are the moments of the size distribution $P(S)$. From (\ref{resmeanfield}) one finds for $n \geq 2$:
\begin{eqnarray} \label{resmeanfield2}
 \frac{\langle S^n \rangle}{\langle S \rangle} = a_n S_m^{n-1}\ .
\end{eqnarray}
The coefficients $a_{n}$ and $b_{n}$ can be calculated explicitly for $n>1$:
\begin{equation}\label{a11}
a_{n} = \left(\frac{\partial}{\partial \lambda} \right)^{n}
\left[\frac{1}{2}\left(1-\sqrt{1-4\lambda} \right)\nonumber -\lambda \right]
\bigg|_{\lambda=0} = 2\,\frac{\left(2n-3 \right)!}{(n-2)!}
\end{equation}
\begin{equation}\label{a12}
b_{n} = \frac{a_{n}}{n!} = 2\,\frac{\left(2n-3 \right)!}{(n-2)!\, n!}
\end{equation}
Although there are some cases where the scale $S_m$ given by (\ref{b188a}) can be calculated, see below, usually it contains a non-universal amplitude. Hence it is interesting to form universal ratios independent of any scale, such as:
\begin{equation}\label{a13}
r_n :=\frac{\left< S^{n-1} \right> \left< S^{n+1} \right>}{\left<
S^{n} \right>^{2}} \ .
\end{equation}
One finds their value at tree level:
\begin{equation}\label{a13b}
r_n = 1+\frac{2}{2n-3}\ ,
\end{equation}
\begin{equation}\label{a14}
r_2 = 3 \ , \qquad r_3 =\frac{5}{3} \ , \qquad r_4 = \frac{7}{5}
\ , \qquad r_5 = \frac{9}{7} \ \dots
\end{equation}

\subsection{Distribution $P(S)$}
\label{sec:distribmf} 

Let us now perform the inverse Laplace tranform of $Z(\lambda)$ to obtain $P(S)$. One must be careful since one expects a form for $P(S)$ which is not normalizable in the absence of a small-$S$ cutoff, hence we work with $S P(S)/\langle S \rangle$ which is normalized.
From (\ref{resmeanfield}) one can write (setting momentarily $S_m \to 1$)
\begin{equation} \label{a18}
\langle S \rangle^{-1} \int_0^\infty \rmd S\, P(S) (\rme^{\lambda S}-1) = \frac{1}{2} (1- \sqrt{1-4 \lambda})\ ,
\end{equation}
where the $S$ integral converges at small $S$. This is {\it equivalent} to
\begin{eqnarray} \label{a18b}
 \langle S \rangle^{-1} \int_0^\infty\rmd S\,   S P(S) \rme^{\lambda S} = \frac{1}{\sqrt{1-4 \lambda}} \ ,
\end{eqnarray}
as one can check by integrating over $\lambda$  on both sides. Using $\int_{-i \infty}^{i \infty} \frac{\rmd \lambda }{2\pi  i}
\rme^{(s-s_{0}) \lambda} = \delta (s-s_{0})$ one has by inverse Laplace
\begin{eqnarray} \label{a18c}
 \langle S \rangle^{-1} S P(S)&=& \int_{-i \infty}^{i \infty}
\frac{\rmd \lambda }{2\pi i} \frac{\rme^{-  \lambda S}}{\sqrt{1-4\lambda }}
\\ \nn
& =& \frac{1}{\pi} \int_{1/4}^{\infty} \rmd \lambda \frac{1}{\sqrt{4\lambda -1}}
\rme^{-\lambda S} = \frac{\rme^{-S/4}}{2\sqrt{\pi}}  S^{-1/2}\ .
\end{eqnarray}
upon deforming the contour around the branch cut. Restoring all factors yields the final result:
\begin{eqnarray} \label{PStree}
P(S) = \frac{\langle S \rangle}{2 \sqrt{\pi}} S_m^{-1/2} S^{-3/2} \rme^{- S/(4 S_m) }
\end{eqnarray}
As discussed in Section (\ref{sec:shocks} ) this is expected to be valid for $S \gg S_{\mathrm{min}}$ in the limit of small $m$, large $S_m$. Note that in general the exponent $\tau$ can be extracted from the behaviour of $Z(\lambda)$ for $\lambda \to - \infty$, which is dominated by small avalanches, and reads, in the scaling regime:
\begin{eqnarray}
Z(\lambda) = \lambda + \hat Z(\lambda) \sim_{} - S_m^{\tau-2} |\lambda|^{\tau-1}  \label{asymptZ} 
\end{eqnarray}
On the other hand, from its definition (\ref{gen2}) and (\ref{b18a}) we see that $\lambda + Z(\lambda)$ must converge to $-1/\langle S
\rangle \sim - S_{\mathrm{min}}^{1-\tau} S_m^{\tau-2}$ at $\lambda \to - \infty$; thus the crossover occurs for
$\lambda \sim 1/S_{\mathrm{min}}$. This estimate is consistent with the relation (\ref{b5ter}). For larger values of $-\lambda$, $Z(\lambda)$ becomes non-universal with respect to UV
details and is out of reach of the present method.

We can now compute the moments for arbitrary real $n>1/2$ by direct integration from (\ref{PStree}). One finds:
\begin{equation}\label{a35}
\frac{\langle S^n \rangle}{\langle S \rangle} = \frac{2^{2 n-2} \Gamma \left(n-\frac{1}{2}\right)}{\sqrt{\pi }} S_m^{n-1}
\end{equation}
which agrees with the analytic continuations of the moments obtained above.

\begin{figure}[tbp]
\begin{center}
\Fig{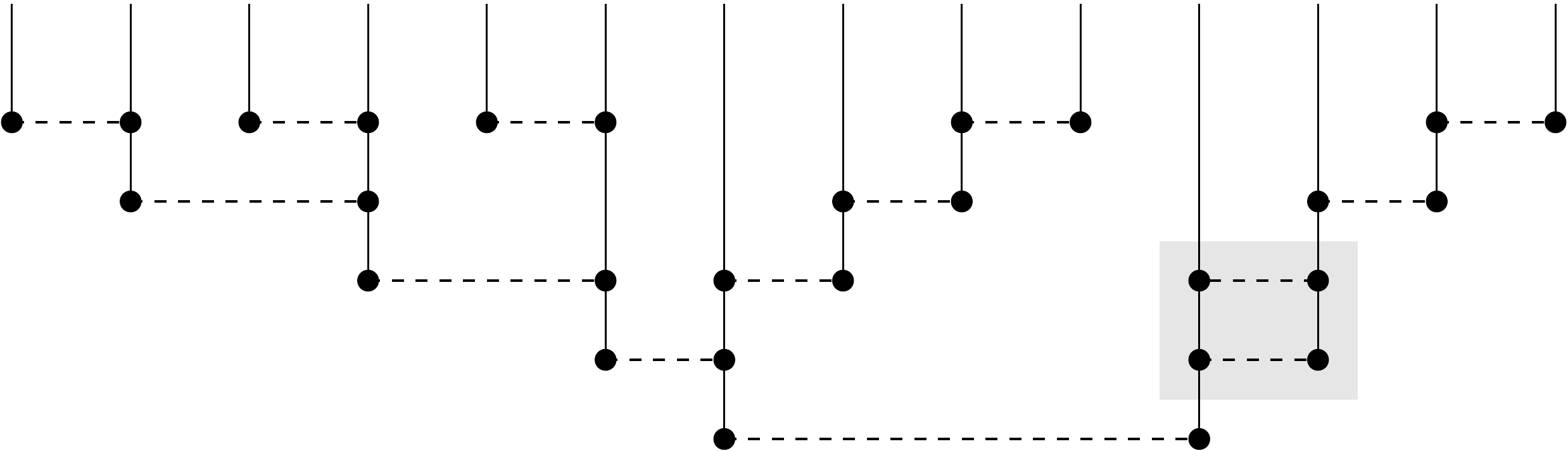}
\caption{Example of a diagram at MF level, as generated by Eq.~(\ref{selfcons}) at $\alpha=0$. It contains a correction to disorder i.e.\ $\Delta'(0^{+})$ at 1 loop (shaded in gray).}
\label{f:resum}
\end{center}
\end{figure}%

\subsection{Discussion of the result: mean-field theory}

The results (\ref{b188a}) and (\ref{PStree})  for $P(S)$ imply the value of the avalanche distribution exponent
\begin{equation}
\tau= 3/2\ .
\end{equation}
It agrees with the results obtained in: (i) a mean-field toy model for dynamic avalanches at depinning \cite{DSFisher1998}, (ii) a mean-field argument given in the context of a non-equilibrium random-field Ising model \cite{DahmenSethna1996}, (iii) mean-field calculations developped for sandpile models \cite{StapletonChristensen2005,DharMajumdar1990}. This form hence seems rather robust as a mean-field result.

Here however, it is derived from first principles using FRG from the elastic model (\ref{model0}). It was obtained by resumming all ``improved tree'' diagrams, i.e.\ trees made of fully dressed vertices $R$. In the standard bare perturbation theory it includes diagrams as represented in Figure \ref{f:resum}. Hence it can also predict the scale $S_m$ given by (\ref{b188a}).

For the elastic model, the result (\ref{b188a}), (\ref{PStree}) for $P(S)$ is valid for dimensions $d \geq 4$.
Convergence to this result as $m \to 0$ requires a 1-loop analysis, as discussed below. The only point to
discuss is thus the scale $S_m$, i.e.\ the $m$ dependence of $\Delta'(0^+)$. For that we use some of the
discussion about elastic manifolds for $d>4$ and the associated (Wilson) FRG flow, given in Appendix H of
\cite{BalentsLeDoussal2004}, and some specific results for the present model, summarized in Appendix
\ref{app:review}. The roughness exponent is $\zeta=0$, and the manifold has a finite width. If bare disorder is
smooth, it should be sufficiently strong for metastability and (typical) shocks to exist (a cusp in $\Delta(u)$
then develops). Alternatively one can consider weak rough disorder (i.e.\ with a cusp), or smooth but with a very
short correlation length. In these cases one has large avalanches with
\begin{eqnarray} \label{sm0}
 S_m = \frac{|\Delta^{* \prime}(0^+)|}{m^4}\ ,
\end{eqnarray}
where for $d>4$, $|\Delta^{* \prime}(0^+)|$ is a non-universal number, see Appendix \ref{app:review}. One can
compare this scale with the fluctuations of the center of mass, such that $L^d
\overline{u(0)^2}=\Delta^{*}(0)/m^4$, also non-universal. In $d=4$ one recovers universality and finds:
\begin{eqnarray} \label{smlarged}
 S_m \approx 8 \pi^2  |\hat \Delta^{* \prime}(0^+)| m^{-4} \left[\ln\left(\frac{m}{m_0}\right)\right]^{1-\zeta_1}
\end{eqnarray}
at small $m$, while, for comparison, $L^d \overline{u(0)^2}= \hat \Delta^{*}(0^+) m^{-4}
[\ln(\frac{m}{m_0})]^{1-2 \zeta_1}$, where $\zeta_1$ and all fixed-point values are given in appendix
\ref{app:review}.

In the end it may not be surprising that mean-field theory is obtainable as a summation of trees. 
However, this had not been done previously within the FRG. It is quite remarkable that the result for the mean-field
generating function $\tilde Z_0(\lambda)$ in (\ref{b16}) is identical to the generating function of the
number of rooted binary planar trees with $n+1$ leaves \footnote{We thank A. Fedorenko for this observation}, also known as 
 the Catalan numbers $C_n \equiv b_{n+1}$. The latter also appear for the number of rainbow-diagrams in RNA folding,  the number of Dyck words of length $2n$, the number $n + 1$ factors can be completely parenthesized,  the number of monotonic paths along the edges of a grid with $n \times n$ square cells, and many more. 
It would be interesting to obtain a more microscopic understanding of the relation between the diagrammatic
trees and the structure of the (static or dynamic) avalanche processes in $d \geq 4$.

This task may be easier to carry out for the dynamics. Intuitively an avalanche starts with a seed and each event may or may not trigger new events.  If the latter do not influence much each other, as expected in mean field, the process indeed looks like a tree. These features are captured by the simple model studied in Ref.\ \cite{DSFisher1998}. There, an
avalanche starting at time $t=0$ and lasting until $t=T$, of total size $S=\sum_{t=0}^\infty S_t$ is a
succession of jumps of size $S_t$ each being the sum of $n_t$ independent events, i.e.\ $S_t=\sum_{i=1}^{n_t}
s_i$. The number of events $n_t$ has a Poisson distribution of average $\langle n_t \rangle= \rho S_{t-1}$, i.e.
depends only on the jump at time $t-1$ (with $n_0=1$, and $S_t=0$ for $t \geq T$). If the sizes of the independent
avalanches are simply $s_i=1$ this model is the famous Galton process \cite{galton}, which describes the evolution of a population of size $S_t$ with poissonian distribution of numbers of offsprings. In this model criticality arises as $\rho$ is increased up to the threshold for an infinite avalanche. Below the threshold the distribution of the avalanche size $S$ is exactly the one obtained here by summing the trees. The $S^{-3/2}$ power law distribution may then be understood as the distribution of return time to the origin of a random walk. Interestingly, there is also a $d=0$ model which captures this mean field physics, and its connection to return time of a random walk. It is the celebrated ABBM model \cite{abbm} for domain wall motion, represented as a particle in a Brownian random force landscape. We recently computed \cite{LeDoussalWiese2008a} the avalanche statistics and the renormalized FRG force correlators for this model, and the similarities with the mean field results obtained here are striking. These analogies involve mostly the zero momentum, $q=0$ structure, but in Appendix \ref {app:spatial} we present a spatial generalization of our tree summation, 
which may help to understand the relation between trees and avalanches in mean-field.

\section{Loop corrections}\label{a22}

\subsection{General method}

Until now the force correlations $\hat C^{(n)}$, and from them the $K^{(n)}$ and the avalanche size moments
$\langle S^n \rangle/\langle S \rangle$, were computed to improved tree level, i.e.\ setting $S^{(n)}=0$ for $n
\geq 3$ in the effective action (\ref{effective}). After Legendre transform this gives each $\hat C^{(n)}$ to
the lowest order in an expansion in powers of $\Delta$ (or $R$), and later, using the fixed-point values for
$\Delta$, to lowest order in $\epsilon$.

It is useful to describe a systematic procedure to compute these quantities to higher order in the expansion in
powers of $R$ in the statics (or $\Delta$ in the dynamics). For each specific calculation, an intuitive
diagrammatic representation is given below.

(i) one first computes all the functions $S^{(n)}$, $n>3$, in the effective action in an expansion in powers of
$R$ \footnote{To be precise, this means in powers of the local part $R(u)$ of the functional $R[u]$, since
higher multilocal parts can themselves be expressed as a function of the local part $R(u)$, see e.g. \cite{LeDoussal2008long}. }.  They are
given by the sum of all 1-particle irreducible ($n$-replica) diagrams with $R$ as vertices, i.e.\ the sum of all loops. For
instance, to leading order to which we restrict below one finds $S^{(n)} \sim R^n$ with a unique 1-loop
integral $I_n$:
\begin{align}\label{b26}
&  I_n = \int_k \frac{1}{(k^2 + m^2)^n} = m^{d-2 n} \tilde I_n \\
& \tilde I_n =  \int_k \frac{1}{(k^2 + 1)^n}
\end{align}
for $d<2 n$, and only the replica combinatorics is non-trivial, e.g.\ for $n=3,4$:
\begin{eqnarray}\label{S3n}
 S^{(3)}_{123} &=& 3 I_3 {\rm sym} [({\sf R}''_{12})^2 {\sf R}''_{13} - \frac{1}{3} {\sf R}''_{12} {\sf R}''_{23} {\sf R}''_{31} ]\qquad  \\
 S^{(4)}_{1234} &=& 3 I_4 {\rm sym} [2 ({\sf R}''_{12})^2 {\sf R}''_{13} (2 {\sf R}''_{14} + {\sf R}''_{24})\nonumber \\
& &-4 {\sf R}''_{12} {\sf R}''_{13} {\sf R}''_{23} {\sf R}''_{14} + {\sf R}''_{12} {\sf R}''_{23} {\sf R}''_{34}
{\sf R}''_{14}] \qquad \label{S4n}
\end{eqnarray}
where here ${\sf R}''(w):=R''(w)-R''(0)$, $R''_{ab}=R''(w_a-w_b)$ and $S^{(3)}_{123}=S^{(3)}(w_1,w_2,w_3)$ and
so on. The formula for general $n$ was given in \cite{ChauveLeDoussal2001} and takes the form for the
$n$-replica term: 
\begin{eqnarray}\label{Smaster}
 S^{(n)}_{a_1,\ldots, a_n} &=& \frac{(n-1)!}{2} I_n \bigg[ \mbox{tr}( W^n )- \sum_{a} R''_{aa_1}... R''_{aa_n}\bigg] \nonumber \\
 W_{ab}&=&\delta_{ab} \sum_{a_1} R''_{aa_1} - R''_{ab}
\end{eqnarray}
i.e.\ it is a trace over replica indices, with the $n+1$ replica term subtracted. This formula yields (\ref{S3n}) and 
(\ref{S4n}), the explicit form for $n=5$ was given in \cite{BalentsLeDoussal2004}.

(ii) one then performs the Legendre transform (\ref{effective}) from $S^{(n)}$ to  $\hat S^{(n)}$. From
them one can then obtain $\hat C^{(n)}(w_1,\ldots, w_n) = (-1)^n \partial_{w_1}\ldots
\partial_{w_n} \hat S^{(n)}(w_1,\ldots, w_n)$. Legendre transformation simply means that each $\hat S^{(n)}$ is a sum
of all ($n$-replica) tree diagrams which can be drawn using $R$ or any of the $S^{(n)}$, $n>3$, at the vertices
of the tree. Since when forming a tree the number of replica can only increase, there are simple exact formulae
for the lowest moments:
\begin{align}
 \hat S^{(3)}(w_{123}) =& S_0^{(3)}(w_{123}) + S^{(3)}(w_{123}) \\
 \hat S^{(4)}(w_{1234}) =& S_0^{(4)}(w_{1234}) + S^{(4)}(w_{1234}) \nn\\
& + \frac{12}{m^2} {\rm sym} R'(w_{14})
\partial_{w_1} S^{(3)}(w_{123}) \label{hatS34}\ ,
\end{align}
as can be checked by explicit Legendre transform \cite{LeDoussal2006b,LeDoussal2008long}. Here the $S_0^{(n)} \sim R^{n-1}$ are
the result of the improved tree approximation, i.e.\ the trees made of only $R$ at the vertices, as described
above, giving (\ref{treecum}) for the two lowest moments. Note that $\hat S^{(3)}$ contains a single $I_3$ loop
integral while $\hat S^{(4)}$ contains both $I_3$ and $I_4$, and so on. Since everywhere we consider uniform
$w$, the trees are at zero external momenta (hence the factor $1/m^2$), and loop momenta only flow inside each
$S^{(n)}$ vertex. In principle using (\ref{hatS34}) and inserting each $S^{(n)}$ to the required order in $R$
allows to compute the $\hat C^{(n)}$ to any desired order in $\Delta$ and in $\epsilon$.

Here we only want $\hat S^{(n)}$ to order $R^n$, hence we do not need all trees made of $S^{(n)}$ and $R$, but
only those trees containing one $S^{(p)}$ and one or more $R$, i.e.\ it with schematically the form:
\begin{equation} \label{schem}
 \hat S^{(n)} = \hat S_0^{(n)} + S^{(n)} + \sum_{p=1}^{n-3} {\rm sym} S^{(n-p)} R^p\ .
\end{equation}
We first perform an explicit calculation of third and fourth moments to one loop, before attempting  the
resummation. Note that to 1-loop order, an equivalent procedure is to perform perturbation theory in the bare
disorder $\Delta_0$, in which case an additional term $I_2$ will appear in each $\hat C^{(n)}$, with trees based
on a single $\Delta_0$ loop. Inserting the 1-loop expression of $\Delta_0$ as a function of $\Delta$ this
term disappears and one recovers the same result as with the  above method.

We need below the momentum integrals $I_n$. One has $\tilde I_{n}  \sim \int_k \int_{0}^{\infty}\rmd s\, \exp (-s-s
k^{2}) s^{n-1}/ (n-1)!$, hence:
\begin{align}\label{In}
&\frac{\tilde I_{n}}{\epsilon \tilde I_{2}} = \frac{\Gamma (n-d/2)}{2 \Gamma (n)
\Gamma (3-d/2)} \nonumber \\
&\quad \stackrel{d\to 4}{\longrightarrow}\quad \frac{\Gamma (n-2)}{2\Gamma (n)}= \frac{1}{2 (n-1) (n-2)}
\end{align}
where $\epsilon \tilde I_{2}$ is discussed in Appendix \ref{app:review}. Note that all these momentum integrals are IR-finite for
$d<6$. This will remain true to any number of loop as all diagrams entering in  $\hat C^{(n)}$ are superficially
IR convergent by power counting for $d\leq 4$. If one uses perturbation theory in the bare disorder, they will
however contain diverging subdiagrams, starting at 2-loop order. These divergences should be removed by the
counterterms for the disorder, i.e.\ the replacement of $\Delta_0$ as a function of $\Delta$. Therefore,
corrections to $\tau$ should come from finite diagrams, and appear when summing {\em all} diagrams. On the other
hand, for small $n$, the moments are interesting since they can be compared to the numerics.

\subsection{Third cumulant}

From the above expressions (\ref{S3n}) and  (\ref{hatS34}),  we obtain
\begin{align}\label{b27}
& \!\!\!\!- \hat C^{(3)}_{\mathrm{1-loop}}(w_1,w_2,w_3) = \nonumber \\
& 6 I_3 {\rm sym} [ \Delta'(w_{12})^2 \Delta'(w_{13})  + (\Delta(w_{12})-\Delta(0))  \nonumber \\
& \qquad \qquad \times (\Delta'(w_{13}) \Delta''(w_{12}) +  \Delta'(w_{12}) \Delta''(w_{13}) \nn\\
& \qquad \qquad \qquad + \Delta'(w_{23})
\Delta''(w_{13})) ]
\end{align}
It is useful to indicate the graphical representation of each term, using the conventions described in
Section \ref{sec:graphrep} (suppressing for clarity below all propagator and momentum loop factors). There are three non
vanishing 1-loop contributions:
\begin{align}\label{a23}
 &\textdiagram{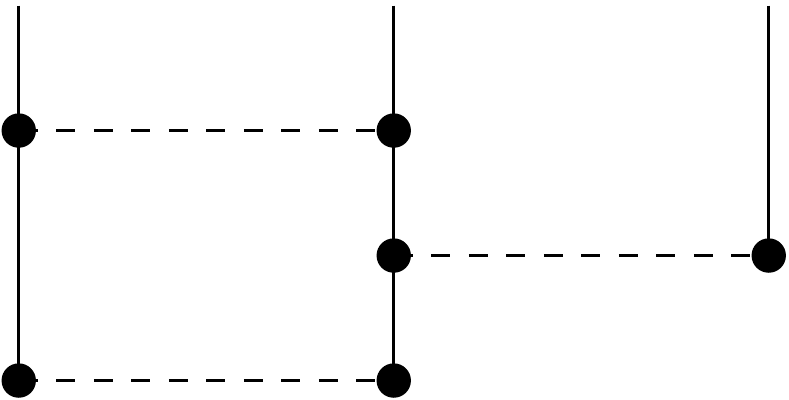}+ \textdiagram{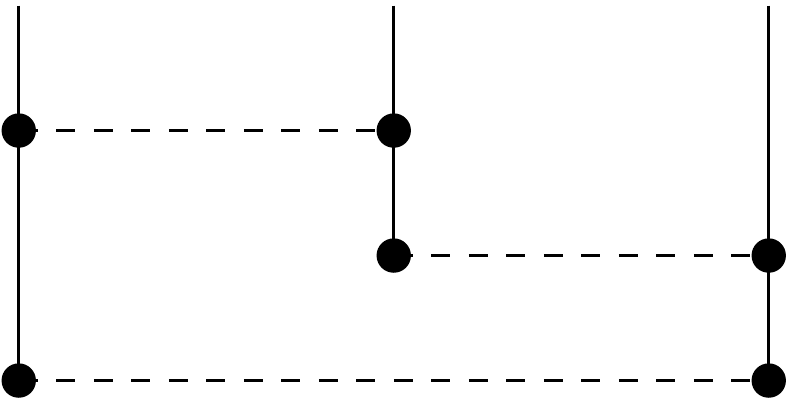} \nonumber \\
&= - 6 \sym \Delta''_{12}\Delta'_{23} \left(\Delta_{12}-\Delta_{13} \right) I_{3}
\end{align}
\begin{equation}\label{a26}
\textdiagram{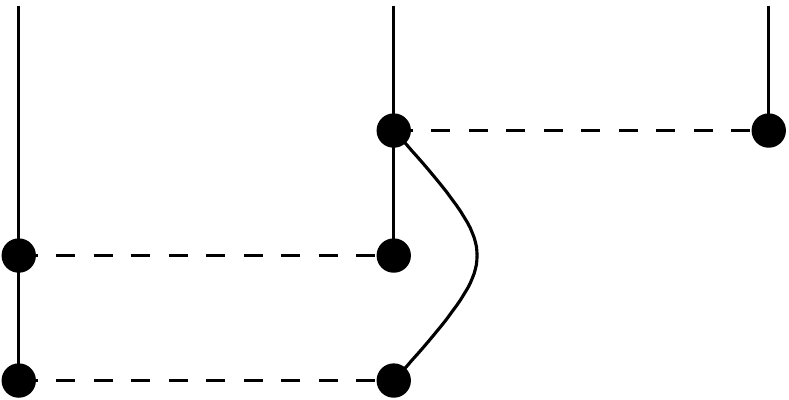}+\textdiagram{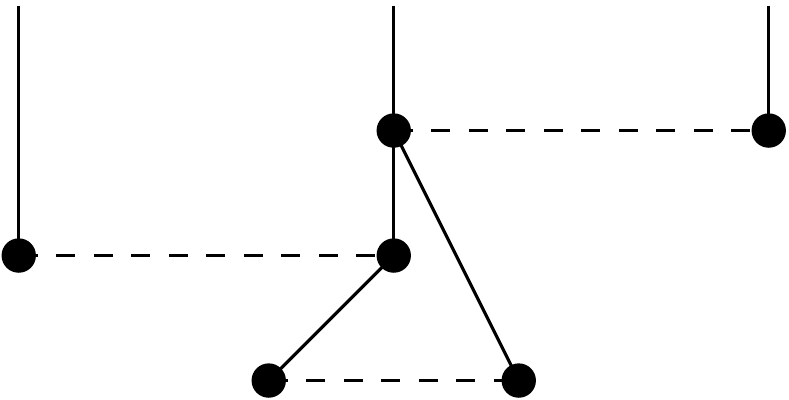} = 6\, (\Delta_{12}-\Delta_{22})\Delta'_{12}\Delta''_{23} I_{3}
\end{equation}
\begin{equation}\label{a28}
\textdiagram{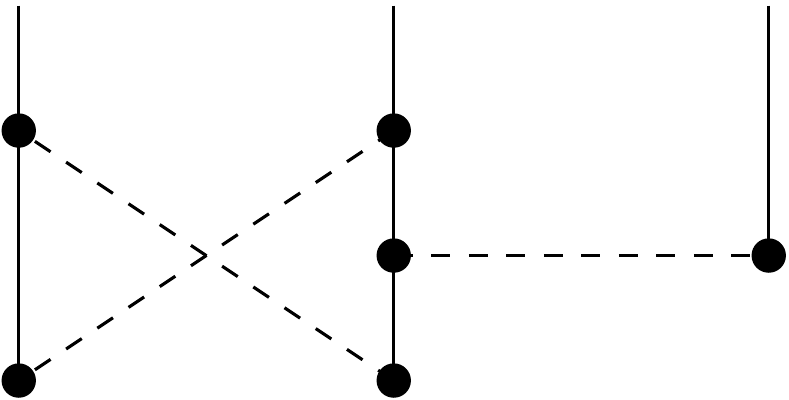} = - 6 [\Delta'_{12}]^{2} \Delta'_{23} I_{3}
\end{equation}
They reproduce the above result (\ref{b27}) after symmetrization. One
additional diagram
\begin{equation}\label{a25}
\sym \textdiagram{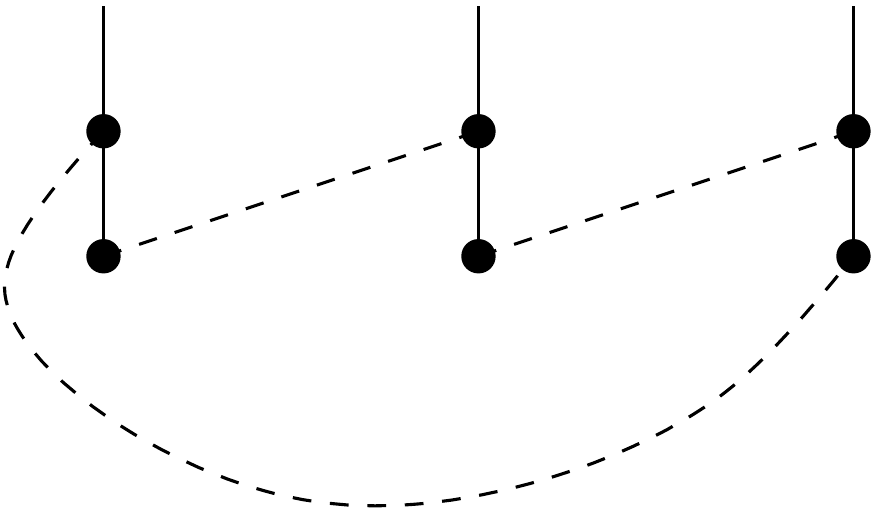} \sim \sym \Delta'_{12}\Delta'_{23}\Delta'_{31} = 0 
\end{equation}
 vanishes after symmetrization in the statics.

Note that the total result for $\hat C^{(3)}$ should be continuous and unambiguous.  
Note that $\hat C^{(3)}(0,0,0) =0$ while this is not
the case for the dynamics (it gives the third cumulant of the critical force).

From there we compute the Kolmogorov cumulant $K^{(3)}(w)$. One finds for each piece:
\begin{align}\label{a24}
&{\cal K} \left[\textdiagram{cumul3-1la}+ \textdiagram{cumul3-1lb}  \right] \nonumber \\
&= 12  I_{3} \left[\Delta (0)-\Delta (w) \right] \Delta' (w) \left[\Delta''
(w)+\Delta'' (0) \right]\nonumber \\
& \approx -24  I_{3} \Delta' (0^{+})^{2}\Delta'' (0) w
\end{align}
\begin{align}\label{a27}
&{\cal K} \left[ \textdiagram{cumul3-1ld} +\textdiagram{cumul3-1lf} \right] \nonumber \\
&= 12 I_3 (\Delta (w)-\Delta (0)) \Delta '(w) \left(\Delta ''(0)-\Delta''(w)
\right)\nonumber \\
&= O\left(w^2\right)
\end{align}
and
\begin{align}\label{a29}
&{\cal K}\ \textdiagram{cumul3-1le} \nonumber \\
&= 12 I_3 \Delta '(w) \left(\Delta
'(0^{+})^2-\Delta '(w)^2\right) \nonumber \\
&= -24 I_3 \Delta '(0^{+})^2 \Delta ''(0) w+O\left(w^2\right)
\end{align}
Together, they give
\begin{align}\label{a30}
&K^{(3)}_{\mathrm{1-loop}} = {\cal
K}\left[\textdiagram{cumul3-1la}+\textdiagram{cumul3-1lb}\right.\nonumber \\
& \ \ \ \ +
\textdiagram{cumul3-1ld}+\textdiagram{cumul3-1lf}+\left.\textdiagram{cumul3-1le}\right] \nonumber \\
&= 12 I_3 \Delta '(w) \left[\Delta '(0^{+})^2-\Delta '(w)^2-2 (\Delta (w)-\Delta
   (0)) \Delta ''(w)\right] \nonumber \\
&= -48 I_3 \Delta '(0^{+})^2 \Delta ''(0) w+O\left(w^2\right)
\end{align}
The final result for the small-$w$ behaviour of $K^{(3)}(w)$ is:
\begin{eqnarray}\label{a31c}
&& K^{(3)}(w) = K^{(3)}_{\mathrm{tree}} + K^{(3)}_{\mathrm{1-loop}} \\
&&  = \frac{12}{m^2} \Delta' (0^{+})^{2} \left[ 1- 4 m^2 I_3 \Delta'' (0^+)  \right] w + \dots
\end{eqnarray}
Hence, using (\ref{relKmom}) we obtain the third moment of avalanche sizes:
\begin{eqnarray}\label{a31d}
&& \frac{\langle S^3 \rangle}{\langle S \rangle} = \frac{12}{m^8} \Delta' (0^{+})^{2} \left[ 1- 4 m^2 I_3
\Delta'' (0^+)  \right] + O(\Delta^4) \nn \\
&& = 12 S_m^2 (1- 4 \tilde \Delta'' (0^+) \frac{\tilde I_{3}}{\epsilon \tilde I_{2}} )
\end{eqnarray}
We now use that
\begin{equation}\label{a33}
\frac{\tilde I_{3}}{\epsilon \tilde I_{2}} = \frac{1}{4} \ , \qquad \tilde \Delta'' (0^+) = \frac{\epsilon
-\zeta}{3} + O(\epsilon^{2})\ ,
\end{equation}
where the second equality is valid at the fixed point (see Appendix \ref {app:review}). Using the exact relation $\langle S^2
\rangle/\langle S \rangle = -2 \Delta' (0^{+})/m^4 = 2 S_m$, we thus obtain the universal ratio to one loop
\begin{equation}\label{a34}
r^{(2)}_{\mathrm{1-loop}} = \frac{\left< S \right>\left< S^{3} \right>}{\left< S^2 \right>^{2} } =3 \left[1-
\tilde \Delta'' (0) \right] = 3 \left[1-\frac{1}{3}\left(\epsilon -\zeta \right) \right]\ .
\end{equation}

\subsection{Fourth and higher cumulants }

Similar calculations, either directly from (\ref{Smaster}), or by calculating the sum of all diagrams at 1-loop order, we find
the corrections for the fourth Kolmogorov cumulant
\begin{align}\label{a36}
&K^{(4)}_{\mathrm{1-loop}}(w) =\nonumber \\
& 12 \Big[-6 \Delta '(w)^2 \left(\Delta '(0^{+})^2+4 (\Delta
   (0)-\Delta (w)) \Delta ''(w)\right)\nonumber \\
& -\Delta '(0^+)^4+2 (\Delta (0)-\Delta (w))^2
   \Delta ''(w)^2+7 \Delta '(w)^4\Big]I_{4} \nonumber \\
& +48 \Big[\left(16 (\Delta (w)-\Delta (0)) \Delta ''(w)-3
   \Delta '(0^+)^2\right) \Delta '(w)^2\nonumber \\
& \qquad +4 (\Delta (0)-\Delta (w))^2 \Delta
   ^{(3)}(w) \Delta '(w)+3 \Delta '(w)^4\nonumber \\
& \qquad  +2 (\Delta (0)-\Delta (w)) \Delta ''(w)\nonumber \\
&\qquad \quad  \times
   \left(\Delta '(0^{+})^2+2 (\Delta (0)-\Delta (w)) \Delta
   ''(w)\right)\Big] I_{3} \nonumber
\end{align}
In the limit of small $w$, 
\begin{eqnarray}\label{a37}
 K^{(4)}_{\mathrm{1-loop}}(w) &=& \left(480 I_{4}+960 \frac{I_{3}}{m^2} \right) \Delta' (0^+)^{3}\Delta'' (0^+)\nn
|w|
\\
&& +O (w^{2})\ .
\end{eqnarray}
Adding the tree-level part, and using (\ref{relKmom}) we obtain the fourth moment of avalanche sizes:
\begin{eqnarray}\label{a31}
 \frac{\langle S^4 \rangle}{\langle S \rangle} &=& \frac{120}{m^{12}} \left| \Delta' (0^{+}) \right|^{3}
\left[1-4\left(m^4 I_{4} +2 m^2 I_{3} \right) \Delta'' (0^+) \right] \nn \\
& =& 120 S_m^3 \left[1 - 4 \frac{\tilde I_4 + 2 \tilde I_3}{\epsilon I_2} \tilde \Delta'' (0^+) \right]
\end{eqnarray}
It yields for the universal ratio:
\begin{eqnarray}\label{a39}
r^{(3)}_{\mathrm{1-loop}} &=& \frac{\left< S^{4}\right>\left<S^{2}
\right>}{\left< S^{3} \right>^{2}} \nonumber \\
&=& \frac{5}{3} \left[1-4 \frac{\tilde I_{4}}{\epsilon \tilde I_{2}} \Delta'' (0) \right]\nn\\
& =&
\frac{5}{3}\left[1- {\tilde \Delta'' (0)}\left(1-\frac{d}{6} \right) \right] \nonumber \\
&=&
\frac{5}{3}\left[1- \frac{(\epsilon -\zeta) (6-d)}{18} \right]
\end{eqnarray}
With Mathematica we have computed $\hat C^{(n)}$ and $K^{(n)}$ up to $n=6$. The expressions are tedious. Let us just note that one finds the limits at zero argument:
\begin{eqnarray}
 \hat C^{(2p)}(0,...,0) = - (2p-1)! I_{2p} \Delta'(0^+)^{2p} \ ,   \label{cumulants0} 
\end{eqnarray}
for even $n \geq 4$ and zero for odd cumulants. This allows to compute the probability distribution of the center of mass fluctuations. The 
calculation and results are given in Appendix \ref{app:u}. It bears some similarities with the one for the distribution of the critical force at depinning in \cite{FedorenkoLeDoussalWiese2006}.

Let us give here only the part of the Kolmogorov cumulants proportional to $I_n$. They read:
\begin{align}
& \hat K^{(5)}(v) = - 240 I_5 \Delta'(v) (- \Delta'(0)^4 + 3 \Delta'(v)^4 \\
& + 4 \Delta(v)^2 \Delta''(v)^2 - 2 \Delta'(v)^2 (\Delta'(0^+)^2 - 8 \Delta(v) \Delta''(v))) \nn
\\
%\end{align}
%\begin{align}
 & \hat K^{(6)}(v) = 240 I_6 (- \Delta'(0)^4 + 31
\Delta'(v)^6 + 4 \Delta(v)^3 \Delta''(v)^3 \nn \\
& - 15 \Delta'(v)^4 (\Delta'(0^+)^2 - 16 \Delta(v) \Delta''(v))) \nn \\
& - 15 \Delta'(v)^2 (\Delta'(0^+)^4 - 8 \Delta(v)^2 \Delta''(v)^2))
\end{align}
Expanding at small $w>0$ one finds, for the part proportional to $I_n$:
\begin{equation}
s\hat K^{(n)}(w) = - A_n I_n (-\Delta'(0^+))^{n-1} \Delta''(0) w + O(w^2)
\end{equation}
with coefficients:
\begin{eqnarray}
 A_2 &=& 6 \ , \quad   A_3 = 48\ , \quad  A_4 = 480 \nn \\
 A_5 &=& 5760\  , \quad  A_6 = 80640\ , \ldots\ .
\end{eqnarray}
Note that $A_{n+1}/A_n = 2 n + 4$. This suggests that
\begin{equation}\label{a45}
A_{n} = 2^{n-2} (n+1)!
\end{equation}
which is supported by a heuristic argument in appendix \ref{a:An}.

\section{Resummation of all diagrams with a single
loop: self-consistent equation} \label{a40}

\subsection{Self-consistent equation}
We can now perform a resummation of all cumulants to 1-loop order. In view of the results of the previous sections
we can absorb all factors of $m$ and $\tilde \Delta'(0^+)$ by using the rescaled generating function $\tilde
Z(\lambda)$ defined in (\ref{b18a}) and (\ref{b188a}). One has
\begin{equation}\label{a42}
\tilde Z (\lambda) = \sum_{n=1}^{\infty} \frac{c_{n}}{n!} \lambda^{n} \ ,
\end{equation}
where the $c_n$ are extracted from the Kolmogorov cumulants at small $w$, and at tree level $c_{n}=a_{n}$.
In the previous section we have obtained the contribution to $c_n$ proportional to $\tilde I_n$,
involving the coefficients $A_n$. Remarkably, the remaining corrections (i.e.\ the terms in $c_n$ proportional to $\tilde
I_p$ with $3 \leq p <n$) can be generated automatically  by taking advantage of the tree structure. This is
illustrated graphically as follows:
$$
\!\!\fig{1}{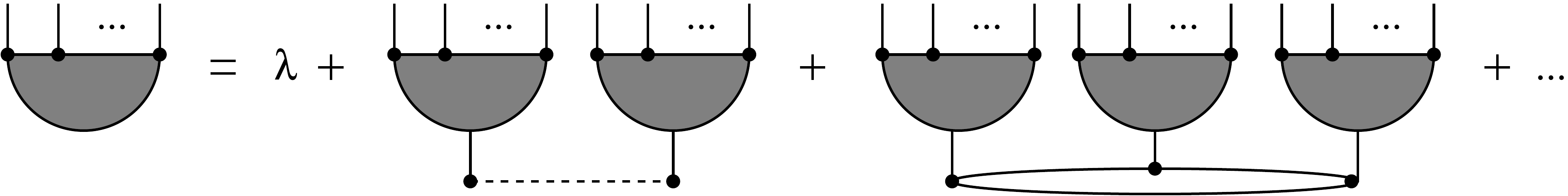}
$$%
Here,  the shaded blob represents $\tilde  Z (\lambda) $. It must contain all sets of trees made with $R$
vertices joined together at their basis either by a $R$ vertex or by a $S^{(n)}$ vertex with $n>3$, i.e.\ a loop,
in the spirit of (\ref{schem}). Expressed as an equation it reads
\begin{equation} \label{selfcons}
\tilde Z (\lambda) = \lambda + \tilde Z (\lambda)^{2}  + \alpha \sum_{n \geq 3} (n+1) 2^{n-2} i_{n} \tilde Z
(\lambda)^{n}\ ,
\end{equation}
where we have defined $i_n:=\tilde I_n/(\epsilon \tilde I_2)$, and substituted the factor $(n+1) 2^{n-2}=A_n/n!$. Here and
below we denote
\begin{equation}
\alpha:=-\epsilon \tilde I_2 m^{-\epsilon} \Delta''(0^{+}) = - \tilde \Delta''(0^{+})\ ,
\end{equation} which is of
order $\epsilon$. It was not strictly necessary to make equation (\ref{selfcons}) self-consistent, one
could with the same accuracy of $O(\epsilon)$ replace all terms $\tilde Z^n$ with $n \geq 3$ by $\tilde Z_0^n$. By
making it self-consistent we include more loops. It is however more convenient in view of the form at tree level. It 
also has a nice interpretation in terms of the generating function of trees with higher-$n$ branching: these occur at a small rate, which is  $O(\alpha)$, and $n$-dependent.

\subsection{Summation of the series and final equation}

From the above self-consistent equation one can compute iteratively $\tilde Z(\lambda)$ with the result:
%\begin{widetext}
\begin{align}\label{a46}
& \tilde Z (\lambda) = \lambda +\lambda ^2+\Big[8 \alpha i_3+2\Big] \lambda
^3+\frac{1}{24} \Big[960 \alpha i_3+480 \alpha i_4\nonumber \\
&\qquad+120\Big] \lambda
^4
 +\frac{1}{120} \Big[23040 \alpha ^2 i_3^2+20160 \alpha i_3\nonumber \\
& \qquad +14400
\alpha i_4+5760 \alpha i_5+1680\Big] \lambda ^5\nonumber \\
& +\frac{1}{720} \Big[1290240 i_3^2 \alpha ^2+806400 i_3 i_4 \alpha ^2+483840 i_3
\alpha \nonumber \\
&\qquad +403200 i_4 \alpha +241920 i_5 \alpha +80640 i_6 \alpha
+30240\Big] \lambda ^6\nonumber \\
&+O\left(\lambda ^7\right)
\end{align}
%\end{widetext}
One recognizes that the results from the third and fourth cumulants of the previous sections are correctly
reproduced. Note that the terms with higher powers of $i_n$ contain more than one (non-overlapping) loops and
higher powers of $\alpha=O(\epsilon)$. The terms of order $\alpha$ are all correctly generated.

We now want to compute the infinite sum in Eq.\ (\ref{selfcons}). For this we use the following representation of
the $i_n$, as in the derivation of (\ref{In}):
\begin{equation}\label{a47}
i_{n} = {\cal N}\, \frac{1}{(n-1)!} \int_{0}^{\infty }\rmd s\, \rme^{-s} s^{-1-d/2} s^{n}\ ,
\end{equation}
with ${\cal N}=1/(2 \Gamma(3-\frac{d}{2}))$. One finds:
\begin{eqnarray}
&& \sum_{n \geq 3} (n+1) 2^{n-2} i_{n} z^{n} \\
&& = {\cal N} \int_{0}^{\infty}\rmd s\, \rme^{-s} s^{-d/2-1} \sum_{n>2} \frac{1}{4} \frac{( n+1)}{(n-1)!} (2 z
s)^{n} \nn
\end{eqnarray}
Let us define:
\begin{equation}\label{a49}
\psi (z):= \sum_{n \geq 3} \frac{1}{4} \frac{( n+1)}{(n-1)!} (2 z)^{n } = z \left(\rme^{2 z} z-3 z+\rme^{2 z}-1\right)
\end{equation}
One can write:
\begin{align}\label{a50}
& \sum_{n \geq 3} (n+1) 2^{n-2} i_{n} z^{n} \nonumber \\
&=  {\cal N}
\int_{0}^{\infty}\rmd s\, \rme^{-s} s^{-d/2-1}  \psi (s z) \nonumber \\
&=\frac{z \left((2-(d+2) z) (1-2
   z)^{\frac{d-4}{2}}+3 d z-6
   z-2\right)}{(d-2)(d-4)}\nonumber \\
&\stackrel{d\to 4}{\longrightarrow} \frac{1}{2} z (2 z+(1-3 z) \log (1-2 z))\nonumber \\
&= 2 z^3+\frac{5 z^4}{3}+2 z^5+\frac{14
   z^6}{5}+\frac{64 z^7}{15}+\frac{48
   z^8}{7}+O\left(z^9\right)
\end{align}
The last line is also obtained by using  $i_{n}= \frac{1}{2 (n-1) (n-2)}$ from Eq.~(\ref{In}).  One can thus
rewrite the self-consistent equation (\ref{selfcons}) as
\begin{align}\label{ZrecFinal}
&\tilde Z (\lambda) =: \tilde Z = \lambda+\tilde Z ^{2} \nonumber \\
&+ \alpha \frac{\tilde Z \left[(2-(d+2) \tilde Z) (1-2
   \tilde Z)^{\frac{d-4}{2}}+3 d \tilde Z-6
   \tilde Z-2\right]}{(d-2)(d-4)}\ .
\end{align}
Close to $d=4$ it becomes
\begin{equation}\label{a51}
\tilde Z  = \lambda+\tilde Z ^{2} + \alpha \frac{1}{2} \tilde Z [2 \tilde Z+(1-3 \tilde Z) \log (1-2 \tilde Z)]\ .
\end{equation}
This equation has to be inverted in order to get both the exponent $\tau$ and the tail of the distribution
$P(S)$ for large $S$. Note that from (\ref{selfcons}) it can also be written as:
\begin{eqnarray}\label{ZrecFinalInt}
\tilde Z (\lambda) &=:& \tilde Z = \lambda+\tilde Z ^{2}\nonumber \\
&& + \frac{\alpha}{\epsilon \tilde I_2}  \int_{k} \bigg[ \frac{\tilde Z^{2}}{(k^{2}+1-2 \tilde Z)^{2}}
 + \frac{\tilde Z}{(k^{2}+1-2 \tilde Z)} \nonumber\\
 &&\hphantom{+ \frac{\alpha}{\epsilon \tilde I_2}  \int_{k} \bigg[ } -\frac{\tilde Z}{k^{2}+1}  
 - 3 \frac{\tilde Z^{2}}{(k^{2}+1)^{2}} \bigg]
\end{eqnarray}
As shown in Appendix \ref{a:An}, this formula has a simple graphical interpretation. The last two terms can be thought of as counterterms which fix the coefficients of $\tilde Z$ and $\tilde Z^2$ according to the choice we implemented here. (They would be absent if we wrote down an expansion in the bare disorder.)

\section{1-loop results for the avalanche-size distribution} 
\label{s:1-loop-res}
\subsection{Avalanche exponent $\tau$}

The formula (\ref{a51}) contains all we need in order to retrieve the avalanche size distribution. We recall that
\begin{eqnarray}
S_m^{-1} \tilde Z(\lambda S_m) = \langle S \rangle^{-1} (\langle \rme^{\lambda S} \rangle - 1)  \label{newtZ} 
\end{eqnarray}
from  (\ref{b18a}), where $S_m$ is given by (\ref{b188a}). Note however that  (\ref{a51}) contains only information about sizes of order $S_m$, from $S \ll S_m$ to $S \gg S_m$, but in all cases much larger than the microscopic cutoff, i.e.\ $S \gg S_{min}$, which has not been included (and was not needed) in the above analysis.

Let us start with the exponent $\tau$. It can be extracted as follows. There is a critical value $p_c<1$ such that all moments $\int \rmd S \, P(S) S^{p}$ diverge for 
$p \leq p_{c}$. Of course these do not strictly diverge, as they should be in the end cutoff by $S_{min}$, but as just discussed we can forget that here to extract the $\tau$ exponent. This exponent is defined from the behavior $P (S)\sim S^{-\tau}$ for $S_{min} \ll S \ll S_m$. Hence we can identify:
\begin{eqnarray}\label{a52}
\tau =1+p_{c}
\end{eqnarray}
We start from the identity
\begin{eqnarray}\label{a52b}
\nn \left< S^{p} \right>  &=& 
 \frac{1}{\Gamma (-p)} \int_{0}^{\infty} \rmd
\lambda \,  \lambda^{-1-p} \, \left< \rme^{-\lambda S}-1 \right>  \\
& =& \frac{\left< S \right>}{S_m}  \frac{1}{\Gamma (-p)} \int_{0}^{\infty} \rmd
\lambda  \,  \lambda^{-1-p} \, \tilde Z(- S_m \lambda) \ , \qquad
\end{eqnarray}
where in the second line we have used (\ref{newtZ}). Since $\tilde Z \sim \lambda$ at small $\lambda$ the integral converges at the lower bound. However, it may diverge at the upper bound. Hence $p_{c}$ is obtained from the power-law tail of $\tilde Z (\lambda)$, for
$\lambda \to - \infty$. The boundary is for  $\tilde Z (-\lambda)\sim
- \lambda^{p_{c}}$, where we have indicated the sign of $\tilde Z$. ($\tilde Z$ is
negative, due to the dominance of the term $-1$ in the definition of $\tilde Z$ for large negative $\lambda$). This is consistent
with the asymptotic behavior (\ref{asymptZ}) and the discussion there.

From  (\ref{a51}) we obtain, for large negative $\tilde Z\equiv \tilde Z(\lambda)$
\begin{equation}\label{a53}
- \lambda \approx  \tilde Z^{2} \left[ 1+\alpha -\frac{3}{2}\alpha \log (- 2
\tilde Z)\right]\approx (- \tilde Z)^{2- \frac{3}{2}\alpha} + O (\alpha^{2})
\end{equation}
This gives $- \tilde Z \approx (-\lambda)^{1/(2-\frac{3}{2}\alpha )}=
(-\lambda)^{p_{c}}$, from which we identify
\begin{eqnarray}\label{taufinal}
\tau &=& 1 + \frac{1}{2-\frac{3}{2}\alpha } \approx \frac{3}{2}
+\frac{3}{8}\alpha\\
&=& \frac{3}{2} -\frac{3}{8} \tilde \Delta'' (0) 
\end{eqnarray}
Using relation (\ref{secder}) of Appendix \ref{app:review}, valid for all classes of disorder, we obtain our main result for the avalanche exponent,
\begin{eqnarray}
\tau &=& \frac{3}{2}- \frac{\epsilon -\zeta}{8} + O(\epsilon^2)\ .
\end{eqnarray}
Let us now discuss the significance of this result. First we note that this formula agrees to first order in $\epsilon$ with the conjecture
\begin{equation}\label{a58}
\tau = \tau_\zeta := 2- \frac{2}{d+\zeta}\ .
\end{equation}
As mentioned in the introduction, this conjecture was put forward in the study of interface {\it depinning} in Ref.\ \cite{NarayanDSFisher1993a} in the absence of a mass term, i.e.\ in the ensemble of fixed applied force $f$. It is based on the {\it assumption} that, as the force $f$ is increased towards the threshold $f_c$, the mean number of avalanches per interval of force increment is not singular. Translated into the present setting it means that
\begin{equation}
\frac{\rmd N_w}{m^2 \rmd w} =\frac{ \rho_0}{ m^{2}} \label{nw}
\end{equation}
remains finite as $m \to 0$. (We use the notations of Section \ref{sec:shocks} where $N_w$ is the number of shocks and  $\rho_0$ the shock number density). Note however that in the present setting, although the increase of the force is $m^2 \rmd w$, it does not bring the system closer to criticality, which is achieved by {\em decreasing} the mass. This is why we had at the start two independent exponents $\rho$ and $\tau$. The STS symmetry provides a first relation (\ref{b3}) while the conjecture of the finiteness of (\ref{nw}) is equivalent, using (\ref{b4}), to the conjecture $\rho=2$ (at least for $\tau>1$). Using (\ref{b4}) again, this implies the value $\tau=\tau_\zeta$. Finally note that a similar conjecture was put forward in the context of sandpiles \cite{DharDhar1997,AgrawalDhar2001}, and if one admits the connection to the depinning of a periodic interface \cite{NarayanMiddleton1994,FedorenkoLeDoussalWiese2008a}  it is equivalent to the conjecture $\tau=\tau_{\zeta=0}$ for the RP class. 

As discussed in Ref.\ \cite{FedorenkoLeDoussalWiesePREP}, we expect the result for $\tau$ for statics (shocks) and for depinning (avalanches) to coincide to one loop. Hence our result indicates that both indeed coincide with the conjecture $\tau=\tau_\zeta$ to one loop. What happens beyond one loop accuracy remains to be elucidated. We know that statics and depinning differ at two loop, e.g.\ for the $\zeta$ exponent. There are various possibilities for either the statics or the depinning to {\em agree}, or {\em disagree} with the conjecture at two loops and for $d=2,3$: Since $\zeta_{\mathrm{dep}}> \zeta_{\mathrm{stat}}$, one possibility is already excluded, that both conjectures are true {\it and} that the exponent $\tau$ is the same in both problems. Some claims that static and dynamic avalanches belong to the same universality classes were indeed put forward in \cite{LiuDahmen2006} for the RF Ising model. Finally note that our result can also be written as
\begin{equation}
\rho=2 + O(\epsilon^2)\ ,
\end{equation}
and a remaining challenge is to compute the higher orders. Until now we have failed to find, within the field theory, a symmetry or a mechanism which would imply $\rho=2$ to all orders. 

Note that exact results for $d=0$ depinning \cite{LeDoussalWiese2008a}, i.e.\ of a particle, yield $\tau=0$ for the so-called Gumbel class which has $\zeta=2$, hence agreement with the conjecture up to logarithmic corrections present in that case. However, the result for the other classes (Weibul and Frechet) have also $\tau=0$ but $\zeta \neq 2$, hence violating the conjecture in its present form. For the statics, exact result for the $d=0$ limit of the RF class, the so-called toy or Sinai Brownian energy landscape, has $\tau=1/2$ and $\zeta=4/3$ which again satisfy the conjecture \cite{LeDoussal2006b}. 

To summarize, for the cases of interest, our result is, with $\epsilon=4-d$ and up to $O(\epsilon^2)$ corrections:
\begin{equation}\label{a59}
\tau = \frac{3}{2}-  \left\{\begin{array}{cc}
\displaystyle \frac{1}{12} \epsilon & \qquad \mbox{RF}\\
0.0989627 ~ \epsilon \rule{0mm}{3ex}&\qquad \mbox{RB}\\
\displaystyle  \rule{0mm}{3.5ex}\frac{1}{8} \epsilon &\qquad \mbox{CDW}
\end{array} \right.
\end{equation}
where we have used the 1-loop values for $\zeta$ in the statics, recalled in Appendix \ref{app:review}. To $O(\epsilon)$ these are the same as for depinning, except that in that case the RB class does not exist, since 2-loop corrections make it flow to RF. 

It is useful to quote the expected result if the conjecture holds. In $d=3$:
\begin{equation}\label{a60}
\tau =\left\{\begin{array}{cc}
\displaystyle \frac{7}{5} & \qquad \mbox{RF}\\
1.37783 \pm  0.00005 \rule{0mm}{3.5ex}&\qquad \mbox{RB}\\
\displaystyle  \rule{0mm}{4ex}\frac{4}{3} &\qquad \mbox{CDW}
\end{array} \right.
\end{equation}
In $d=2$:
\begin{equation}\label{a60b}
\tau =\left\{\begin{array}{cc}
\displaystyle \frac{5}{4} & \qquad \mbox{RF}\\
1.17987 \pm 0.00037 \rule{0mm}{3.5ex}&\qquad \mbox{RB}\\
\displaystyle  \rule{0mm}{4ex} 1 &\qquad \mbox{CDW}
\end{array} \right.
\end{equation}
In $d=1$:
\begin{equation}\label{a60c}
\tau =\left\{\begin{array}{cc}
\displaystyle 1 & \qquad \mbox{RF}\\
\frac{4}{5} \rule{0mm}{3.5ex}&\qquad \mbox{RB}\\
\displaystyle  \rule{0mm}{4ex} 0 &\qquad \mbox{CDW}
\end{array} \right.
\end{equation}
For the RB class we have used the 2-loop result $\zeta=0.208298042 \epsilon + 0.0068582 \epsilon^2 + O(\epsilon^3)$ and constructed three different Pade approximants, using the constraint that $\zeta_{d=1}=2/3$. This gives $\zeta^{\mathrm{RB}}_{d=3}=0.21454 \pm 0.00028$ and 
$\zeta^{\mathrm{RB}}_{d=2}=0.43864 \pm 0.0011$. The error bars denote the $1\sigma$ spread of the three Pade approximants. For RF we use $\zeta=(4-d)/3$ and for RP $\zeta=0$, hence there is no uncertainty. 
These are the values for the statics. For depinning, if the conjecture holds, the values for the RP (CDW) class are the same as given here. For non-periodic interfaces $\zeta=\frac{1}{3} \epsilon(1+0.14331\epsilon)+ O(\epsilon^3)$ and 
one gets $\tau_{d=3}=1.409 \pm 0.001$, $\tau_{d=2}=1.31 \pm 0.01$ and $\tau_{d=1}=1.2 \pm 0.1$ (in the latter inserting the numerically measured value for $\zeta=1.25$ yields instead $\tau_{d=1}=1.11$).

\subsection{Distribution of avalanche sizes}
\begin{figure*}[tb]
\fig{0.98}{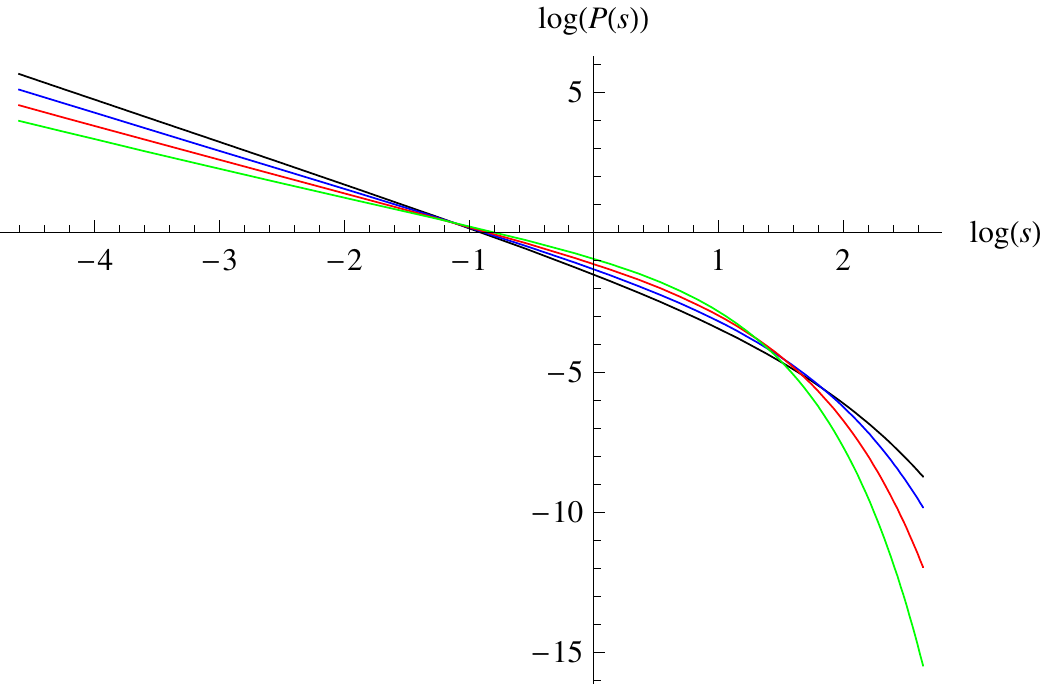}\hfill\fig{0.98}{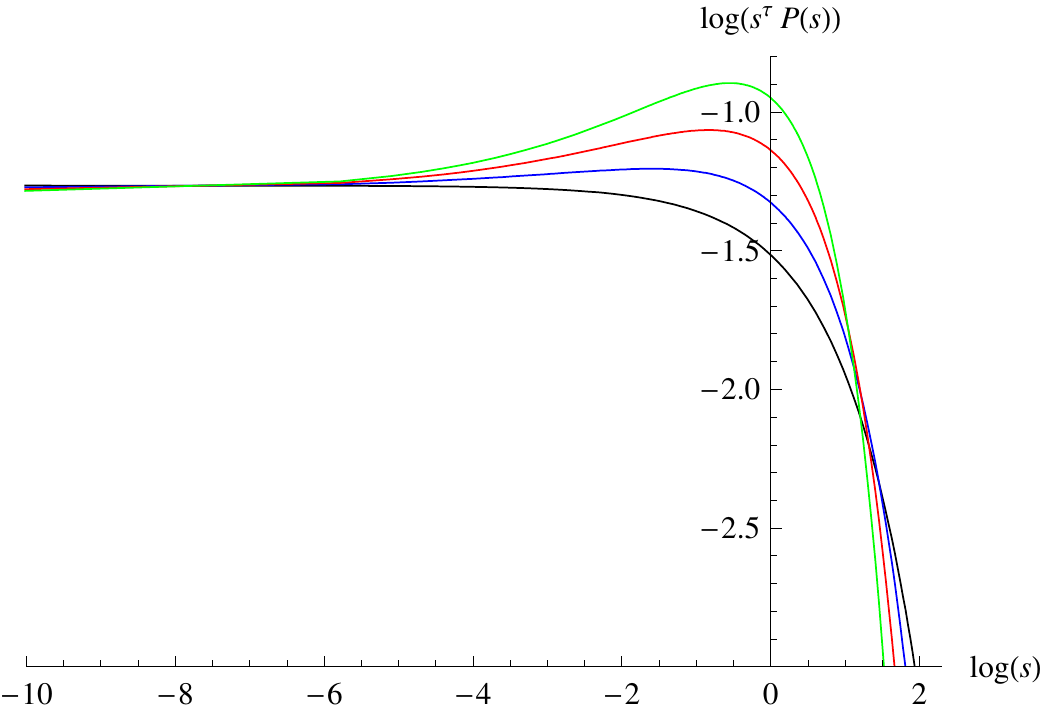}
\caption{$P(s)$, both for mean-field (black), and $\epsilon =1,2,3$
(blue, red, green).}
\label{PofS}
\end{figure*}
We now compute the  avalanche-size distribution. To do so, we have to invert
(\ref{a51}). To first order in $\alpha$ this can be done order by order,  by first inserting in the term
proportional to $\alpha$ the MF solution, i.e.\ the solution of
(\ref{a51}) at $\alpha =0$. Solving for $\tilde Z$ then yields
\begin{widetext}
\begin{equation}\label{a61}
\tilde Z (\lambda) = \frac{1}{2} \left(1-\sqrt{1-4 \lambda }\right)+\frac{\left(\left(3
\lambda +\sqrt{1-4 \lambda }-1\right) \log (1-4 \lambda )-2 \left(2
\lambda +\sqrt{1-4 \lambda }-1\right)\right) \alpha }{4 \sqrt{1-4
\lambda }}+O\left(\alpha ^2\right)\ .
\end{equation}
%\end{widetext}
Expanding in $\lambda$ gives
\begin{equation}\label{a62}
\tilde Z (\lambda)=\left(\lambda +\lambda ^2+2 \lambda ^3+5 \lambda ^4+14
\lambda ^5+42 \lambda ^6+O\left(\lambda ^7\right)\right) %\nonumber \\ && 
+\left(2
\lambda ^3+\frac{35 \lambda ^4}{3}+54 \lambda ^5+\frac{3472 \lambda
^6}{15}+O\left(\lambda ^7\right)\right) \alpha %\nonumber \\ && 
+O\left(\alpha
^2\right)\ ,
\end{equation}
consistent with our previous result (\ref{a46}) if one uses the values of the $i_n$ given above. 
The challenge is to find a distribution $P(S)$ which generates all terms in (\ref{a61}). 
By trial and error, one arrives at the following integral representation:
%\begin{widetext}
\begin{eqnarray}\label{a63}
\tilde Z (\lambda) = \frac{1}{2\sqrt{\pi}} \int_{0}^{\infty} &\rmd
S& \left(\rme^{\lambda S} -1\right) \rme^{-S/4} S^{-3/2} \nonumber \\
& & \times
\left[1+\frac{1}{16} \left(\log (S) S+\gamma S+4 S-8 \sqrt{\pi }
\sqrt{S}-6 \log (S)-6 \gamma +4\right) \alpha +O\left(\alpha ^2\right)
\right]\qquad
\end{eqnarray}
\end{widetext}
To leading order in $\alpha$, using the definition (\ref{newtZ}), this
yields our final result for the avalanche-size distribution at one loop:
\begin{equation}\label{finalS}
P(S) = \frac{\left<S\right>}{2 \sqrt{\pi}} S_m^{\tau-2} A S^{-\tau} \exp\!\left(C \sqrt{\frac{S}{S_m}} -
\frac{B}{4} \left[\frac{S}{S_m}\right]^\delta\right)
\end{equation}
for $S \gg S_{\mathrm{min}}$, with coefficients:
\begin{eqnarray} \label{amp} 
&& A = 1+ \frac{1}{8} (2-3 \gamma_{\mathrm{E}} ) \alpha \quad , \quad B = 1-\alpha \left(1+\frac{\gamma_{\mathrm{E}}}{4}\right) \nonumber  \\
&& C=- \frac{1}{2} \sqrt{\pi} \alpha \quad , \quad \alpha= - \frac{1}{3} (1 - \zeta_1) \epsilon\ .
\end{eqnarray}
$\gamma_{\mathrm{E}}=0.577216$, 
and the exponents read:
\begin{eqnarray}\label{a64}
&& \tau = \frac{3}{2} + \frac{3}{8} \alpha =  \frac{3}{2} - \frac{1}{8} (1 - \zeta_1) \epsilon + O(\epsilon^2)
\\
&& \delta=1 - \frac{\alpha}{4} = 1 + \frac{1}{12} (1 - \zeta_1) \epsilon\ .
\end{eqnarray}
Note that the decay of large avalanches becomes stretched exponential, faster than exponential, with an exponent $\delta>1$. For the RF class $\zeta_1=1/3$, hence $\delta=1+\epsilon/18 + O(\epsilon^2)$. This goes in the right direction to match the exact result $\delta=3$ for the RF class in $d=0$, but the magnitude seems underestimated by the 1-loop formula, which suggests large higher-loop corrections for this exponent. Note the pronounced bump in the plot of $S^\tau P(S)$ on figure \ref{PofS}. This bump is indeed seen in numerical simulations \cite{LeDoussalMiddletonWiese2008}.

\subsection{Normalization and scales in the distribution of sizes} 

One can check that the final formula (\ref{finalS}) is properly normalized, i.e.\ $\int_0^\infty \rmd S S P(S)=\langle S \rangle$, to leading and first order in $\epsilon$. Note that since $\tau>1$, this formula does not give information about typical avalanches, which are of the order of the cutoff $S_{\mathrm{min}}$, but about larger ones which control the moments $\langle S^p \rangle$ with $p>p_c=\tau-1$. It is then universal, i.e.\ it does not depend on the details of the small-scale cutoff $S_{\mathrm{min}}$. However the average avalanche size $\langle S \rangle$, which appears as a factor in the distribution (\ref{finalS}), is non-universal and cannot be computed from this theory. In Section \ref{sec:distribmf} we showed that it behaves as $\langle S \rangle \sim S_{\mathrm{min}}^{\tau-1} S_m^{2-\tau}$ but this is all one knows about it. In comparing with numerics or experiments, one replaces $\langle S \rangle$ by its actual measured value. Then there is only one free parameter left, the global scale $S_m$.  Note  that this scale is given by $S_m=\langle S^2 \rangle/2 \langle S \rangle$, hence it can also easily be measured. We recall that its value is predicted by field theory via the  exact relation
\begin{equation}
 S_m = \frac{|\Delta'(0^+)|}{ m^{4} } = \frac{ K^{d/2}}{m^{d+\zeta}} \frac{|\tilde \Delta'(0^+)|}{ \epsilon \tilde I_2}\ .
\end{equation}
$\epsilon \tilde I_2=2 (4 \pi)^{-d/2} \Gamma(3-\frac{d}{2})$ is a $d$-dependent number. A question is how predictive this formula,
and how universal this scale is. The answer depends on the universality class, RP, RF or RB. 

For the RP class, $\zeta=0$ and the fixed point for $|\tilde \Delta'(0^+)|/a$, where $a$ is the period, is a universal number. To two loops, this number is equal in the statics and at depinning and reads $\epsilon/6 + \epsilon^2/9 + O(\epsilon^3)$. In general $K$ and $m$ will experience small corrections for any e.g.\ a lattice model which does not exactly satisfy the STS symmetry. They can however be measured from large scale measurements on the system. Once they are extracted, then $S_m$ can be predicted. Alternatively, one can construct ratios which, for the RP class, are universal, for instance, 
\begin{equation}\label{176}
 r_{\mathrm{RP}}=\frac{a S_m}{\overline{u^2} L^d} =\frac{a |\Delta'(0^+)|}{\Delta(0)} = \frac{a |\tilde \Delta'(0^+)|}{\tilde \Delta(0)}= 6 \ ,
\end{equation} 
where $u$ is the center of mass and $\overline{u^2}$ the variance of its fluctuations. Its definition is $\overline{u^2} L^d=\sum_x \overline{u(0)u(x)}$. Note that in the statics there is no higher order correction, since the fixed point for $R(u)= a + b[u(1-u)]^2$ with some constants $a$ and $b$ which drop out from (\ref{176}). Replacing by the depinning values gives $r=6 + 2 \epsilon+ O(\epsilon^2)$. This universal ratio can thus be used to distinguish statics from depinning.

For the RF class, $\zeta=\epsilon/3$ in the statics, as recalled in Appendix  \ref{app:review}. The fixed point values contain a
 scale $\xi$ (in the direction of $u$), which can be fixed if one knows the amplitude of the random field $\sigma=\int_0^\infty \rmd u\, \Delta(u)$. $\sigma$ can be retrieved from large distances in  experiments, and is a parameter of simulations. There are various amplitude combinations which can be studied, depending on whether one is willing to measure $m$ and $K$, see  Appendix \ref{app:review} and Ref.\ \cite{LeDoussalWieseToBePublished}.
  The nicest universal ratio is: 
\begin{eqnarray}
 r_{\mathrm{RF}}&=&\frac{\sigma S_m}{(m^2\overline{u^2} L^d)^2} =\frac{\sigma |\Delta'(0^+)|}{\Delta(0)^2} \nn\\
 &=& \gamma_1 + \epsilon\left(\frac29 \gamma_1+ \gamma_2\right) + O(\epsilon^2) \end{eqnarray}
with $\gamma_1=0.775304245188$ and $\gamma_2=-0.13945524$, where one can check from Appendix \ref{app:review} that all dependence on $m$, $K$ and the scale $\xi$ cancel (thanks in part to the strict equality $\zeta=\epsilon/3$). 

Finally, for the RB class (and for the RF class for depinning) there is the least universality: $|\tilde \Delta'(0^+)|$ is non-universal, hence measuring $K$ and $m$ will not be enough. One has,  for instance, also  to measure $\overline{u^2} L^d$ to obtain the combination:
\begin{equation}
\frac{S_m^2}{\overline{u^2} L^d} =  \frac{\Delta'(0^+)^2}{m^{4}\Delta(0)} = \frac{m^{-d} K^{d/2}}{ \epsilon \tilde I_2}  \frac{[\tilde \Delta'(0^+)]^2}{\tilde \Delta(0)} 
\end{equation}
with the universal ratio  
\begin{equation}
 \frac{\tilde \Delta'(0^+)^2}{\tilde \Delta(0)} = 0.583405 \epsilon + 0.294205  \epsilon^2 + O(\epsilon^3)\ .
\end{equation}
Of course, in all these cases there are other possibilities for interesting universal ratios. 
If one measures for instance the function $\Delta(u)$ as in Ref.\ \cite{LeDoussalMiddletonWiese2008}, one can in some cases get rid of measuring $K$ and $m$. Eventually, a direct numerical test of the relation between $S_m$ and the cusp would also be welcome. Analytical solution of toy models in $d=0$ \cite{LeDoussalWiese2008a} has also successfully tested this relation.

\subsection{Moments and universal ratios}

Having obtained $P(S)$ we can compute its moments. This is useful for comparison with numerics.
Direct integration of equation (\ref{a64}) gives
\begin{align}
&\frac{\left< S^{n}\right>}{\left<S\right> S_m^{n-1}}=\nn 
\frac{4^{n-1} \Gamma \left(n-\frac{1}{2}\right)}{\sqrt{\pi }} + \frac{\epsilon (1-\zeta_1) }{3} 
\frac{ 4^{n-2}}{\sqrt{\pi}} \Big\{ 4 \sqrt{\pi } \Gamma (n)
\\ &- \Gamma
   \left(n-{\textstyle\frac{1}{2}}\right)\Big[ \Big( \psi(n-{\textstyle\frac{1}{2}})+\log(4)+\gamma \Big) (n-2) +4 n \Big]\Big\}\nn\\&+O\left(\epsilon ^2\right)
\end{align}
a formula valid for any fixed real $n>1/2$. From this we can extract the universal dimensionless ratios:
\begin{eqnarray}\label{rn}
r_n:&=& \frac{\langle S^{n+1} \rangle \langle S^{n-1} \rangle}{ \langle S^{n} \rangle^{2}} = \frac{2 n-1}{2 n -3} \\
&& - \frac{\epsilon}{3}(1-\zeta_1) \frac{ n \Gamma(n-\frac{3}{2}) + \sqrt{\pi} \Gamma(n-1)}{(2 n - 3)^2
\Gamma(n-\frac{3}{2}) }  + O(\epsilon^2)\ ,\nonumber
\end{eqnarray}
for any real $n>3/2$, with $\zeta_1=1/3$ for RF, $\zeta_1=0$ for RP and $\zeta_1= 0.208298042 $ for RB.
The lowest-order integer ones are:
\begin{eqnarray}
 r_2&=& 3 - \epsilon (1-\zeta_1) \\
 r_3 &=& \frac{5}{3} - \frac{5}{27} \epsilon (1-\zeta_1)\ .
\end{eqnarray}

Another useful form when comparing to numerics for $n$ near $\tau$, is to isolate the simple pole divergence which occurs in any dimension:
\begin{eqnarray}\label{rn}
&& r_n = \frac{A_d}{n-\tau} + B_{n,d} \\
&& A_d =  1 - \frac{1+\pi}{12} \epsilon (1-\zeta_1) + O(\epsilon^2) \\
&& B_{n,d} = 1 + \epsilon (1-\zeta_1) \frac{\pi \Gamma[n-1/2] - \sqrt{\pi} \Gamma(n-1)}{6 (2 n - 3) \Gamma(n-\frac{1}{2})}
 \nonumber \\
 && + O(\epsilon^2)\ ,\nonumber
\end{eqnarray}

Note that for comparison with numerics it is useful to estimate the corrections due to the small scale cutoff,
assuming it just cuts the previous result at $S_{min}$. The correction to the above result for the dimensionless quantity $\frac{\left< S^{n}\right>}{\left<S\right> S_m^{n-1}}$ are thus of order $(S_{min}/S_m)^{n-\tau+1}/(n-\tau+1)$ and 
an approximation to $r_n$ is thus:
\begin{eqnarray}\label{rn}
&& r_n = A_d \frac{1 - (S_{min}/S_m)^{n-\tau}}{n-\tau} + B_{n,d} 
\end{eqnarray}

For the RF class one finds either from a direct expansion, or using the $(0,1)$-Pade: $r_2=\{2.333,2.45455\}$, $r_3=\{1.5432,1.55172\}$ ($d=3$), $r_2=\{1.667,2.07692\}$, $r_3=\{1.41975,1.45161\}$ ($d=2$);
$r_2=\{1,1.8\}$ and $r_3=\{1.2963,1.36364\} $ ($d=1$) and $r_2=\{0.3333,1.58824\}$ and $r_3=\{1.17284,1.28571\}$ ($d=0$).

This can be compared with the exact result \cite{LeDoussal2006b} for RF disorder in $d=0$:
\begin{align}
& r_2= 1.2978 \\
& r_3 = 1.17776
\end{align}
Using the exact result as a constraint yields three Pad\'e aproximants and the results:\begin{align}
& r_2= 2.404 \pm 0.009 \quad d=3 \\
& r_2=1.935  \pm 0.021 \quad d=2 \\
& r_2=1.571  \pm 0.022 \quad d=1
\end{align}
and
\begin{align}
& r_3=1.5427 \pm 0.0012 \quad d=3 \\
& r_3=1.419  \pm 0.003 \quad d=2 \\
& r_3=1.297  \pm 0.003 \quad d=1
\end{align}
where averages are over the three Pad\'e approximants and error bars are the corresponding one sigma deviations.

\section{Spatial structure of avalanche distributions and self-consistent equation at non-zero momentum}
\label{s:spatial}

In this section we introduce generating functions which encode for the spatial correlations in the
avalanches. An explicit calculation is performed at the level of the improved tree approximation (mean field).
It exhibits an interesting connection to instanton calculations in a cubic field theory.

\subsection{Generating function}

To obtain information about the structure of avalanches in internal space one may define for avalanche $i$
\begin{equation}
 S_i^\phi =  \int_x \phi(x) S_i^x
\end{equation}
where $\phi(x)$ is a given function. One recovers the standard definition of size for $\phi(x)=1$, i.e.\ $S_i=S_i^1$. One would like to compute averages such as \footnote{The case of a vanishing $\int_x \phi(x)$ has to be considered separately.}
\begin{eqnarray}
 Z^{\phi}(\lambda) &=& \frac1{\int_x \phi(x)} \overline{ \sum_i (\rme^{\lambda \int_x \phi(x) S_i^x} -1) \delta(w-w_i) } \nn\\
& =&  \frac1{\int_x \phi(x)}  \int_0^\infty \rmd S^\phi (\rme^{\lambda S^\phi} -1) \rho_\phi(S^\phi)\ , 
\end{eqnarray}
where $\rho_\phi(s) =  \rho_0 P_\phi(s)$ is the density of avalanches with $S_\phi=s$, and $P_\phi(s)$ the normalized distribution of $S_\phi$.  Note that STS implies $\int_x \phi(x) \overline{(u_x(w)-w)}=0$ for any $\phi(x)$ and $w$. Hence, taking a derivative w.r.t $w$ one obtains
\begin{align}
& \int_0^\infty \rmd S^\phi S^\phi \rho_\phi(S^\phi) = \int_x \phi(x) \ .
\end{align}
Note that $\rho_0$, the total density of avalanches is independent of $\phi$, hence one also has the
exact relation 
\begin{equation}\label{EINS}
\left< S^\phi \right> = \left< S \right> L^{-d} \int_x \phi(x)\ .
\end{equation}

Extending the arguments leading to (\ref{twopoint2}) and (\ref{derw}) we write
\begin{align}  \label{derw2} 
& \partial_w \overline{\rme^{\lambda (\int_x \phi(x) ( u_x(w)-w - u_x(0)))}}\Big|_{w=0^+}\nn \\
& \qquad = \int \rmd S^\phi \rho_\phi(S^\phi) (\rme^{\lambda S^\phi}-1 - \lambda S^\phi) \ .
\end{align}
Hence one needs to compute the generating function
\begin{eqnarray}  \label{derw2b} 
 G^\phi(\lambda) &=& \frac1{\int_x \phi(x)}\, \overline{\rme^{\lambda (\int_x \phi(x) ( u_x(w)-w - u_x(0)))}} \qquad \\
 G^\phi(\lambda)&=&\hat Z^\phi(\lambda) w + O(w^2) \\
 \hat Z^\phi(\lambda) &=& \frac{1}{\langle S^\phi \rangle} ( \langle \rme^{\lambda S^\phi} \rangle -1 - \lambda \langle S^\phi \rangle) 
\end{eqnarray}
for $w>0$, with again $Z^\phi(\lambda)=\lambda + \hat Z^\phi(\lambda)$. Note that here we consider only a uniform $w_x=w$.

\subsection{The self-consistent equation}

Let us now study this quantity in the improved tree approximation (also called mean-field above). We can show that it is given by
\begin{equation}  \label{integr}
 Z^\phi(\lambda) = \frac1{\int_x \phi(x)}\int_x  Z_x^\phi(\lambda) \ ,
\end{equation}
where $Z_x^\phi(\lambda)$ is solution to the self-consistent equation
\begin{equation} \label{s4}
Z^{\phi}_x(\lambda) = \lambda \phi(x) + |\Delta'(0^+)| \int_{zy} g_{x-y} g_{x-z} Z^{\phi}_y(\lambda) Z^{\phi}_z(\lambda)\ .
\end{equation}
$g_{x}=\int_k g_k \rme^{i k x}$ is the free (elastic) propagator. One can check that for $\phi(x)=1$, $Z_x(\lambda)=Z(\lambda)$ and one recovers the tree-level recursion given in the text for $Z(\lambda)$, and, upon replacing $\Delta'(0^+) \to 1$ and $g_k \to \tilde g_k=g_k|_{m=1}$, the one for the rescaled function $\tilde Z(\lambda)$. A derivation, resumming all diagrams, is given in appendix \ref{a:der-rec}.

We now go to rescaled quantities, using (\ref{b18a}), i.e.\ $Z(\lambda)= \tilde Z(\lambda S_m)/S_m$. Then   we can  write for an arbitrary function $\phi(k)$ in Fourier space:
 \begin{equation} \label{s2bis}
\tilde Z_k(\lambda) =  \lambda \phi(k) + \int_{q} \frac{m^2}{q^2+m^2} \frac{m^2}{(k-q)^2+m^2}  \tilde Z_q(\lambda)  \tilde Z_{k-q}(\lambda)
\end{equation}
In real space, this is
\begin{equation} \label{s2tri}
\tilde Z_z(\lambda) =  \lambda \phi(x) + m^4  \int_{xy} g_{z-x} g_{z-y} \tilde Z_x(\lambda) \tilde Z_y(\lambda)\ .
\end{equation}
We note that this self-consistent equation is simplified by defining $Y_k(\lambda):= m^2 g_k \tilde Z_k(\lambda)$, or in real space 
\begin{eqnarray}\label{defY1}
Y({z m} ,\lambda)&:=&m^2 \int_x g_{z-x} \tilde Z_x(\lambda) \\
\varphi(m z) &:=& \phi(z) \label{defY2}
\ . 
\end{eqnarray} This results in (suppressing from now on the explicit dependence on $\lambda$ when convenient)
\begin{eqnarray} \label{s1}
\left(-{\nabla^2} + 1 \right) Y(x)  = \lambda \varphi(x) + Y(x)^2\ .
\end{eqnarray}

\subsection{Solution for $\varphi$ localized on a codimension one hyper plane}

The function $Z(\lambda)$ will be qualitatively different, depending on whether $\varphi$ is extended on the scale of the inverse mass (absorbed in $x$), or is a $\delta$-distribution. We now study one special case, namely $\varphi(x)=\delta(x_1)$. The function $Y(x) $ will then be constant along the directions $x_2, \ldots, x_d$, and for simplicity of notation we will denote $x_1\to x$, and suppress $x_i$, $i>1$. Thus we effectively consider a 1-dimensional problem.

Eq.\ (\ref{s1}) can then be integrated analytically. Consider first the homogenous equation ($\lambda=0$)
\begin{equation} \label{206}
Y''(x) =  Y(x) - Y(x)^2 \ .
\end{equation}
Multiplying with $Y'$ and integrating once gives
\begin{equation}
[Y'(x)]^2 = \mbox{const} -  \frac2 3 Y(x)^3 + Y(x)^2 \ .
\end{equation}
If  (\ref{206}) is viewed as the equation of motion of a particle, then setting $\mbox{const}\to 0$ will be the solution which has zero kinetic energy at the saddle-point $Y=0$ of the potential $V(Y)= \frac13 Y^3-\half Y^2$. It is the unique solution which decays (exponentially fast) to $Y=0$ for $x\to \pm \infty$.  The other solutions are either oscillating or unbounded. Integrating once more  for $\mbox{const} = 0$ gives $-{2\,  \mbox{arctanh} \left(\frac{\sqrt{3 -2  y}}{\sqrt{3} }\right)} = x$, or equivalently
\begin{equation}
Y(x) =Y_0(x): =  \frac3{\cosh( x) +1} \ ,
\end{equation}
where the center of the solution has been chosen to be at $x=0$.

A  symmetric solution $Y(x)=Y(-x)$ of (\ref{s1}) with $\varphi(x)=\delta(x)$ can now be constructed as follows:
\begin{equation}\label{209}
Y(x,\lambda) = \frac{3}{\cosh(x+x_0(\lambda)) +1} \quad \mbox{for} \quad x>0\ .
\end{equation}
Inserting into (\ref{s1}) and integrating from $-\delta$ to $\delta$ gives
\begin{equation}
-2\partial _\delta Y(\delta,\lambda) = \lambda + O(\delta) \ .
\end{equation}
Thus, in the limit of $\delta\to 0$, 
\begin{equation}\label{f37}
\frac{\sinh (x_0(\lambda))}{(1+\cosh (x_0(\lambda)))^2} = \frac{\lambda}{6}\ .
\end{equation}
Let us consider  the unique real branch, s.t.\ $x_0(\lambda)\to \infty$ when $\lambda \to 0$, which gives $Z^\phi(\lambda)\to 0$ in the same limit. Increasing $\lambda$ from $0$, the solution breaks down, when 
$\lambda$ reaches $\lambda_c$ with 
\begin{equation}
\lambda_c = \frac{2}{ \sqrt 3} \ .
\end{equation}
such that $\cosh(x_0(\lambda_c))=2$. 
Now we need $Z^\phi (\lambda)$ defined in (\ref{integr}),  or the rescaled version $\tilde Z^\phi( \lambda)= \frac1{\int_x \phi (x)} \int_x \tilde Z_y^\phi ( \lambda)$: 
\begin{eqnarray}
\tilde Z^\phi ( \lambda) &=&\frac1{\int_x \phi (x)} \int_x \tilde  Z_x^\phi (\lambda) \nn\\
&=&\frac1{\int_x \varphi (x)} \int_x (-\nabla^2 + 1) Y(x ,\lambda) \ ,
\end{eqnarray}
where from the first to the second line we switched to dimensionless variables. 
Inserting (\ref{s1}) and using that $\varphi(x)=\delta(x)$ yields
\begin{eqnarray}
\tilde Z^\phi( \lambda)&=&  \int_x \lambda \varphi(x) + Y^2(x,\lambda) \nn\\
&=& \lambda +  \int_{0}^\infty \rmd x\,  \frac{18  }{[1+\cosh( x+ x_0(\lambda) )]^2} \nn\\
&=&  \lambda + 12   \frac{1+3 e^{x_0(\lambda)}}{ [1+e^{x_0(\lambda)}]^3}  \nn\\
&=& \frac{12}{1+\rme^{x_0(\lambda)}}\ ,
\end{eqnarray}
where in the last line  (\ref{f37}) was used. Therefore $x_0(\lambda)$ can be expressed in terms of $\tilde Z^\phi(\lambda)$, and inserted into (\ref{f37}), with the result
\begin{equation} \label{resphi}
\lambda = \frac{\tilde Z^\phi (\tilde Z^\phi-6)(\tilde Z^\phi-12)}{72}\ .
\end{equation}
\begin{figure}
\fig{1}{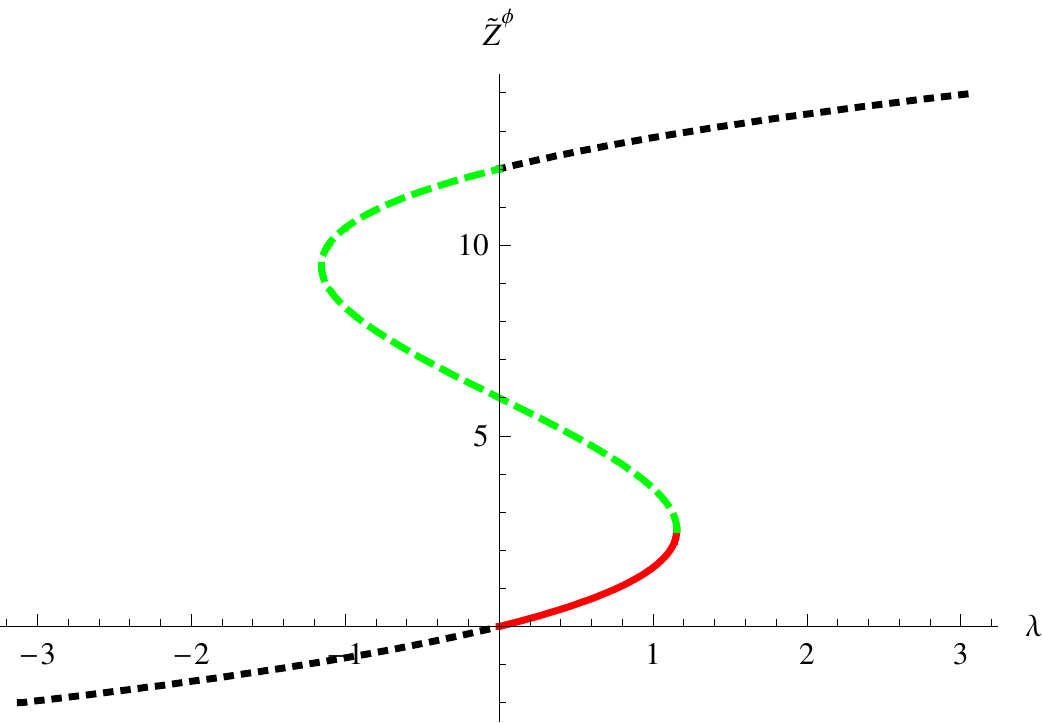}
\caption{$Z^\phi(\lambda)$ for $\phi(x)=\delta(x)$, as explained in the main text.}
\label{Zlambda}
\end{figure}%
Note that this result was derived for $\lambda \in [0, \lambda_c]$, $\tilde Z^\phi \in [0,6-2 \sqrt 3]$, which corresponds to the red (solid) branch in figure \ref{Zlambda}, and then (implicitly) continued analytically to all $\lambda$ and all $\tilde Z^\phi$. We note that the green part of the curve (long dashes) can also be obtained analytically, using the second point for which (\ref{f37}) is satisfied, and the two solutions for negative $\lambda$. One easily checks that the symmetry $Y(x)=Y(-x)$ implies point-reflection symmetry around the point $(\lambda, \tilde Z^\phi)=(0,6)$. This already strongly suggests that $\lambda$ is a third order polynomial in $\tilde Z^\phi$, symmetric around the point  $\tilde Z^\phi=6$ (in our chosen units).

For large negative $\lambda$, $|\tilde Z^\phi(\lambda)| \sim \sqrt[3]{|\lambda|}$, thus from our discussion following (\ref{a52}), $p_c=\frac13$, and the avalanche exponent $\tau$ is 
\begin{equation}\label{tauphi}
\tau^\phi = \frac 4 3\ .
\end{equation}
Note that this value is consistent with a generalized conjecture that we can put forward, $\tau^\phi = 2 - \frac{2}{d_\phi + \zeta}$ where $d_\phi=d-1$ for a codimension  one subspace, inserting $d=4$ and $\zeta=0$ since the above result was derived at mean-field level. Again, much work remains to validate or invalidate this generalized conjecture (e.g. a two loop calculation). 

Interestingly, the probability distribution is non-trivial and different from the standard mean field one (i.e for $\phi=1$). It takes the form, for $S^\phi \gg S^\phi_{\mathrm{min}}$:
\begin{align}
P^\phi(S^\phi) = \frac{1}{S_m} p^\phi\Big(\frac{S^\phi}{S_m}\Big) 
\end{align}
Here the avalanche sizes are defined as $S^\phi_i=\frac{1}{m} \int_{x_2,..x_d} S^i_{x_1=0,x_2,..x_d} \sim m^{-d-\zeta}$ and $S_m$ is the same quantity as defined in 
(\ref{b188a}). The distribution $p^\phi(s)$ can be calculated as follows:
\begin{align}
&{\left<s\right>}^{-1}  p^\phi(s) = \int_{-i \infty}^{i \infty} \frac{\rmd \lambda }{2\pi i} \rme^{-\lambda s} \tilde Z^\phi (\lambda) \nn\\
&  \qquad = \int _{-i \infty}^{i \infty} \frac{\rmd Z }{2\pi i }\frac{\rmd \lambda(Z)}{\rmd Z} \rme^{-\lambda(Z) s} Z  \nn\\
& \qquad=\frac{18}\pi \int_0^\infty \rmd x \, (3 x^2+1)\nn\\
& \qquad \ \qquad \times \!\left[x \sin(3 s x
  (x^2+1))-\cos(3 s x
  (x^2+1)) \right]\!\!\nn\\
  & \qquad =\frac{2 K_{\frac{1}{3}}\!\!\left(\frac{2 s}{\sqrt{3}}\right)}{\pi  s}\ .
\label{217}
\end{align}
Here $s=S^\phi/S_m$ and $\left<s\right>=\left<S^\phi\right>/S_m$, and $K_{\frac13}$ is a Bessel function. The steps of the derivation are: a change of variables from $\lambda$ to $Z =\tilde Z^\phi(\lambda)$; a change of variables $Z = 6 i (x-i)$ and combining the integrand for $x$ and $-x$. 
For large $s$ the asymptotic behavior is similar to the standard mean-field result $P(s)\sim s^{-3/2} \rme^{-\lambda_c s}$: 
\begin{equation}\label{218}
p^\phi(s) = \left<s\right> e^{-\frac{2 s}{\sqrt{3}}} \left(\frac{\sqrt[4]{3}
   \left(\frac{1}{s}\right)^{3/2}}{\sqrt{\pi }}-\frac{5
   \left(\frac{1}{s}\right)^{5/2}}{48 \left(\sqrt[4]{3} \sqrt{\pi
   }\right)}+\ldots \right)\ .
\end{equation}
This was expected, since in both cases the solution for $Z(\lambda)$ ends at a $\lambda_c$ with a square-root singularity. 
For small $s$ the asymptotic behavior is in accordance with (\ref{tauphi}):
\begin{equation}
\left<s\right>^{-1} p^\phi(s) =\frac{\sqrt[6]{3} \Gamma \left(\frac{1}{3}\right)}{\pi  s^{4/3}}+\frac{\Gamma
   \left(-\frac{1}{3}\right)}{\sqrt[6]{3} \pi  s^{2/3}}+\frac{\sqrt[6]{3}
   \Gamma \left(\frac{1}{3}\right) s^{2/3}}{2 \pi }+\ldots
\end{equation}
Finally, from (\ref{217}) one obtains the moments:
\begin{equation}
\frac{\left<(S^\phi)^n\right>}{\left< S^\phi \right> } = \frac{3^{n/2} \Gamma \left(\frac{n}{2}-\frac{1}{6}\right) \Gamma
   \left(\frac{n}{2}+\frac{1}{6}\right)}{2 \pi }\  S_m^{n-1}
\end{equation}
for $n>1/3$, from which the universal ratios $r_n$ can be computed. In particular that $\left<(S^\phi)^2\right>/\left< S^\phi \right>=\frac{1}{2} S_m$ instead of $2 S_m$ for $\phi=1$. 

Note that for $\phi$ localized on a hyperplane, (\ref{EINS}) implies 
\begin{equation}\label{ZWEI}
\left< S^\phi\right> = \frac 1{L m} \left<S\right> \ .
\end{equation}
This is understood from the observation, that only a fraction $1/(Lm)$ of all avalanches leads to an advance of the interface (avalanche) in the hyperplane. Thus, if one were to define an avalanche distribution by considering only the hyperplane, one would naturally choose a different normalization.

\subsection{Diagrammatic expansion}

Here we have solved the codimension one case, i.e. $d_\phi=d-1$. More generally one may consider the case $d_\phi=d-d'$, i.e. a function $\varphi(x)$ localized on a $d' \leq d$ dimensional hyper-plane. This amounts to study Eq. (\ref{s1}) in dimension $d'$. This can be done graphically, as (\ref{s1}) can be used to define, in any dimension $d'$, a  diagrammatic expansion:
\begin{align}\label{221}
Z(\lambda) &= \lambda + \lambda^2 \textdiagram{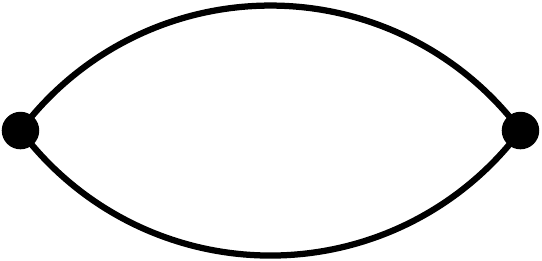} + 2 \lambda^3 \textdiagram{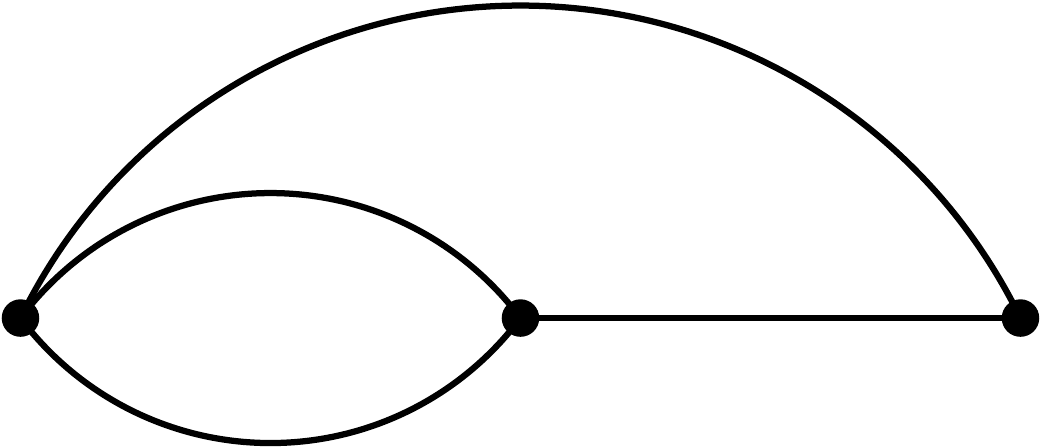}\nn\\
& + \lambda^4 \left[\textdiagram{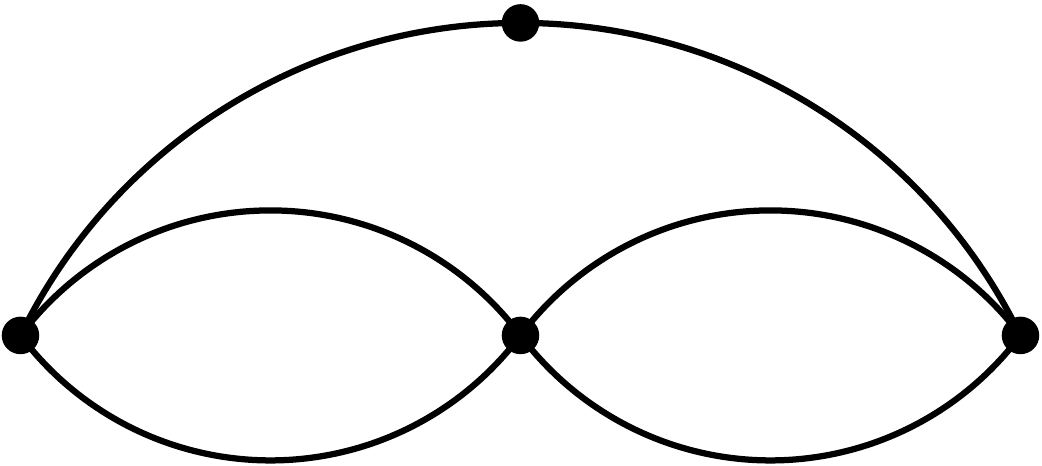}+4 \textdiagram{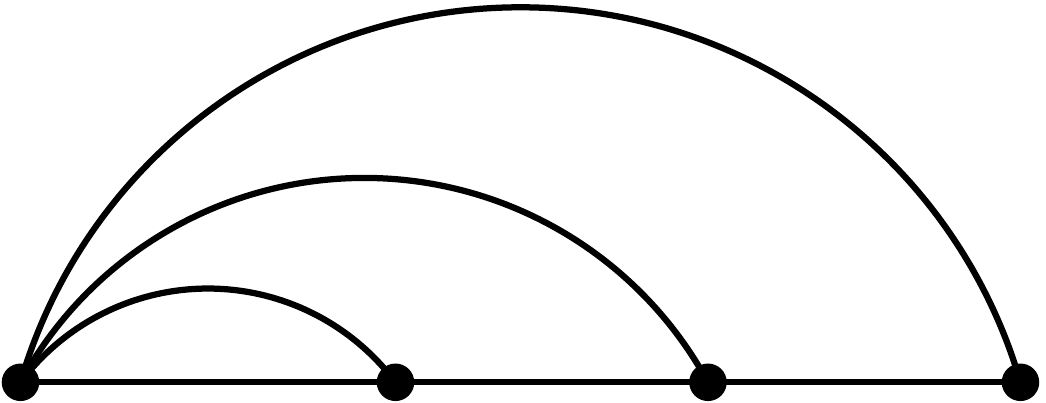}  \right] \nn\\
& + O(\lambda^5)\ ,
\end{align}
where independent $d'$-dimensional momenta flow in each loop, the solid lines are propagators $\tilde g_k=1/(k^2+1)$ and each loop contains a momentum integration. Note that this expansion is valid, i.e. IR- and UV-convergent, in all dimensions $d'<4$ (hence any $d>0$ for the original problem). In $d'=0$ it reproduces (\ref{b16}).
In $d'=1$ the integrals are easily computed by switching from Fourier to real space where $\tilde g_x=\frac{1}{2} e^{- |x|}$, and one obtains
\begin{equation}
Z(\lambda) = \lambda + \frac1{4} \lambda^2 + \frac19{\lambda^3} + \left[\frac{11}{864} + 4\frac{83}{6912} \right] \lambda^4 + \ldots \ .
\end{equation}
This is in agreement with the analytical result (\ref{resphi})
\begin{eqnarray}
Z(\lambda)&=& \lambda +\frac{\lambda ^2}{4}+\frac{\lambda ^3}{9}+\frac{35 \lambda ^4}{576}+\frac{\lambda^5}{27}+\frac{1001 \lambda^6}{41472}+\frac{4 \lambda ^7}{243}\nn\\
&& +\frac{46189 \lambda ^8}{3981312}+\frac{55 \lambda ^9}{6561}+\frac{5311735 \lambda^{10}}{859963392}+O\left(\lambda ^{11}\right)\nn\\
 \end{eqnarray}
It is amazing that the quite non-trivial series (\ref{221}) is indeed resummed by the instanton solution,  (\ref{209}) in the case $d'=1$. Study of
higher $d'$ and extension to the loop expansion is left for the future.

\section{Non-local elasticity and avalanche distributions for a contact line}
\label{s:non-local}

The calculations performed in this paper can be extended to a broader class of elasticity functions $g_q^{-1}$ in  (\ref{model0g}). One possible generalization is
\begin{eqnarray}
g_q^{-1} = |q|^\gamma + \mu^{\gamma} \ ,  \label{firstchoice}
\end{eqnarray}
where $\gamma=2$ corresponds to the choice studied above of local elasticity, while the case $\gamma=1$ is relevant for long-range elasticity as in contact lines of fluids. In fact a more realistic form in that case is
\begin{eqnarray}  \label{contactel} 
g_q^{-1} = \sqrt{q^2 + \mu^2}\ .
\end{eqnarray}
The mass is provided by capillarity, i.e.\ the interplay of surface tension and gravity.
We keep the notation $m^2 = g_{q=0} = \mu^{\gamma}$ for the curvature of the quadratic well, while distinguishing it from $\mu$ which is an inverse characteristic internal  length in the interface (the inverse capillary length). One parametrization which contains these three cases is
\begin{eqnarray}
g_q^{-1} = \mu^{\gamma} g(q/\mu)  \label{gfunct}
\end{eqnarray}
for some function $g(y) \sim y^\gamma$ at large $y$, and $g(0)=1$. Of course other scales may be present in a realistic problem, and the form (\ref{gfunct}) assumes that $\mu^{-1}$ is the single largest internal length scale which cuts off the avalanches. 

The upper critical dimension $d_c$ in all cases is given by the UV divergence of
\begin{eqnarray}
I_2 = \int_q g_q^2 = C_{d,\gamma} \mu^{-\epsilon} \frac{1}{\epsilon}\ ,
\end{eqnarray}
where $\epsilon=d_c-d$, $d_c=2 \gamma$. The constant $C_{d,\gamma}=\epsilon \tilde I_2$ with $\tilde I_2=\int_q g_q^2|_{\mu=1}$ remains finite at $d_c$. 

With these choices all calculations of this paper are easily extended. Appendix \ref{app:review} can still be used, up to trivial changes, e.g.\ replacing $m$ by $\mu$ in all final formula there, including in the definition (\ref{defresc}) of the rescaled disorder (we can set $K=1$). One has again
\begin{equation}
\alpha = - \tilde \Delta''(0^+) = - \frac{\epsilon}{3} (1- \zeta_1) + O(\epsilon^2) 
\end{equation}
to one loop,  with the same values for $\zeta_1$, independent of $\gamma$ (and $g(y)$), and of course now $\epsilon=d_c-d$ everywhere. For more details, including the calculation of the exponent $\zeta$ (and $z$ for the depinning) to two loops for the non-local elasticity see Ref.\ \cite{LeDoussalWieseChauve2002,LeDoussalWieseChauve2003}. Here we compute the avalanche distribution to one loop.

To do that one starts again from Eq.\ (\ref{ZrecFinalInt}). It generalizes as follows:
\begin{align}
&\tilde Z (\lambda) =: \tilde Z = \lambda+\tilde Z ^{2} +\alpha {\cal J}_{\gamma}
\\
& {\cal J}_{\gamma} =\frac{1}{\epsilon \tilde I_2} \int_{k} \frac{\tilde Z^{2}}{(\tilde g_k^{-1} -2 \tilde Z)^{2}}
 + \frac{\tilde Z}{\tilde g_k^{-1} -2 \tilde Z}- \tilde Z \tilde g_k - 3 \tilde Z^2 \tilde g_k^2 \label{general}
\end{align}
where $\tilde g_k=g_k|_{\mu=1}$. Note that the definition of the rescaled $\tilde Z$ is still given by
(\ref{newtZ}) with now
\begin{equation}
 S_m = m^{-4} \Delta'(0^+) = \mu^{-2 \gamma}  \Delta'(0^+) = (\epsilon \tilde I_2)^{-1} \tilde \Delta'(0^+) \mu^{-d+\zeta} .\end{equation}
The calculations are performed in Appendix \ref{app:nonlocal}. For the choice of elasticity (\ref{firstchoice}) we find again, for any $\gamma$, the distribution $P(S)$ given
by (\ref{final}), with amplitudes
\begin{eqnarray} \label{amp2} 
&& A = 1+ \frac{1}{4 \gamma} (2-3 \gamma_{\mathrm{E}} ) \alpha \quad , \quad B = 1-\alpha(1+\frac{\gamma_{\mathrm{E}}}{2 \gamma}) \nonumber  \\
&& C=- \frac{1}{\gamma} \sqrt{\pi} \alpha\ ,
\end{eqnarray}
and exponents
\begin{eqnarray}
 \tau&=&\frac{3}{2} + \frac{3}{4 \gamma} \alpha = \frac{3}{2} + \frac{1}{4 \gamma} (1-\zeta_1) \epsilon + O(\epsilon^2) \label{taug} \\
 \delta &=& 1 - \frac{\alpha}{2 \gamma} = 1 + \frac{1}{6 \gamma}  (1-\zeta_1) \epsilon + O(\epsilon^2)\ .
\end{eqnarray}
Of course everywhere $\epsilon=2 \gamma-d$. Note that the 1-loop result for the avalanche exponent $\tau$ is now compatible  to $O(\epsilon)$ with the generalized conjecture
\begin{eqnarray}
 \tau= 2 - \frac{\gamma}{d + \zeta} \ .
\end{eqnarray}
This conjecture can again be reexpressed as
\begin{eqnarray}
 \rho= 2 \ ,
\end{eqnarray}
given that the scaling form (\ref{b2c}) generalizes into
\begin{equation}\label{b2k}
 \rho(S)=L^d m^\rho S^{- \tau} \tilde \rho(S \mu^{d+\zeta})\ ,
\end{equation}
and that the STS relation (\ref{sts2}) still holds, hence now the exponent
relation (always true) reads: $\rho=2 (2-\tau)(d+\zeta)/\gamma$.

In the case $\gamma=1$ and for the form (\ref{contactel}) more suitable to describe the contact
line, one finds again the same exponent $\tau$ (\ref{taug}), however the shape of the distribution is different.
We find, see Appendix \ref{app:nonlocal},
\begin{eqnarray}\label{finalnonloc} \nn
P(S) &=& A' \frac{\left<S\right>}{2 \sqrt{\pi}} S_m^{-2}  \left[\left(\frac{S}{S_m}\right)^{-\tau}  + D'\right] \\
&& \times  \exp\!\left(C' \sqrt{\frac{S}{S_m}} -
\frac{B'}{4} \left[\frac{S}{S_m}\right]^{\delta'}\right)
\end{eqnarray}
for $S \gg S_{\mathrm{min}}$, with amplitudes
\begin{eqnarray} \label{amp3} 
&& A' = 1+ \frac{1}{4} (8-3 \gamma_{\mathrm{E}} ) \alpha \quad , \quad B' = 1- \frac{3}{2} \alpha(\gamma_{\mathrm{E}}-2) \nonumber  \\
&& C'=- \frac{3}{2} \sqrt{\pi} \alpha \quad , \quad D'=  \frac{1}{8} \sqrt{\pi} \alpha\ .
\end{eqnarray}
The exponent $\delta$ is now given by
\begin{eqnarray} \label{amp4} 
&& \delta'=  1 - \frac{3 \alpha}{2} = 1 + \frac{1}{2}  (1-\zeta_1) \epsilon + O(\epsilon^2)\ .
\end{eqnarray}
Note that the presence of the constant $D'$ suggests that the pre-exponential power law at large $S$ is not $S^{-\tau}$ any more.

\section{Conclusion}

In conclusion we have succeeded in computing from the FRG the distribution of shock or jump sizes which occur in the minimal energy configuration of an interface pinned in a random potential and tied to a spring of varying position. These are the static analog of the avalanches which occur in the dynamics if the interface is instead pulled from a metastable configuration to the next one. Hence it opens the way to the same calculation in the dynamics, performed in \cite{FedorenkoLeDoussalWiesePREP}, which yields very similar results. Shocks in the statics are often called static avalanches as there are many analogies, as well as  some differences. 

We believe that this is an important achievement. First because the FRG has been around for a while but it was not understood previously how to extract the avalanche statistics in a controlled way. In the sandpile literature this is still an open question despite many exact results for other quantities. It turns out to be conceptually simple (a posteriori) to extract these distributions from the FRG. In fact the beautifully simple relation (\ref{secmom}) between the cusp of the FRG function $\Delta(u)$ and the avalanche-size's second moment, unveiled in this work, gives a very transparent physical picture of the cusp. Similar relations hold for all moments, and the challenge is to sum them. This part is a priori technically difficult but some surprising simplifications occur in the calculation, which in the end lead to a simple self-consistent equation, with an interpretation in graph theory. This suggests a, yet to be discovered, simpler and presumably more powerful structure behind the present state of the art of the theory. 

The 1-loop result for the probability distribution  is equally striking. First the avalanche exponent $\tau$ is found equal to order $O(\epsilon)$ to a conjectured form, put forward for the depinning transition. Note that our result is a first-principle derivation of this exponent. Hence the conjecture is confirmed to one loop. It is then of high interest to look for possible deviations to two loop, both in the statics and dynamics. Alternatively, if the conjecture is true in general, it would be interesting to derive this from first principles in the field theory. Besides the exponent, we obtained the general formula for the distribution $P(S)$. This includes the precise way in which it is cut off for  large avalanches, at a cutoff scale $S_m$. These large and rare events are  the important ones in terms of moments of distributions, and, in real life, if one is interested in e.g.\ earthquakes. We predict the scale $S_m$ and the distribution, which is insensitive to details of the model at {\it short} scales (up to a single and measurable global factor). Of course the distribution depends on the details of the model at {\it large} scales, here we use mainly a quadratic well to cut the size off, but this dependence can explicitly be computed within the theory. We illustrate this point by computing it for contact-line depinning which has a more complicated elastic energy.

Some progress was made to study the spatial structure of avalanches. Additional definitions of local avalanche sizes was given which integrate information about jumps within a subspace of the interface. It was shown that to obtain their probability distribution at mean field level one must solve an instanton problem in a cubic field theory. An explicit solution was found for a subspace of co-dimension one, leading to a novel exponent $\tau=4/3$ at mean-field level, i.e. in $d=4$, and a novel size distribution involving a Bessel function. It is different from the 
usual mean field size distributions, recovered here, with $\tau=3/2$, and an exponential function.

There are many open  interesting questions which can now be addressed, both within the statics and the dynamics. First, one would like to know more about the spatial structure of the avalanches, and their correlations. A first step is to extend the instanton calculation to regions of arbitrary shapes and co-dimension, and to study it within the loop expansion. The problem can also be extended to a manifold with $N$ components. Consequences for hysteresis loops of magnets deserve a new study. Applications to  earthquakes are also of interest, especially if avalanche correlations can be handled (pre- and after-shocks). Plastic avalanches can now be studied, e.g. in the framework of Ref.~\cite{LeDoussalMarchettiWiese2008}.

\acknowledgements

We specially thank Andrei Fedorenko for many useful and stimulating discussions and ongoing collaborations, as well as Alan Middleton and Alberto Rosso. We are grateful to Yoshua Feinberg for useful discussions. 
This work was supported by ANR under program 05-BLAN-0099-01.
We thank KITP for hospitality and NSF for partial support under grant number
PHY05-51164. We are indebted to Air France for multiple flight cancellations during a strike, which provided ample free time for
crucial advances in the early stage of the work. 

%\vspace*{3cm}

%\newpage

\appendix

\section{Shock-size distributions}

\label{app:generalizations}

\subsection{Generating function}
We start again from the shock decomposition Eq.\ (\ref{shockdec2}). The field $\rme^{\lambda L^d [u(w)-w]}$ has jump discontinuities at each shock position, where its value is multiplied by a factor $\rme^{\lambda S_i}$. Hence one can write
\begin{align}
& (\partial_w + \lambda L^d) \,\rme^{\lambda L^d [u(w)-w]} \nn \\
& = \sum_i (\,\rme^{\lambda S_i} -1) 
\,\rme^{\lambda L^d [u(w_i^-)-w]} \delta(w-w_i) \label{eqs1} \\
& = \sum_i (1 - \rme^{- \lambda S_i})
\,\rme^{\lambda L^d [u(w_i^+)-w]} \delta(w-w_i) \label{eqs2}
\end{align}
where $L^d u(w_i^-) = \sum_{j<i} S_j$ and $L^d u(w_i^+) = \sum_{j \leq i} S_j$.
The labels of the shocks are ordered according to their spatial position $w_i$.
Let us multiply these equations by the field $\rme^{\lambda L^d [u(w)-w]}$ taken at a different point, i.e.\ consider
\begin{align}
& (\partial_{w_1} + \lambda  L^d ) \,\rme^{\lambda L^d [u(w_1)-w_1 - (u(w_2)-w_2)]}
\nn \\
& = \sum_i (\rme^{\lambda S_i} -1)
\,\rme^{\lambda L^d [u(w_i^-)-w_1]} \delta(w_1-w_i) \nn \\
&\qquad \ \ \ \times \rme^{- \lambda L^d [u(w_2)-w_2]}\ ,
\label{twopoint1}
\end{align}
where $w_1$ and $w_2$ are generic points and different from the $w_i$ (even though the notation might suggest otherwise). If we consider the limit $w_2=w_1^-$ in (\ref{twopoint1}) and average over disorder the term containing $u(w_i^-)$ cancels and one obtains:
\begin{align}
& \lim_{w_2 \to w_1^-} (\partial_{w_1} + \lambda  L^d) \,\overline{\rme^{\lambda L^d [u(w_1)-w_1 - (u(w_2)-w_2)]}} \nn\\
&\qquad= \overline{\sum_i (\rme^{\lambda S_i} -1) \delta(w_1-w_i)}\nn\\
&\qquad= \int \rmd S \rho(S) (\rme^{\lambda S}-1)\ .
\label{twopoint2}
\end{align}
A similar equation can be derived starting from (\ref{eqs2}) and considering the limit $w_2=w_1^+$. Using translational invariance one finally obtains
\begin{align}  \label{derw} 
& \partial_w \overline{\rme^{\lambda L^d [u(w)-w - u(0))]}}\Big|_{w=0^+} = \int \rmd S \rho(S) (\rme^{\lambda S}-1 - \lambda S) \\
& \partial_w \overline{\rme^{\lambda L^d [u(w)-w - u(0))]}}\Big|_{w=0^-} = \int \rmd S \rho(S) (1 - \rme^{- \lambda S} - \lambda S)\ ,
\end{align}
where we have used (\ref{sts}). One thus recovers Eq.\ (\ref{gen1}) in the text.
Another method to derive the part proportional to $|w|$ is to multiply (\ref{eqs1})  at two different points (and setting $\lambda\to -\lambda$ for the second) and extract the single shock contribution:
\begin{align}
& (\partial_{w_2} + \lambda  L^d) (\partial_{w_1} + \lambda  L^d)\, \overline{\rme^{\lambda L^d [u(w_1)-w_1 - (u(w_2)-w_2)]}} \nn\\
&\qquad= \delta(w_1-w_2) \overline{ \sum_i (\rme^{\lambda S_i} -1)(\rme^{-\lambda S_i} -1) \delta(w_1-w_i)}\nn\\
& \qquad  \hphantom{=\ } + \text{smooth}\ .
\end{align}
This yields, using translational invariance:
\begin{align}
&  (- \partial^2_w + \lambda^2  L^{2 d})\,  \overline{\rme^{\lambda L^d (u(w)-w - u(0)))}} \nn\\
&\qquad = 2 \delta(w) \int \rmd S \rho(S) [1 - \cosh(\lambda S)] + \text{smooth}
\label{twopoint3}\ .
\end{align}
Integrating twice one recovers Eq.\ (\ref{gen1}), but only the coefficient of
$|w|$ is determined. To determine the coefficient of $w$ one needs another equation, e.g.\ as above, or has to use the symmetry under $w\to -w$.

Note that this can be generalized as follows: Consider a set of $\lambda_k$ with $\sum_{k=1}^n \lambda_k = 0$, then one has:
\begin{align}
& \prod_{k=1}^n (\partial_{w_k} + \lambda  L^d)\, \overline{\rme^{L^d \sum_{k=1}^n \lambda_k (u(w_k)-w_k)}}\nn\\
&\quad = \delta(w_1-w_2) \dots  \delta(w_1-w_n) \nn\\
& \qquad \times \int \rmd S \rho(S) \prod_{k=1}^n (\rme^{\lambda_k S} -1)  + \text{less singular}
\end{align}

\subsection{Multi-shock size distribution}
The multi-shock size distributions are encoded in the higher pinning force
cumulants (\ref{chatdef}). For instance the third cumulant can be written as:
\begin{align}
&- \partial_{w_1}\partial_{w_2}\partial_{w_3} \hat C(w_1,w_2,w_3)\\
& = m^6 L^{2 d} \overline{(u'(w_1)-1)(
u'(w_2)-1) (u'(w_3)-1)}^c \nn \\\nn
& = m^{6} L^{-d} \int \rmd S S^3 \rho(S) \delta(w_1-w_2) \delta(w_1-w_3) \\\nn
&  + m^{6} L^{-d} \bigg[ \int \rmd S_1 \rmd S_3 S_1^2 S_3 \rho_c(S_1,S_3,w_1,w_3) \delta(w_1-w_2) \nn\\ \nn
& \qquad \qquad \qquad +\text{2 perm} \bigg] \\
& +  m^{6} L^{-d} \int \rmd S_1
\rmd S_2 \rmd S_3 S_1 S_2 S_3 \rho_c(S_1,S_2,S_3,w_1,w_2,w_3) \nonumber
 \\
& \rho_c(S_1,\ldots,S_n,w_1,\ldots,w_n) = \sum_{i_a {\mathrm{~all~ distinct}}}\nn\\
&\  \overline{ \delta(S_1-S_{i_1}) \cdots
\delta(S_n-S_{i_n}) \delta(w_1-w_{i_1}) \cdots \delta(w_n-w_{i_n})}^c
\end{align}
which is easily generalized to any cumulant. Note that the connected parts and the $-1$
substractions conspire, using (\ref{sts}), to give the correct
connected shock distributions (such that scaling with volume is
straighforward). Such formula show explicitly the structure of the sigularities expected in the force  $n$-cumulants as $p$ points are brought together, as a consequence of the assumption of a finite density of shocks (dilute shocks). In principle it allows to check this assumption by computing all moments within the $\epsilon$ expansion.

The generating function generalizing $Z$, and  allowing to extract  $\rho_c(S_1,S_2,w_1,w_2)$, which is a function of $w_1-w_2$, is constructed in analogy with (\ref{gen1}) and (\ref{gen2}):  \begin{widetext}
\begin{eqnarray}
G(\lambda_1,\lambda_2,w_1,w_2,\delta_1,\delta_2) &=&  Z(\lambda_1,\lambda_2,w_1-w_2)\delta_1 \delta_2 + \ldots\\
G(\lambda_1,\lambda_2,w_1,w_2,\delta_1,\delta_2) &=& L^{-d} \overline{ \left(\rme^{\lambda _1 L^d [u(w_1+\delta_1)-u(w_1)-\delta_1]}-1\right)
\left(\rme^{\lambda _2 L^d[ u(w_2+\delta_2)-u(w_2)-\delta_2]}-1\right)
}^c
\end{eqnarray}
for $\delta_1,\delta_2>0$
\end{widetext}

%\end{widetext}

\section{Review of basic FRG results}
\label{app:review}

We review basic equations for functional RG, mostly at 1-loop order, making explicit  the dependence on the elastic coefficient $K$, which usually is set to one. Details (mostly at $K=1$) can be found in \cite{LeDoussalWieseChauve2002, LeDoussalWieseChauve2003}. 
The FRG equation for the function $\Delta(u)$ to one loop is:
\begin{equation}
- m \partial_m \Delta(u) = - \frac{1}{2} (- m \partial_m I_2) [(\Delta(u)-\Delta(0))^2]''\ ,
\end{equation}
where $I_n=\int_k (K k^2+m^2)^{-n}$, with an implicit UV cutoff $\Lambda$. One has $- m \partial_m I_2=4 m^2 I_3$. One should distinguish $d<6$ and $d>6$. The latter is dominated by the UV cutoff and is briefly discussed at the end.
 For $d<6$, in the limit $m \ll \Lambda$ one has $I_3=\tilde I_3 m^{d-6} K^{-d/2}$ with $\tilde I_n=\int_k
(k^2+1)^{-n}$ (defined with infinite UV cutoff whenever convergent). Thus $\tilde I_3=(\epsilon \tilde
I_2)/4$ with $\tilde I_2 = (4 \pi)^{-d/2} \Gamma(2-\frac{d}{2})$. Hence for $d<6$:
\begin{eqnarray}
- m \partial_m I_2=4 m^2 I_3 = m^{d-4} (\epsilon \tilde I_2) K^{-d/2}
\end{eqnarray}
where the combination $\epsilon \tilde I_2=2 (4 \pi)^{-d/2} \Gamma(3-\frac{d}{2})$ is well defined for all $d<6$ (the pole at $d=4$ is suppressed). One defines the rescaled (dimensionless) function $\tilde \Delta(u)$ through
\begin{eqnarray}  \label{defresc}
\Delta(u) = \frac{K^{d/2} }{\epsilon \tilde I_2} m^{\epsilon - 2 \zeta} \tilde \Delta(u m^\zeta)
\end{eqnarray}
with $\epsilon \tilde I_2=1/(8 \pi^2)$ in $d=4$. It satisfies the dimensionless FRG equation:
\begin{eqnarray}
 - m \partial_m \tilde \Delta(u) &=& (\epsilon - 2 \zeta) \tilde \Delta(u) + \zeta u \tilde \Delta'(u)\nn \\
&& - \frac{1}{2} \left[(\tilde \Delta(u)-\tilde  \Delta(0))^2\right]'' \qquad \label{resc}
\end{eqnarray}
This equation is valid for all $d<6$ and yields a fixed point $\tilde \Delta^*(u)=O(\epsilon)$ for $d<4$ with $\zeta=\zeta_1 \epsilon$ and a few universality classes. The two loop equation, not reproduced here was also analyzed. We recall the main results:

\smallskip

(a) random-bond class:
\begin{align}
& \zeta = 0.208298042 \epsilon +  0.006858 \epsilon^2+O(\epsilon ^3) \\
 &\frac{[- \tilde \Delta^{* \prime}(0^+)]^2}{{ \tilde \Delta^*(0)}} =0.583405 \epsilon +0.294205 \epsilon ^2+O(\epsilon ^3)\ . 
\end{align}
Since (\ref{resc}) is invariant under $\tilde \Delta(u) \to \xi^{-2} \tilde \Delta(\xi u)$, 
$\tilde \Delta(u)$ contains one non-universal scale, and only the above ratio is universal.
\smallskip

(b) random-field class, with $R(u)=-\sigma |u|$ at large $u$:
\begin{align}
& \zeta = \frac{\epsilon}{3} \\
& \tilde \Delta^{*}(0) = \frac{\epsilon}{3} \xi^2 \quad , \quad  - \tilde \Delta^{* \prime}(0^+) = \frac{\epsilon}{3} \xi \left[1+\frac {2\epsilon}9 + O(\epsilon^2)\right] \nn
\end{align}
where $\xi$ is the non-universal scale introduced above. Integrating $\Delta(u)$ from $0$ to $\infty$ yields \cite{LeDoussalWieseChauve2002}, Eq.\ (4.39): 
\begin{align}
 K^{-d/2} (\epsilon \tilde I_2) \sigma=\frac{\epsilon}{3} \xi^3 \left[ \gamma_1 + \epsilon \gamma_2 +O(\epsilon^2)\right]
\end{align}
  with $\gamma_1=\int_0^1 \rmd y\,
\sqrt{2(y-1-\ln y)}=0.775304245188$, $\gamma_2=-0.13945524$ (see \cite{LeDoussalWieseChauve2002}, Eq.\ (4.41)).

\smallskip

(c) periodic class with period $u=a$:
\begin{eqnarray}
 \zeta &=& 0 \\
 \tilde \Delta^*(0) &=&{a^2}\left[ \frac{\epsilon }{36}+\frac{\epsilon ^2}{54}+O(\epsilon ^3)\right] \\
 - \tilde \Delta^{* \prime}(0^+) &=& {a} \left[\frac{\epsilon }{6}+\frac{\epsilon ^2}{9}+O(\epsilon ^3)\right]
\end{eqnarray}
For all classes:
\begin{eqnarray}
 \tilde \Delta^{* \prime \prime}(0^+) = \epsilon (1-\zeta_1)/3 + O(\epsilon^2) \label{secder} 
\end{eqnarray}
In dimension $d=4$ one defines: 
\begin{equation}
 \tilde \Delta(u) = \hat \Delta(u \ln(m_0/m)^{-\zeta_1}) (\ln(m_0/m))^{-1+2\zeta_1}
\end{equation}
and setting $\zeta=0$ and $\epsilon=0$ in (\ref{resc}) one finds:
\begin{eqnarray}\nn
 \ell \partial_\ell \hat \Delta(u) &=& (1 - 2 \zeta_1) \hat \Delta(u) + \zeta_1 u \hat \Delta'(u) \\
&& - \frac{1}{2} \left[(\hat \Delta(u)- \hat \Delta(0))^2\right]''
\end{eqnarray}
where $\ell=\ln(m_0/m)$. Hence $\hat \Delta(u)$ converges to the same fixed point $\hat \Delta^*(u) = \tilde
\Delta^*(u)/\epsilon$ as for $\epsilon>0$. This yields for the original function $\Delta(u)$ in $d=4$: 
\begin{equation}
 \Delta(u) = 8 \pi^2 K^2 \hat \Delta^*(u \ln(m_0/m)^{- \zeta_1}) (\ln(m_0/m))^{-1+2 \zeta_1}
\end{equation}
up to subdominant terms of the order of $1/\ln(m_0/m)$. Here $1/m_0$ is a non-universal scale, presumably (at least) of the order of the Larkin scale. In dimension $4<d<6$, we set again $\zeta=0$. The equation for $\tilde \Delta''(0)$ reads:
\begin{eqnarray}
&& - m \partial_m \tilde \Delta''(0) = - (d-4) \tilde \Delta''(0) - 3 \tilde \Delta''(0)^2
\end{eqnarray}
for an analytic disorder ($\Delta''(0)<0$). Hence if the bare disorder is smooth one needs $|\tilde
\Delta_0''(0)| > (d-4)/3$ to generate a cusp and metastability. Hence for sufficiently strong bare smooth
disorder, or with (even weak) rough bare disorder, a cusp is generated. Eventually as $m \to 0$ the flows
converges back to the attractive fixed point $\tilde \Delta(u)=0$ as:
\begin{eqnarray}
 - m \partial_m \tilde \Delta(u) &=& -(d-4) \tilde \Delta(u) \\
 \tilde \Delta(u) &=& \left(\frac{m}{m_0}\right)^{d-4} \tilde \Delta^*(u) \\
 \Delta(u) &=& \frac{K^{d/2}}{ \epsilon \tilde I_2 m_0^{d-4}} \tilde \Delta^*(u)
\end{eqnarray}
where now $\tilde \Delta^*(u)$ is {\it non-universal} and depends on details of the FRG flow at intermediate stages. It has a cusp, unless one starts from smooth weak disorder. Note that for $d \geq 4$, if we make the natural assumption (as for $d<4$) that no other strong-disorder fixed point exist, the asymptotic behaviour is exactly given by the 1-loop FRG equation (in $d=4$) and by perturbation theory {\it in the renormalized disorder}, for $d>4$, i.e:
\begin{equation}
 \Delta(u) = \Delta_0(u) - \frac{1}{2} I_2 \left[(\Delta_0(u){-}\Delta_0(0))^2\right]'' + O(\Delta_0^3)  \label{pert}
\end{equation}
with $I_2 \approx (\epsilon \tilde I_2) (\Lambda^{d-4}-m^{d-4})/(d-4)$ (setting from now on $K=1$ for
simplicity). This equation is valid if we choose a model with weak and rough bare disorder, i.e. whose correlator exhibits a 
cusp. Of course, if we choose smooth bare disorder this perturbation formula fails, and one must run the RG to
determine if the system is in the weak-smooth disorder phase where dimensional reduction holds (possibly up to rare events)
or in the non-analytic phase. Finally, for $d>6$ the situation is qualitatively similar up to additional dependence in the UV cutoff.

\section{Graphical interpretation  of ${\cal K} \Gamma[w]$ at 1-loop order}
\label{a:An}
In this appendix, we give an intuitive derivation of the self-consistent equation (\ref{ZrecFinalInt}) at 1-loop order. Two classes of diagrams contribute, ${\cal C}_{1}$ and ${\cal C}_{2}$. We use the diagrammatics explained in Section \ref{sec:graphrep} which can be used both for statics and dynamics. Here we chose to show explicity the arrows of causality. At this order, no difference is expected between static and dynamic shocks. We do not distinguish here between $Z$ and its rescaled version $\tilde Z$ and use loose notations. 
\smallskip

\leftline{\underline{Class ${\cal C}_{1}$:}}

These are the diagrams in the calculation of $Z$ which, before expansion in $w$, are proportional to $\Delta(w)-\Delta(0)$. Therefore in order to get the term of order $w$,  $\Delta(w)-\Delta(0)$ has to be expanded, and all other disorder vertices will be taken in the limit of $w\to 0$.
For the $n$-th cumulant, we obtain (with $\Delta(w)-\Delta(0)$ sitting at the bottom, and $\Delta''$ at the top):
\begin{align}\label{a65}
&{\cal K} \textdiagram{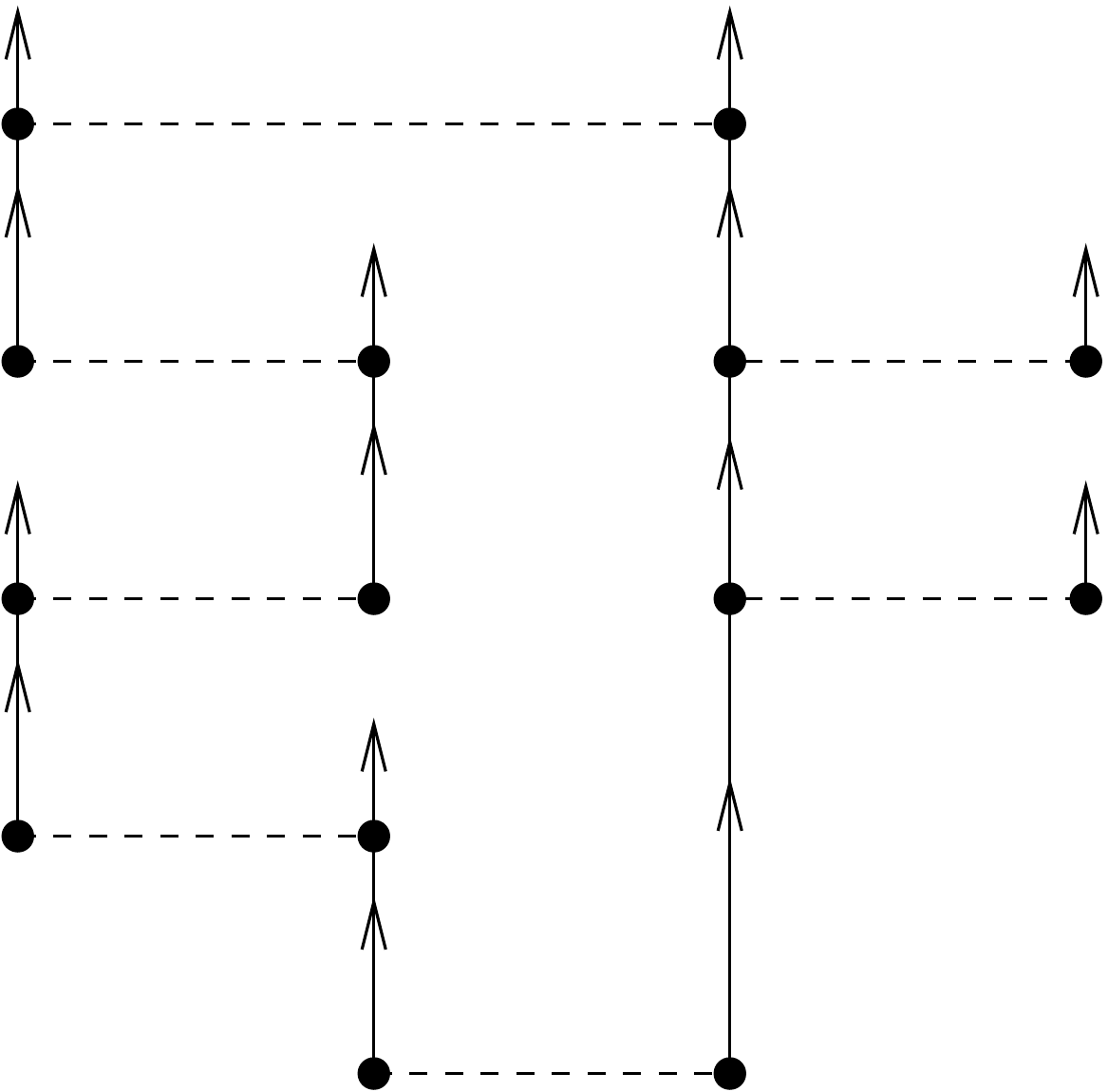} \nonumber \\
&= n! \frac{1}{2}\sum_{l_{1}\ge 0,l_{2}\ge 0}
\delta_{n-2 =l_{1}+l_{2}}
\left[- 2 \Delta' (w) \right]^{l_{1}}\times  \left[- 2 \Delta' (w)
\right]^{l_{2}} \nonumber \\
&\qquad \times \left[\Delta'' (w)+\Delta'' (0) \right]
\left[\Delta (w)-\Delta (0) \right] \nonumber \\
&=  n!\, (n-1)\, 2^{n-3} [- \Delta' (w)]^{n-2} \left[\Delta'' (w)+\Delta'' (0) \right]\nonumber \\
&\qquad \times
\left[\Delta (w)-\Delta (0) \right]\nonumber \\
&=  -  n!\, (n-1)\, 2^{n-2} |\Delta'(0^{+})|^{n-1} \Delta'' (0) w \ .
\end{align}
$l_{1}$ and $l_{2}$ denote the number of $\Delta'$ in the left
and right leg respectively. Rescaling by $S_m$ amounts to set $ |\Delta'(0^{+})| \to 1$. Hence we find that the coefficient $A_n$ defined in the text reads, for this class:
\begin{equation}
A_{n}^{{\cal C}_{1}} = n! \times 2^{n-2} (n-1)\ .
\end{equation}
Since in the self-consistent equation each outgoing line in the above is branched to a $Z$, this implies that the contribution of this class of diagrams to the r.h.s. of the self-consistent equation can be summed into:
\begin{equation}\label{A3}
\sum_{n=2}^{\infty} \frac{A_{n}^{{\cal C}_{1}}}{n!} I_{n} Z^{n} = {Z^{2}}\int_{k} \frac{1}{(k^{2}+m^{2}-2Z)^{2}}\ .
\end{equation}
In the end one sets $m^{2}\to 1$ for rescaled quantities. Starting the sum instead  at $n=3$, as we did in the main text, yields the additional subtraction $-{Z^{2}}\int_{k} \frac{1}{(k^{2}+m^{2})^{2}}$. This procedure identifies the latter term as counter-term of the disorder renormalization, thus is the correct expansion in terms of the renormalized disorder, whereas (\ref{A3}) is the correct result in terms of the bare disorder. 

We can now give a graphical interpretation to the dressed propagator $1/(k^2 + m^2 - 2 Z)$ which appears here.
In the above formulae, the left and right legs are effectively dressed propagators, which we represent as a double line:
\begin{equation}\label{}
\diagram{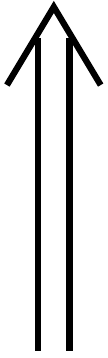} := \diagram{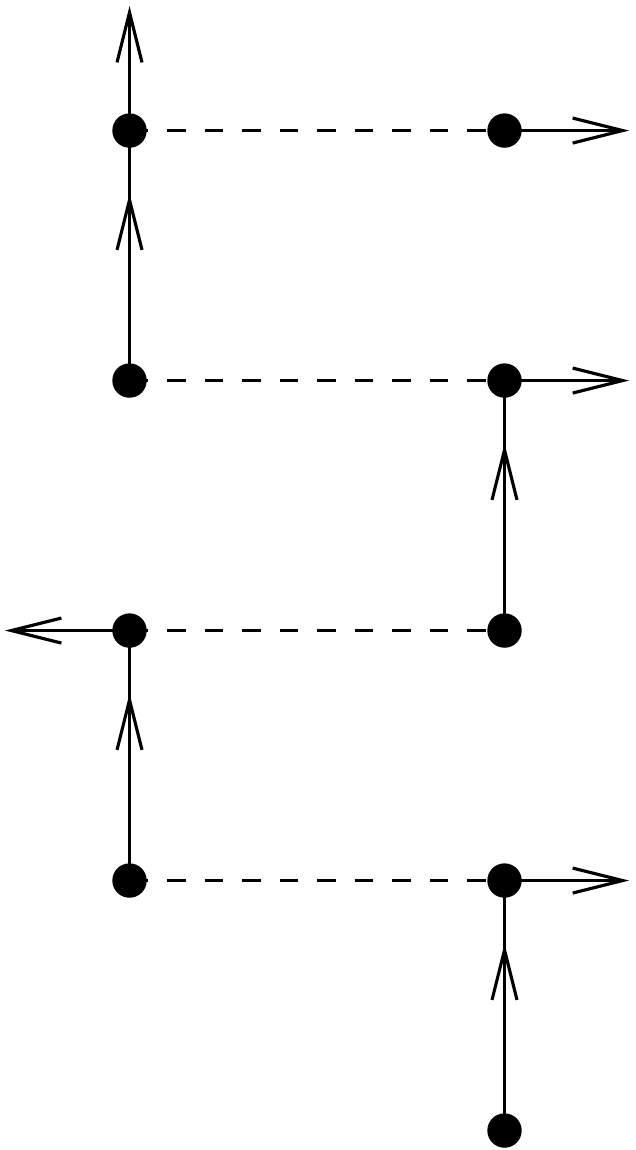}
\end{equation}
The notation on the r.h.s. of the equation is as follows:  The left vertex of each
disorder is at $0$, the right one at $w$ (the choice being dictated so as to have a non-vanishing vertex  under application of the ${\cal K}$ operator). Arrows pointing to the right
or left are going out -- they will be branched to an external point
sitting at $0$ (left) or $w$ (right). From each disorder vertex $\Delta'(w)$, there are 2 outgoing lines, resulting in a combinatorial
factor of 2 for continuing either with the left ($0$) or right ($w$) vertex. Also note that the object exists,
independently of whether the in- and out-going lines are at $0$ or
$w$. Now class ${\cal C}_{1}$ can be written as:
\begin{align}\label{A6}
\displaystyle  {\cal C}_{1} &= \displaystyle \diagram{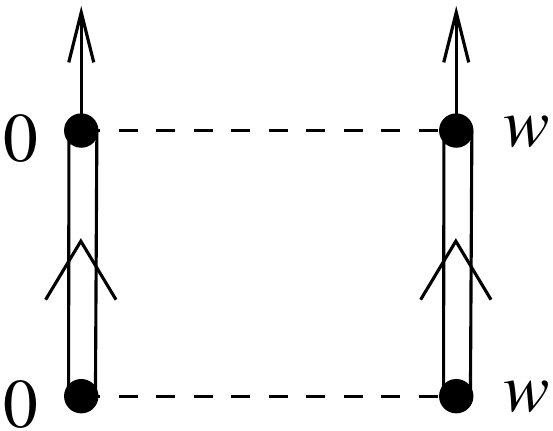}+  \diagram{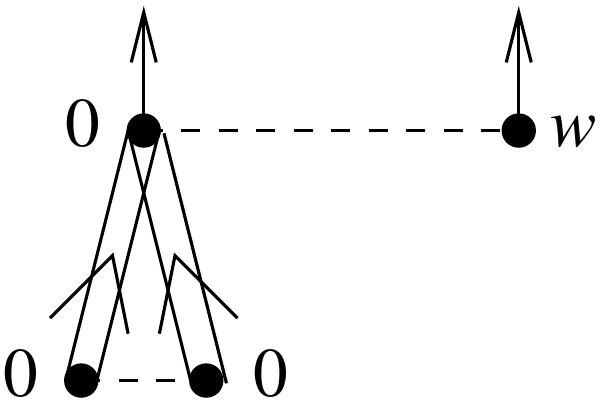}\nonumber \\
& =\displaystyle
\Delta'' (0)[\Delta (w)-\Delta (0)] \int_{k}\frac{Z^{2}}{(k^{2}+m^{2}-2 Z)^{2}} +{\cal O}(w^{2})\ .
\end{align}
(Actually, the graphical notation is a little sloppy, since the upper vertex can either be $\Delta''(w)$ or $\Delta''(0)$, but there is an additional combinatorial factor of $\frac12$ in eq.\ (\ref{a65}), first line).
The result is in agreement with (\ref{A3}).

\smallskip

\leftline{\underline{Class ${\cal C}_{2}$:}}
This class of diagrams looks like a correction to the critical force
\begin{align}\label{A7}
\displaystyle  {\cal C}_{2} &= \displaystyle \diagram{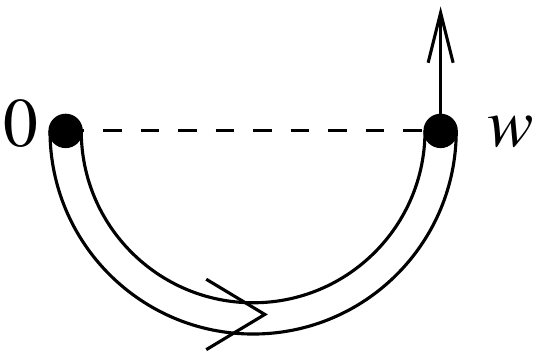}+ \diagram{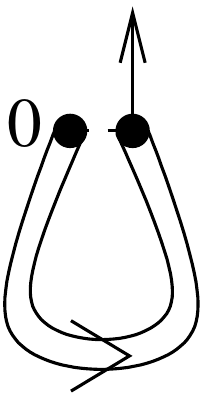} \nonumber \\
& =\displaystyle
[\Delta' (w)-\Delta' (0^{+})] \int_{k}\frac{Z}{(k^{2}+m^{2}-2 Z)}-\frac{Z}{k^{2}+m^{2}}\ .
\end{align}
Indeed it should be viewed as a circle of $\Delta'$, of which exactly one is expanded in $w$, leading to a circle with one marked vertex. This is the vertex drawn above. Note that the double line needs {\em at least one} $\Delta'(w)$ otherwise it cannot start at 0 and go to $w$ as indicated. This leads to the last term in (\ref{A7}) being subtracted.
If one wants the expression in terms of the renormalized disorder, one again has to subtract the contribution proportional to $\int_k \frac 1{(k^2+m^2)^2}$, giving an additional term $-[\Delta' (w)-\Delta' (0^{+})] \int_k \frac {2 Z}{(k^2+m^2)^2}$.

(\ref{A6}) and (\ref{A7}) (with the proper subtraction) together give (\ref{ZrecFinal}).

\section{Moments from $\tilde Z(\lambda)$}

A direct series expansion of the formula (\ref{a61}) in powers of $\lambda$, e.g.\ using mathematica, yields the following formula for the moments ($S$ expressed in units of $S_m$)
\begin{align}
\frac{\left< S^{n}\right>}{\left<S\right>} = \hspace{7mm}&\hspace{-7mm}
\frac{(-4)^{n-1} \sqrt{\pi }}{\Gamma \left(\frac{3}{2}-n\right)} \nn\\
+ \alpha\Big[& -\frac{3 (-1)^n 4^{n-2} \sqrt{\pi } \psi\left(\frac{3}{2}-n\right) (n-1)}{\Gamma
   \left(\frac{3}{2}-n\right)}\nn\\&
   +\frac{(-1)^n 2^{2 n-3} \sqrt{\pi } (n-1)}{\Gamma \left(\frac{3}{2}-n\right)}\nn\\
   &-\frac{3 (-1)^n
   4^{n-2} \sqrt{\pi } (-2+\gamma +\log (4)) (n-1)}{\Gamma \left(\frac{3}{2}-n\right)}\nn\\
   &-4^{n-1} \Gamma (n)-\frac{(-1)^n 2^{2
   n-3} \sqrt{\pi } \psi \left(\frac{1}{2}-n\right)}{\Gamma \left(\frac{1}{2}-n\right)}\nn\\
   &+\frac{(-1)^n 2^{2 n-1}
   \sqrt{\pi }}{\Gamma \left(\frac{1}{2}-n\right)}\nn\\
   &-\frac{(-1)^n 2^{2 n-3} \sqrt{\pi } (-2+\gamma +\log (4))}{\Gamma
   \left(\frac{1}{2}-n\right)}
\Big] \label{huge}
\end{align}
which yields the universal ratios:
\begin{equation}
r_n =1+\frac{2}{2 n-3}+\frac{\alpha  \left(n+\frac{(-1)^n \Gamma \left(\frac{5}{2}-n\right) \Gamma (n-1)}{\sqrt{\pi }}\right)}{(2 n-3)^2}
\end{equation}

We note that these formula make sense only for $n$ integer, and in fact one can check that for integer $n=1,2,..$ they give the same result as the formula in the text obtained from $P(S)$. Because of the factor $(-1)^n$ they are not suited for analytical continuation to any real $n$. However using reflection identities of $\Gamma(x)$ and $\psi(x)=\Gamma'(x)/\Gamma(x)$ functions one can eliminate the factors $(-1)^n$ and get back the formula of the text which are real for all real $n$.

\section{Calculations for non-local elasticity}

\label{app:nonlocal} 

Let us first recover the result for local elasticity $\gamma=2$, $d_c=4$ with $\tilde g_k^{-1}=k^2+1$. One can rewrite the momentum integral defined in the text as:
\begin{eqnarray} \label{J2}
 {\cal J}_{2} &=&  N_{2} \int_{0}^{\infty} ds s^{{d/2-1}} \bigg[ \frac{\tilde Z^{2}}{(s{+}1{-}2 \tilde Z)^{2}}
 + \frac{\tilde Z}{s{+}1{-}2 \tilde Z} \nonumber \\
 && \hphantom{ N_{2} \int_{0}^{\infty} ds s^{{d/2-1}} \bigg[}-\frac{\tilde Z}{s {+}1} - 3 \frac{\tilde  Z^2}{(s+1)^2} \bigg] \ .
 \end{eqnarray}
The constant $N_2$ is defined as
$$
N_{2} = \frac{1}{\epsilon} \frac{1}{ \int_{0}^{\infty} ds  s^{{d/2-1}}\frac{1}{(s+1)^{2}} }
\ .
$$
Explicit calculation yields back (\ref {ZrecFinal}) and for $d \to 4$:
\begin{equation} \label{J2old} 
{\cal J}_{2} = \frac{1}{2} \tilde Z (2 \tilde Z+(1-3 \tilde Z) \log (1-2 \tilde Z))+O(\epsilon)\ ,
\end{equation}
i.e.\ one recovers (\ref{a51}) by a different method, which is also a check of (\ref{general}). 

Let us now study the case of general $\gamma$ using the model (\ref{firstchoice}).
\begin{align}\label{J1a1}\nn
&{\cal J}_{1}^{a}= \frac{1}{\epsilon \tilde I_2}  \int_{k}\bigg[ \frac{\tilde Z^{2}}{(|k|^\gamma+1-2 \tilde Z)^{2}}
 + \frac{\tilde Z}{|k|^\gamma+1-2 \tilde Z}\\
 &\hphantom{{\cal J}_{1}^{a}= \frac{1}{\epsilon \tilde I_2}  \int_{k}\bigg[}-  \frac{\tilde Z}{|k|^\gamma+1} -  3 \frac{\tilde Z^2}{(|k|^\gamma+1)^2}\bigg] \ .
\end{align}
Introducing  $N_{1}^{a} = \frac{1}{2 \gamma-d} \big( \int_{0}^{\infty} \rmd s\,  s^{{d/\gamma-1}}\frac{1}{(s+1)^{2}} \big)^{-1}$, this is computed as:
\begin{align}\label{J1a2}
{\cal J}_{1}^{a}&= N_{1}^{a} \int_{0}^{\infty} ds 
s^{{d/\gamma-1}} \bigg[ \frac{\tilde Z^{2}}{(s{+}1{-}2\tilde Z)^{2}}
 + \frac{\tilde Z}{s{+}1{-}2 \tilde Z} \nn \\
 & \hphantom{= N_{1}^{a} \int_{0}^{\infty} ds 
s^{{d/\gamma-1}} \bigg[}
  -\frac{\tilde Z}{s {+}1} -  3 \frac{\tilde Z^2}{(s+1)^2} \bigg]  \nn\\
  &= \frac{1}{\gamma} \tilde Z (2 \tilde Z+(1-3 \tilde Z) \log (1-2 \tilde Z))+O(2 \gamma-d)
\end{align}
i.e.\ a function identical to (\ref{J2old}) at the critical dimension up to the global multiplication by $2/\gamma$.
Hence the distribution of avalanche sizes to one loop with this choice of non-local elasticity will be exactly given by (\ref{final}) with the replacement of $\alpha \to 2 \alpha/\gamma$ in 
(\ref{amp}) and (\ref{a64}) as detailed in the text.

Let us now study the case $\gamma=1$ and the form (\ref{contactel}) suitable to describe the contact
line. We need to compute:
\begin{align}\label{J1b1}
{\cal J}_{1}^{b}&= N_1^b 
\bigg[  \int_{0}^\infty k^{d-1} \rmd k  \frac{\tilde Z^{2}}{(\sqrt{k^{2}+1}-2 \tilde Z)^{2}}
 \\
 &\hphantom{= N_1^b 
\bigg[  \int} + \frac{\tilde Z}{\sqrt{k^{2}+1}-2 \tilde Z}  -\frac{\tilde Z}{\sqrt{k^{2}+1}} - 3 \frac{\tilde Z^2}{k^{2}+1} \bigg] \nn
\end{align}
where $N_1^b=\big( \epsilon  \int_{0}^\infty \frac{k^{d-1} d k}{k^2+1} \big)^{-1}$, $\epsilon = 2-d$. We will use the following integral representation:
\begin{equation}\label{e:gen}
\frac{1}{\sqrt{1+k^2}-2 \tilde Z}= \int_{\alpha=0}^{\infty}\int_{t=0}^{\infty}\frac{\rme^{-\left(1+k^2\right) \alpha  t^2+2 \tilde Z t-\frac{1}{4 \alpha }}}{2 \sqrt{\pi } \alpha ^{3/2}}
\end{equation}
This yields:
\begin{align}
&N_1^b \int_{k} \bigg[\frac{1}{\sqrt{1+k^2}-2 \tilde Z}-\frac{1}{\sqrt{1+k^2}}\bigg]
=\frac{2 \tilde Z \left(1-4 \tilde Z^2\right)^{\frac{d}{2}-1}}{2-d}\nn\\
& -\frac{\Gamma(\frac{1}{2}-\frac{d}{2}) \Gamma(\frac{d}{2}-1) \,
  [ _2F_1(1,\frac{1-d}{2};\frac{1}{2};4 \tilde Z^2) -1] \sin \left(\frac{d \pi }{2}\right)}{2 \pi ^{3/2}}\ .
\end{align}
Taking a derivative w.r.t.\ $\tilde Z$ yields
\begin{align}
&\int_{k}\frac{1}{(\sqrt{1+k^2}-2 \tilde Z)^{2}}=
\frac{\left(1-4 \tilde Z^2\right)^{\frac{d-4}{2}} \left(4 (d-1) \tilde Z^2-1\right)}{d-2}\nn\\
&\ -\frac{4 \tilde Z \,\Gamma(\frac{3}{2}-\frac{d}{2}) \Gamma
  (\frac{d}{2}-1) \, _2F_1(2,\frac{3}{2}-\frac{d}{2};\frac{3}{2};4 \tilde Z^2) \sin \left(\frac{d \pi }{2}\right)}{\pi ^{3/2}}\ .
\end{align}
Together this gives
\begin{align}\label{J1b9}
{\cal J}_{1}^{b}=& \frac{ \tilde Z^2 \left(4 (d+1) \tilde Z^2-3\right) \left(1-4 \tilde Z^2\right)^{\frac{d-4}{2}}}{d-2} - \frac{3 \tilde Z^2}{d-2} \nn\\
&+\frac{\tilde Z\, \Gamma \left(\frac{1}{2}-\frac{d}{2}\right) \left[\,
   _2F_1 (2,\frac{1}{2}-\frac{d}{2};\frac{1}{2};4 \tilde Z^2)-1\right]}{2 \sqrt{\pi } \Gamma \left(2-\frac{d}{2}\right)} \\
   =& \frac{2  \tilde Z^{3}}{1-2 \tilde Z}-3 \tilde Z^{2} \log (1-2 \tilde Z)+O(2-d)
\label{J1b2}
\end{align}
This result (\ref{J1b2}) has to be compared to (\ref{J1a2}) for $\gamma=1$. One sees that the leading behavior for $\tilde Z\to -\infty$, from which is extracted the exponent $\tau$, is identical hence the avalanche size exponent $\tau$ is still given by (\ref{taug}). The form of the avalanche distribution however will be different. To see this, first invert
\begin{equation}
\tilde Z= \lambda + \tilde Z^{2} + \alpha \left[\frac{2 \tilde Z^{3}}{1-2 \tilde Z}-3 \tilde Z^{2} \log (1-2 \tilde Z)\right]
\end{equation}
This gives
\begin{align}
&\tilde Z_{\gamma=1}^{b}(\lambda)= \frac{1}{2} \left(1-\sqrt{1{-}4 \lambda }\right)\nn\\
&-\alpha
\frac{(\sqrt{1{-}4 \lambda }-1)^2 \left[\frac{3}{2} \sqrt{1{-}4 \lambda } \log (1{-}4
   \lambda )+\sqrt{1{-}4 \lambda }-1\right]}{4(1-4\lambda)}
\end{align}
One can now verify that to first order in $\alpha$ one can write (setting $S_m \to 1$, restored in the text):
\begin{equation}\label{}
\tilde Z_{\gamma=1}^{b}(\lambda)= \int_{0}^{\infty}\rmd S\, \left[\rme^{\lambda S}-1\right] P_{\gamma=2}^{b}(S)
\end{equation}
with
\begin{align}
&P_{\gamma=2}^{b}(S)=\frac1{2\sqrt{\pi}}\left[1+\frac{\alpha}{4} (8-3 \gamma_{\mathrm{E}} ) \right]  \left[S^{-\frac{3}{2}-\frac{3
   \alpha }{4}}+\frac{\sqrt{\pi } \alpha }{8}\right]\nn\\
&\qquad\times\exp\left[{-\frac{1}{4} \left(1-\frac{3}{2} (\gamma_{\mathrm{E}}-2 ) \alpha \right)
   S^{1-\frac{3 \alpha }{2}}-\frac{3}{2} \sqrt{\pi } \alpha  \sqrt{S}}\right]
\end{align}
which is the form given in the text. Note that to first order in $\alpha$ it can equally well be written as
\begin{align}
P_{\gamma=2}^{b}(S)=&\frac1{2\sqrt{\pi}}\left[1+\frac{\alpha}{4} (8-3 \gamma_{\mathrm{E}} ) \right]  S^{-\frac{3}{2}-\frac{3
   \alpha }{4}}\nn\\
&\times\exp\bigg[-\frac{1}{4} \left(1-\frac{3}{2} (\gamma_{\mathrm{E}}-2 ) \alpha \right)
   S^{1-\frac{3 \alpha }{2}}\nn\\
   &~~\qquad\quad-\frac{3}{2} \sqrt{\pi } \alpha  \sqrt{S}+\frac{\sqrt{\pi } \alpha }{8}S^{\frac32}\bigg]\ .
\end{align}
This would suggest an abrupt jump in the effective stretched exponential decay exponent $\delta$
to the value $3/2$. However since the above results are valid for {\it fixed} $S$ at first order in $\alpha$ one cannot conclude on this point at this stage.

%\end{widetext}

\section{More on spatial structure of avalanche distributions}
\label{app:spatial}
\subsection{Derivation of the recursion relation}
\label{a:der-rec}
We want to show in Fourier space, using  $\phi(x)=\cos(q x)$:
\begin{eqnarray} \label{s3}
Z_k(\lambda) &=& \lambda \frac{1}{2} \left[ \delta_{k+q} +  \delta_{k-q} \right] \nn\\
&& +  |\Delta'(0^+)| \int_{p} g_p g_{k-p}
Z_p(\lambda) Z_{k-p}(\lambda)\ ,\qquad
\end{eqnarray}
where $\delta_q=(2 \pi)^d \delta^d(q)$. 

The calculation of $G^{\phi}(\lambda)$ is a generalization of the one of Section \ref{sec:tree} to non-zero momentum graphs. One still takes 
$w$ uniform and now computes the spatially dependent correlations (for $q_i$ all non-vanishing):
\begin{align}
& g_q^{-2} \overline{u_{q}(w_1) u_{q'}(w_2)}^c  =  - R''_{q}[w_1-w_2] \delta_{q+q'}  \\
&  \overline{u_{q_1}(w_1) \cdots u_{q_n}(w_n)}^c  =(-1)^n \overline{\hat V_{q_1}'[w_1] \cdots \hat V'_{q_n}[w_n]}^c \nn\\
& ~~=  (-1)^n \hat C^{(n)}_{q_1,..q_n}(w_1,\ldots ,w_n)  \delta_{q_1+\cdots q_n}\ ,
\end{align}
where $\delta_q=(2 \pi)^d \delta^d(q)$. We use Fourier space versions of the definitions in (\ref{deffunct}). In particular we have defined
\begin{equation}
R''_{q}[w] :=\int_x \rme^{i q x} R''_{0x}[w] \quad , \quad 
R''_{xy}[w]:=\frac{\delta^2 R[w]}{\delta w_x \delta w_y}|_{w_x=w}  \label{rder}
\end{equation} 
using translational invariance for a uniform $w$ and the definition of the $R[w]$ functional \cite{LeDoussal2006b} $\overline{\hat V[w_1] \hat V[w_2]}^c=R[w_1-w_2]$. The $q=0$ limit is recovered using $u_{q=0}(w)=L^d u(w)$. From these $\hat C$ one computes the associated momentum-dependent Kolmogorov moments, applying the operator ${\cal K}$. The choice
$\phi(x)=\cos(q x)$ amounts to choose $q_i=\pm q$ on the external legs of the tree diagrams of Section \ref{sec:tree}, as for the calculation of $|S_q|^{2 n}$, hence we need $q_i=\pm q$.

We extend the previous tree-level analysis and associate $a_{m,m'}$ to the box with $m$ times $+q$ and $m'$ times $-q$ entering from the top, and a
moment $(m-m') q$ exiting from the bottom of the tree diagram. Pasting together the trees as explained in Section \ref{sec:tree} we obtain the recursion relation:
\begin{equation} \label{s11}
a_{m,m'} = \sum_{p+l=m;\,p'+l'=m'} a_{p,p'} a_{l,l'} \frac{m!}{p! l!} \frac{m'!}{p'! l'!} g_{(p-p')q} g_{(l-l')q}
\end{equation}
Introducing $b_{m,m'}:=a_{m,m'}/m! m'!$ one obtains
\begin{align} \label{s1c}
& b_{m,m'} \rme^{i z (m-m') q}= \int_{xy} g_x g_y  \\
& \qquad \times\sum_{p+l=m;p'+l'=m'} b_{p,p'} b_{l,l'} \rme^{i (x+z) (p-p')q} \rme^{i
(y+z) (l-l')q} \ . \nonumber
\end{align}
By definition:
\begin{equation} \label{s1d}
Z_z(\lambda) := \sum_{m,m'} b_{m,m'}  |\Delta'(0^+)|^{m+m'+1} \rme^{i z (m-m') q} \lambda^{m+m'}\ .
\end{equation}
In the sum (\ref{s11}) the terms $1,0$ and $0,1$ are left out, so one finds in real space:
\begin{equation} \label{s2}
Z_z(\lambda) =  \lambda \cos(q z) +  |\Delta'(0^+)| \int_{xy} g_{z-x} g_{z-y} Z_x(\lambda) Z_y(\lambda)
\end{equation}
The Fourier transform of this relation gives (\ref{s3}) and it is a particular case of (\ref{s4}). The general case is obtained by attaching $\phi(x)$ to each external leg. At the end of the calculation one wants $Z_{k=0}(\lambda)$ since no momentum flows from the lower vertex, hence formula (\ref{integr}). Note that we have neglected everywhere the non-local component (i.e.\ the momentum dependence) of the $\Delta'(0)$ vertex which we have taken at zero momentum. Near $d=4$ this should be sufficient, as discussed again below.

\subsection{Generalized cusp-moment relation and check on the second order}

Let us start from the general relation:
\begin{align}
\overline{(u_x(w_1) - w_1)(u_y(w_2) - w_2)}^c =- \int_{zz'} g_{xz} g_{yz'} R''_{xy}[w_1-w_2]
\end{align}
Consider now the expansion to second order $\lambda^2$ of  (\ref{derw2b}). It yields:
\begin{equation}
 G^\phi(\lambda)|_{\lambda^2} = \frac{1}{\int_x \phi(x)}  \int_{xyzz'} \phi(x) \phi(y) g_{xz} g_{y z'}
(R''_{zz'}[w]-R''_{zz'}[0]) 
\end{equation}
Taking a derivative w.r.t. $w$ we obtain:
\begin{equation}
 \hat Z^\phi(\lambda)|_{\lambda^2} = \frac{1}{\int_x \phi(x)}  \int_{xyzz't} \phi(x) \phi(y) g_{xz} g_{y z'} R'''_{zz't}[w=0^+]
\end{equation}
Comparing with its definition, and using the relation $\left<S^\phi\right>=\left<S\right> L^{-d} \int dx \phi(x)$ 
yields the result for the second moment of the avalanche amplitude at different points in internal space:
\begin{equation} \label{nonlocrel}
 \frac{\langle S^x S^y \rangle}{\langle S \rangle} =  2 L^{-d} \int_{zz't} g_{xz} g_{y z'} R'''_{zz't}[w=0^+]
\end{equation}
an exact relation which is the non-local generalization of (\ref{secmom}). 

To leading order in $\epsilon$ the third derivative vertex is local i.e.\ $R'''_{zz't}[w]=\delta_{zz't} R'''(w)$, hence we find simply:
\begin{eqnarray}
\frac{\langle S^x S^y \rangle}{\langle S \rangle} =  2 L^{-d} \int_z g_{xz} g_{y z} R'''(0^+)
\end{eqnarray}
or in $q$-space:
\begin{eqnarray}
 \frac{\langle S^{q} S^{q'} \rangle}{\langle S \rangle} =  2 |\Delta'(0^+)| L^{-d} g_{q}^2 \delta_{q+q'}
\end{eqnarray}
This is what is predicted by the self-consistent equation 
\begin{eqnarray}
\langle (S^\phi)^2 \rangle/\langle S \rangle=2 |\Delta'(0^+)| L^{-d} \int_x (\int_y g_{xy} \phi(y))^2
\end{eqnarray}
a simple generalization of  (\ref{secmom}), valid to lowest order in $\epsilon$ (or in the locality expansion).

\section{Distribution of the center-of-mass fluctuations}
\label{app:u}
One defines the generating function for the probability distribution of the center-of-mass fluctuations as
\begin{equation}
f(\lambda) = L^{-d} \ln \overline{ \exp( \lambda L^d  u(0) ) } = L^{-d} \ln \overline{ \exp( \lambda \int_x u_x(0) ) } \ .
\end{equation}
The result (\ref{cumulants0}) implies that, to one-loop accuracy,
\begin{eqnarray}
 f(\lambda) &=& \frac{\lambda^2}{2}  \frac{\Delta(0)}{m^4} \ -  \int_k \sum_{p=2}^\infty  \frac1{2p} \frac{\Delta'(0^+)^{2p} \lambda^{2p}}{m^{4 p} (k^2+m^2)^{2p}} \nn
\\
&=&  \frac{\lambda^2}{2} \frac{\Delta(0)}{m^4} \\
&& + \frac12 \int _k \left[ \ln \left( 1-\frac{\Delta'(0^+)^{2}\lambda^2}{m^4 (k^2{+}m^2)^{2}} \right) + \frac{\Delta'(0^+)^{2}\lambda^2}{m^4 (k^2{+}m^2)^{2}} \right]  \nonumber 
\end{eqnarray}
Note that the term $p=1$ is contained in the renormalized $\Delta(0)$. 
Since we work to one-loop order, we can now compute the integral in the second term in $d=4$. The calculation yields
\begin{eqnarray}\nn
 f(\lambda) &=& \frac{\lambda^2}{2}  \frac{\Delta(0)}{m^4} \ +  S_4 m^d g(\lambda m^{-4} |\Delta'(0^+)|) \\
 g(b)&=& \frac{1}{8} \big( (1+b)^2 \ln(1+b) + (1-b)^2 \ln(1-b) - 3 b^2 \big) \nonumber \\
& =& -\frac{b^4}{48} - \frac{b^6}{240} - \frac{b^8}{672} - \frac{b^{10}}{1440} + O(b^{12}) \label{G3}
\end{eqnarray}
with $S_4=\epsilon \tilde I_2=1/(8 \pi^2)$ in $d=4$. 
We note that the scale $S_m=m^{-4} |\Delta'(0^+)| = O(\epsilon m^{-d-\zeta})$ naturally appears. 
Indeed we can rewrite the 
result for the generating function near $d=4$, using the rescaled correlator  from appendix \ref{app:review} as
\begin{eqnarray}
 && \overline{ e^{ \lambda  \int_x u_x(0)/S_m } } = \exp\bigg( (m L)^{d} (\epsilon \tilde I_2) \Big[ \frac{\lambda^2}{2} \frac{\tilde \Delta(0) }{\tilde \Delta'(0^+)^2} + g(\lambda) \Big] \bigg) \nonumber \\
 && =  \exp\bigg( \frac{(m L)^{d}}{8 \pi^2}  \Big[ \frac{\lambda^2}{2} \Big(\frac{1}{\epsilon (1- 2 \zeta_1)} + g_2 \Big) + g(\lambda) \Big] \bigg)\ . \label{resu}
\end{eqnarray}
$\lambda$ is now a dimensionless number. The total factor in the exponential is obtained correctly to both orders $O(1/\epsilon)$ and $O(1)$. In the last equation we have used the FRG fixed-point equation for $\tilde \Delta(0)$. The term $g_2$ is zero if one uses the FRG equation to one loop. However to get the $\lambda^2$ term to $O(1)$ requires injecting the ratio $\frac{\tilde \Delta(0) }{\tilde \Delta'(0^+)^2}$ to two loop. These ratio are given in the Appendix and it is easy from there to find the number $g_2$ for each universality class. 

The result (\ref{resu}) seems to indicate \footnote{The integration path for the inverse Laplace transform should be chosen s.t.\ the integrand remains real. For $\epsilon=0$ this is achieved by taking $\lambda=-u/2+i s$, $s\in \mathbb{R}$. For $\epsilon>0$, two cuts appear, from $-1$ to $-\infty$, and from $1$ to $\infty$. Taking $u$ large, only the left one matters. At leading order in $\epsilon$, one can still follow almost the same contour, except that the integral from $-u/2$ to $-1$ below the cut and from $-1$ to $-u/2$ above the cut has to be added. Its non-vanishing part comes from the $2\pi i$ discontinuity of the $\ln$ across the cut, thus does not give a $\ln$ contribution. The final result at leading order is then obtained by replacing $\lambda$ by $\lambda-u/2$, doing the (almost) Gaussian integral in $\lambda$ in imaginary direction, with the result that $P(u) \approx (\ref{resu})|_{\lambda\to -u/2}$. Exponentiating the $\ln$ leads to the quoted result for $\delta$. } that the tail of $P(u)$ decays as $\exp(-|u|^\delta)$, with $\delta=2-\epsilon (1-2\zeta_1) /2<2$, contrary to the $d=0$ result $\delta=3$. It remains to be understood whether $\delta$ first decreases and then increases again for $d\to 0$, or whether the functional form is different, and the assumption that the $\ln$ can be re-exponentiated is unfounded. A larger tail ($\delta<2$) seems to be counter-intuitive, since all connected moments in (\ref{G3}) above the second are {\em negative}.

The main result (\ref{resu}) can be interpreted as a nice example of the central-limit theorem. There are effectively $m L^d$ independent regions \footnote{Note that the above result is for $m L \gg 1$, the result for arbitrary $m L$ can in principle be obtained by modifying the integrals into periodic sums. The opposite limit $m L \to 0$ was studied in \cite{FedorenkoLeDoussalWiese2006}.}. In each region the generating function is given by the same formula  (\ref{resu}) setting the $m L^d$ factor to unity. The final distribution of $u(0)$ is obtained as a convolution of these $m L^d$ identical ones. In each region one can write $u(0)= \sqrt{\epsilon} u_1 + \epsilon u_2 = S_m (\frac{1}{\sqrt{\epsilon}} \tilde u_1 + \tilde u_2)$ where $u_1,u_2$ (resp. $\tilde u_1,\tilde u_2$) are independent random variables of order one, $u_1$ (resp. $\tilde u_1$) has a gaussian distribution, and $u_2$ (resp. $\tilde u_2$) has a non trivial distribution encoded in the function $g(\lambda)$. Hence near $d=4$ the center-of-mass fluctuations are gaussian with a small correction encoded in $g(\lambda)$.

\vfill

%\bibliography{citation}

\tableofcontents

\end{document}